\documentclass{aa}  
\usepackage{natbib}
\usepackage[breaklinks=true]{hyperref} 
\bibpunct{(}{)}{;}{a}{}{,} 
\usepackage{txfonts}
\usepackage{tasks}
\usepackage{tabularx}
\usepackage{float}
\usepackage{graphicx}
\usepackage{amsmath}

\begin{document} 

   \title{VLTI-PIONIER imaging of the red supergiant V602 Carinae\thanks{Based on observations made with the Very Large Telescope Interferometer at the Paranal Observatory under program IDs 097.D-0286, 60.A-9138 (NAOMI science verification) and 2103.D-5029.}}

   \author{J.B. Climent
          \inst{1}\thanks{Email: j.bautista.climent@uv.es}
          \and M. Wittkowski \inst{2}
           \and A. Chiavassa \inst{3}
          \and F. Baron \inst{4}
          \and J.M. Marcaide \inst{1}\thanks{Visiting Professor at the Department of Quantum Physics and Astrophysics, and the Institute of Cosmos Sciences of the University of Barcelona}
          \and J. C. Guirado \inst{1}\fnmsep\inst{5}
          \and B. Freytag \inst{6}
          \and S.~H\"{o}fner \inst{6}
          \and X. Haubois \inst{7}
          \and J.~Woillez \inst{2}
          }

         \institute{Departament d’Astronomia i Astrof\'isica, Universitat de Val\`encia, C. Dr. Moliner 50, 46100 Burjassot, Val\`encia, Spain
         \and
         European Southern Observatory, Karl-Schwarzschild-Str. 2, 85748 Garching bei M\"unchen, Germany
         \and 
         Universit\'e C\^ote d’Azur, Observatoire de la C\^ote d’Azur, CNRS, Lagrange, CS 34229 Nice, France
         \and
         Department of Physics and Astronomy, Georgia State University, PO Box 5060 Atlanta, GA 30302-5060
         \and
         Observatori Astron\`omic, Universitat de Val\`encia, Parc Cient\'ific, C. Catedr\'atico Jos\'e Beltr\'an 2, 46980 Paterna, Val\`encia, Spain
         \and Theoretical Astrophysics, Department of Physics and Astronomy, Uppsala University, Box 516, SE-751 20 Uppsala, Sweden
         \and European Southern Observatory, Casilla 19001, Santiago 19, Chile
        }    

   \date{Received ...; accepted ...}
 
  \abstract
   {Red supergiant stars possess surface features and extended molecular atmospheres. Photospheric convection may be a crucial factor of the  levitation of the outer atmospheric layers. However, the mechanism responsible  is still poorly understood.}
   {We   image the stellar surface of V602 Carinae (V602 Car)   to constrain the morphology and contrast of the surface features and of the extended atmospheric layers.}
   {We observed V602 Car with the Very Large Telescope Interferometer (VLTI) PIONIER instrument (1.53-1.78 $\mathrm{\mu}$m) between May and July 2016, and April and July 2019 with different telescope configurations. We compared the
   image reconstructions
   with 81 temporal snapshots of 3D radiative-hydrodynamics (RHD) CO$^5$BOLD
   simulations in terms of contrast and morphology, 
   using the Structural Similarity Index.}
   {The interferometric data are compatible with an overall spherical disk of angular diameter 4.4$\pm$0.2\,mas, and an extended molecular layer. In 2016, the reconstructed image reveals a bright arc-like feature toward the northern rim of the photospheric surface. 
   In 2019, an arc-like feature is seen at a different orientation and a new peak of emission is detected on the opposite side. 
   The contrasts of the reconstructed  surface images are 11\%$\pm$2\% and 9\%$\pm$2\% for 2016 and 2019, respectively.
   The morphology and contrast of the two  images are consistent with 3D RHD simulations, within our achieved spatial resolution and dynamic range. The extended molecular layer contributes 10--13\% of the total flux with an angular diameter of 6--8 mas. It is present but not clearly visible in the reconstructed images because it is close to the limits of the achieved dynamic range. The presence of the molecular layer is not reproduced by the 3D RHD simulations.}
   {3D RHD simulations predict  substructures similar to the observed surface features of V602 Car at two different epochs.
   We interpret the structure on the stellar surface as being related to instationary convection. This structure is further convolved to larger observed patches on the stellar surface with our observational spatial resolution. 
   Even though the simulations reproduce the observed features on the stellar surface, convection alone may not be the only relevant process that is  levitating the atmosphere.  
   }


   \maketitle
%
\section{Introduction}
\label{sec:intro}
Red supergiants (RSGs) are cool evolved massive stars before their transition toward Wolf-Rayet (WR) stars and core-collapse supernovae. Their characterization and their observed location in the Hertzsprung-Russell (HR) diagram is important in order  to calibrate stellar evolutionary models for massive stars and to understand their further evolution toward WR stars and supernovae \citep[e.g.,][]{2013MNRAS.433.1745D,2013A&A...558A.131G,2014A&A...564A..30G,2014ARA&A..52..487S,2015A&A...575A..60M}. Moreover, red supergiants are of importance in stellar synthesis models because of their high luminosities and high masses \citep[e.g.,][]{2013A&A...552A..92M}. 

The structure and morphology of the close circumstellar environment and wind regions, including the atmospheric molecular layers and dusty envelopes, are currently a matter of intense debate \citep[e.g.,][]{2010ApJ...717L..62Y,2012MNRAS.419.2054W}. Knowledge of the circumstellar envelope and fundamental parameters is important to understand the matching of supernova (SN) progenitors to the different types of core-collapse SNe \citep{2003ApJ...591..288H,2013A&A...558A.131G}. The mass loss from red supergiants is, as well, one of the most important sources for the chemical enrichment of the interstellar medium.

The study of fundamental parameters and atmospheric extensions of RSGs in our neighborhood \citep{2012A&A...540L..12W,2013A&A...554A..76A,2014A&A...566A..88A,2015A&A...575A..50A,2017A&A...597A...9W} has shown that extended molecular atmospheres, with extensions comparable to Mira variable asymptotic giant branch (AGB) stars, are a common feature of RSGs stars and that, unlike for Miras, this phenomenon is not predicted by 3D radiative-hydrodynamics (RHD) or 1D pulsation models \citep{2015A&A...575A..50A}.  

The onset of the mass-loss process, that is the levitation of the outer atmospheric layers to radii where dust can form, is currently not understood for RSG stars. The most commonly proposed mechanism has been an interplay of pulsation and convection \citep[e.g.,][]{2010ApJ...717L..62Y}. \citet{2007A&A...469..671J} suggested that a decrease in the effective gravity, caused by convective motions, combined with radiative pressure on molecular lines, may initiate the mass loss in RSG stars. It was also suggested that magnetic fields could contribute to the heating of the outer atmosphere and to the mass loss \citep{2010A&A...516L...2A}. \citet{2015A&A...575A..50A} showed that current 1D and 3D radiative-hydrodynamics models of pulsation and convection alone cannot levitate the molecular atmospheres of RSGs to observed extensions. They observed a correlation of atmospheric extension with luminosity, which may support a scenario that includes radiative acceleration on Doppler-shifted molecular lines. 
However, there are alternative mechanisms such as magnetic fields and Alfv\'en waves \citep[e.g.,][]{2010ApJ...723.1210A,2011ApJ...741...54C,2012MNRAS.422.1272T,2019ApJ...882...37R,2019ApJ...879...77Y}, differential rotation \citep{2018A&A...613L...4V}, or the presence of giant dominating hot spots (recently observed by \citealt{2016A&A...588A.130M}).
Although the processes that initiate the mass loss from RSG stars are not currently known, it is well established that RSG stars show mass-loss rates between 2 $\times$ $10^{-7}$ M$_{\rm \odot}$yr$^{-1}$ to 3 $\times$ $10^{-4}$ M$_{\rm \odot}$yr$^{-1}$ \citep{2010A&A...523A..18D}.

In this work our aim is to characterize the effects of convection on the stellar surface and to investigate the role that convection may play in the mass-loss process of RSGs. We compare VLTI-PIONIER image reconstructions of the stellar surface of V602 Carinae (V602 Car) 
with predictions by 3D simulations of stellar convection.
We chose V602 Car as our main target, which was part of our previous sample of VLTI-AMBER studies, i.e., targets for which we had already established the fundamental stellar parameters and the presence of extended molecular atmospheres. \citet{2015A&A...575A..50A} reported for V602 Car a radius of 1050$\pm$165\,R$_\odot$, an effective temperature of 3432$\pm$280\,K, a surface gravity $\log g$ = -0.30$\pm$0.16, and an initial mass of 20-25\,M$_\odot$ corresponding to a current mass of 10-13\,M$_\odot$.

\begin{table}[b]
    \centering
    \caption{Observation log of V602 Car with the instrument PIONIER.}
    \label{table:logs}
    \begin{tabular}{cccrr} 
    \hline\hline
    Date & Stations & Conf.$^{a}$ & Seeing & Coh. time \\
         &          &                     & (\arcsec) & (msec)  \\
    \hline
    2016-04-07 & A0/G1/J2/J3 & L & 0.47 & 7.6\\
    2016-05-23 & A0/B2/C1/D0 & S & 0.63 & 4.4\\    
    2016-05-24 & A0/B2/C1/D0 & S & 0.44 & 5.4\\
    2016-05-25 & A0/B2/C1/D0 & S & 0.47 & 3.3\\
    2016-05-31 & D0/G2/J3/K0 & M & 0.66 & 3.5\\
    2016-06-01 & A0/G1/J2/J3 & L & 0.66 & 2.5\\
    2016-06-27 & A0/B2/C1/D0 & S & 0.62 & 3.1\\\hline
    2019-04-29 & A0/D0/G1/J3 & L & 0.75 & 5.3\\
    2019-05-02 & A0/G1/J2/J3 & L & 0.57 & 6.7\\
    2019-05-03 & A0/G1/J2/J3 & L & 0.46 & 13.5\\
    2019-05-04 & A0/G1/J2/J3 & L & 0.52 & 4.9\\
    2019-05-10 & A0/B2/C1/D0 & S & 1.09 & 2.2\\
    2019-05-30 & A0/B2/C1/D0 & S & 0.77 & 2.5\\
    2019-05-31 & A0/B2/C1/D0 & S & 0.70 & 2.6\\
    2019-07-07 & A0/G2/J2/J3 & L & 0.47 & 6.3\\
    2019-07-08 & D0/G2/J3/K0 & M & 0.46 & 6.9\\
    \hline
    \end{tabular}
    \begin{flushleft}
    \footnotesize{\textbf{Notes.} $^{a}$Short configuration (S): AT stations A0/B2/C1/D0, ground baselines 10-40 m; Medium configuration (M): D0/G2/J3/K0, 40-100 m; Long configuration (L): A0/G1/J2/J3, A0/G2/J2/J3, and A0/G2/J2/J3, 60-140m.}
    \end{flushleft}
\end{table}
    
\section{Observations and data reduction}\label{sect:observations}

\begin{figure}
    \centering
    \includegraphics[width=\linewidth]{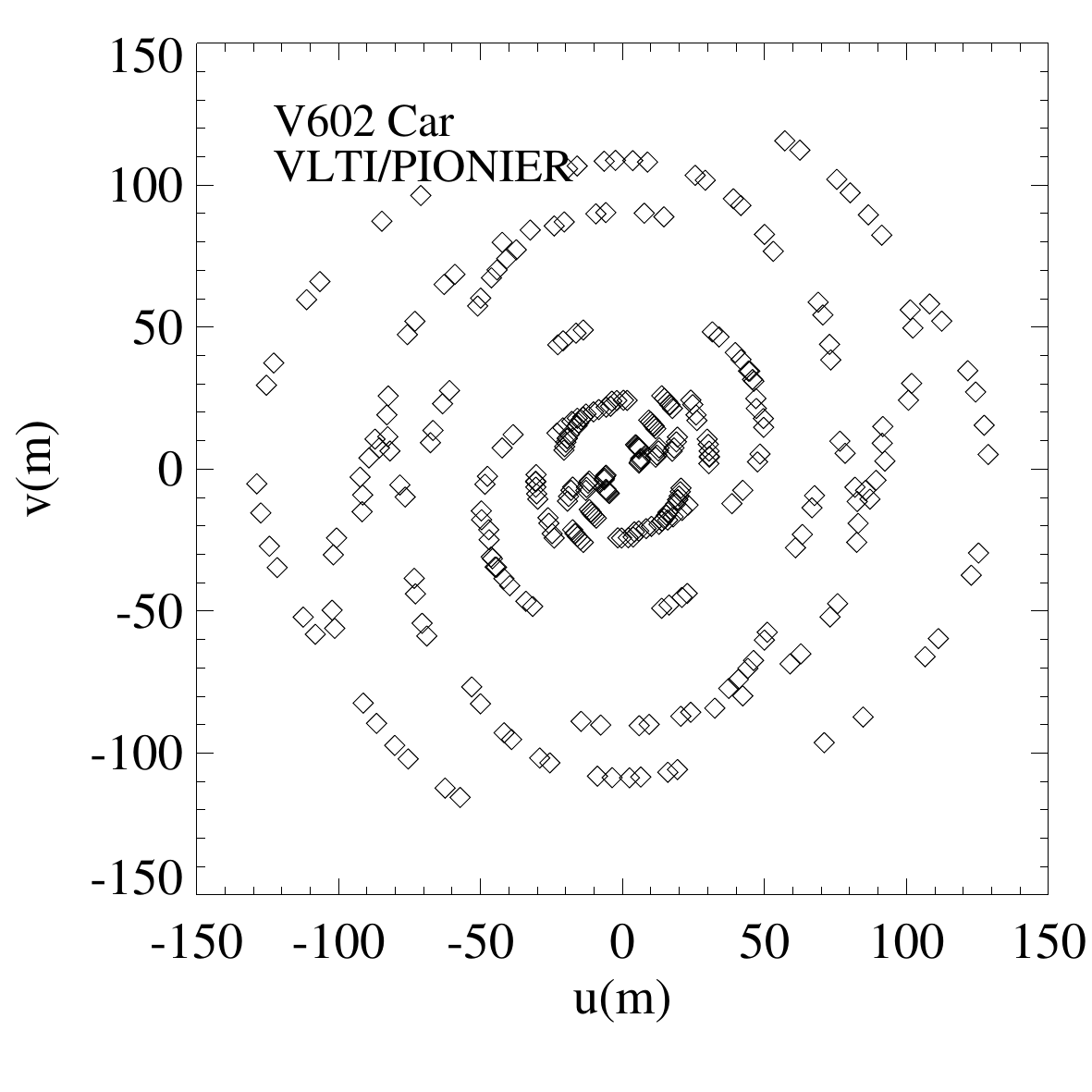}
    \includegraphics[width=\linewidth]{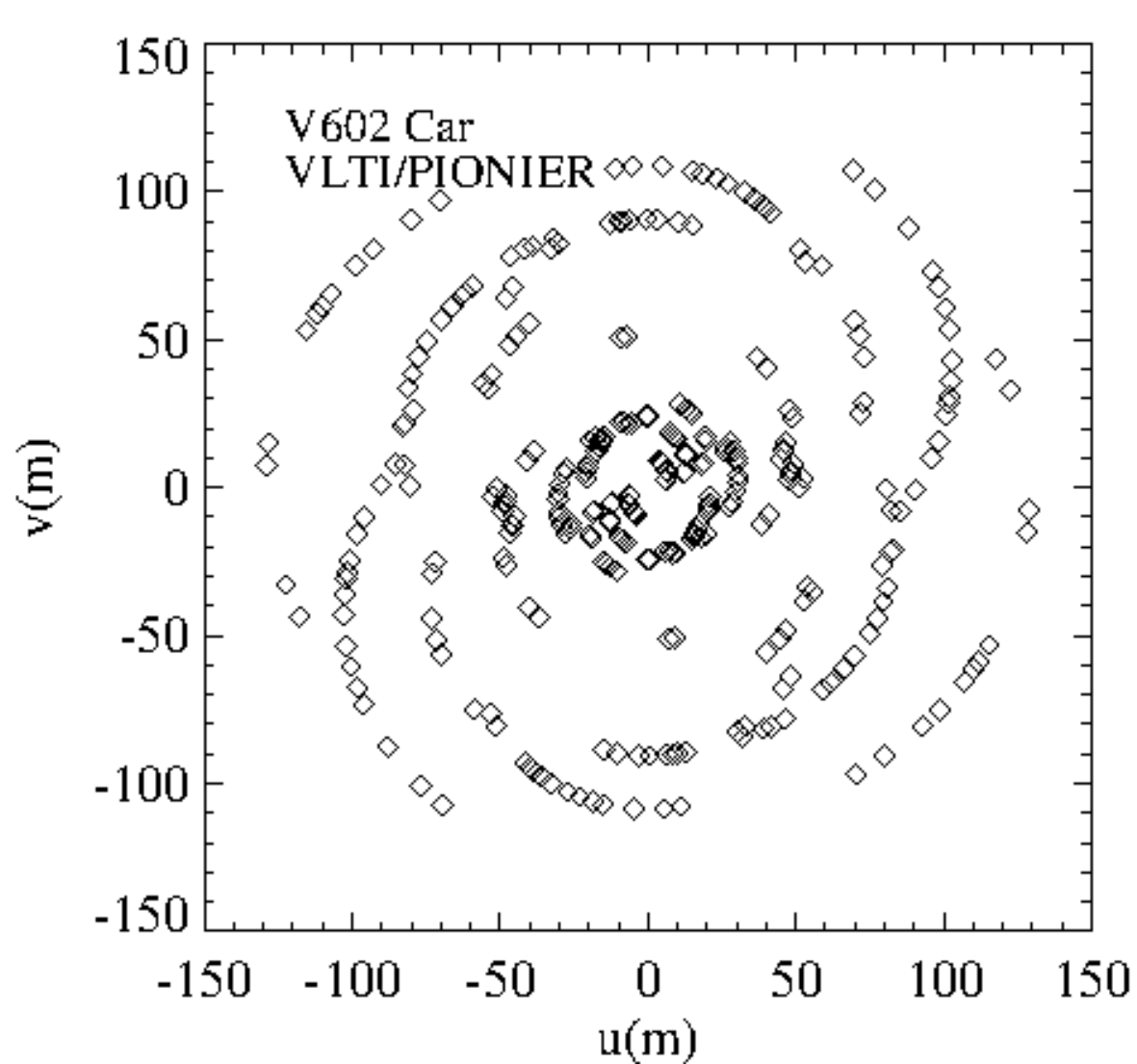}
    \caption{
    Map of the $uv$ coverage of our PIONIER observations of V602 Car (upper, 2016; lower, 2019), where $u$ and $v$ are the spatial coordinates of the baselines projected on sky.
    }
    \label{fig:uvplane}
\end{figure}

We obtained interferometric observations of V602 Car employing the PIONIER instrument \citep{2011A&A...535A..67L} of the Very Large Telescope Interferometer (VLTI) and its four auxiliary telescopes (ATs). The ATs were placed in three different effective configurations: short, medium, and long (see Table \ref{table:logs}). Observations were taken using the  ESO service mode between 7 April 2016 and 27 June 2016, and between 29 April 2019 and 8 July 2019. The 2019 data were taken using the new NAOMI adaptive optics system \citep{2019A&A...629A..41W} at the ATs during NAOMI science verification, providing an improved precision and accuracy compared to 2016 (see Appendix~\ref{sec:compNAOMI} for details). The data were dispersed over six spectral channels with central wavelengths 1.53\,$\mu$m, 1.58\,$\mu$m, 
1.63\,$\mu$m, 1.68\,$\mu$m, 1.72\,$\mu$m, 1.77\,$\mu$m and widths of $\sim$0.05\,$\mu$m. Observations of V602~Car were interleaved with observations of the interferometric calibrator HD~96566 with spectral type G8III and an angular uniform disk diameter of 1.50~$\pm$~0.11 mas \citep{2010yCat.2300....0L}. A log of our observations can be found in Table~\ref{table:logs}. We initially divided the 2016 observation dates into three
sub-epochs, where each epoch lasted not more than 9 days, because the V602~Car is a semi-regular variable. However, an analysis of the different sub-epochs showed, within our accuracy and
spatial resolution, that there was no significant variability of the visibility data over the sub-epochs, so that in the following we analyzed the data of all sub-epochs together. We did not repeat the exact $uv$ coverage within the full epoch so that variability on small scales might be present and would be smeared by combining the data.
The 3D convection models of RSGs by \citet{2009A&A...506.1351C} showed time variations of surface structures on timescales  of one month in the H-band. However, we show  in Sect.~\ref{sect:comparison_morphology} that snapshots of convection simulations are similar at our spatial resolution, albeit not identical, on timescales of about 3 months, justifying our approach of combining the sub-epochs.
The same reasoning was applied for the 2019 observations. 
The total $uv$ coverage that we obtained for our observations is very similar for 2016 and 2019, as can be seen in Fig. \ref{fig:uvplane}. 

\begin{figure*}
    \centering
    \includegraphics[width=0.49\linewidth]{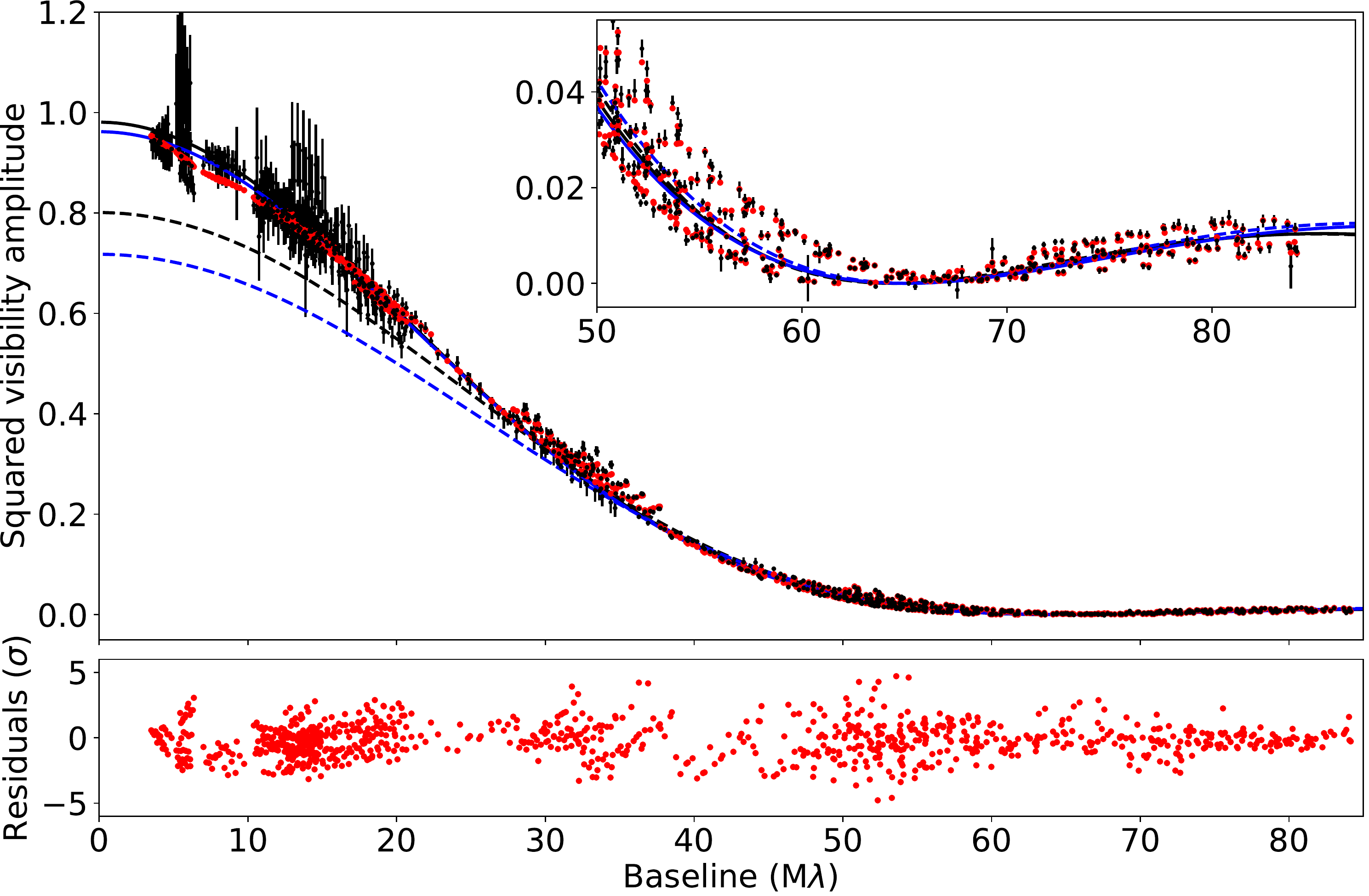}
    \includegraphics[width=0.49\linewidth]{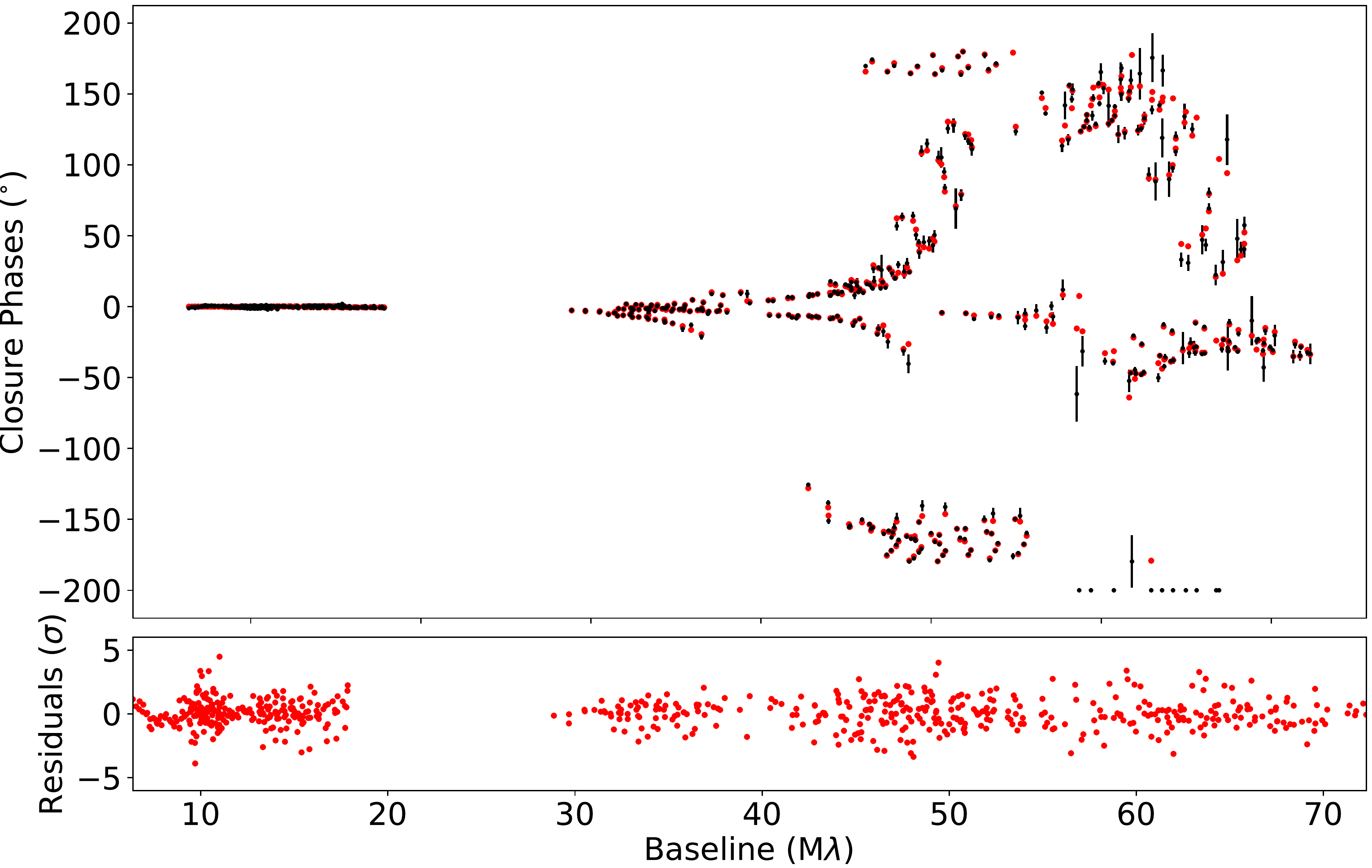}
    \includegraphics[width=0.49\linewidth]{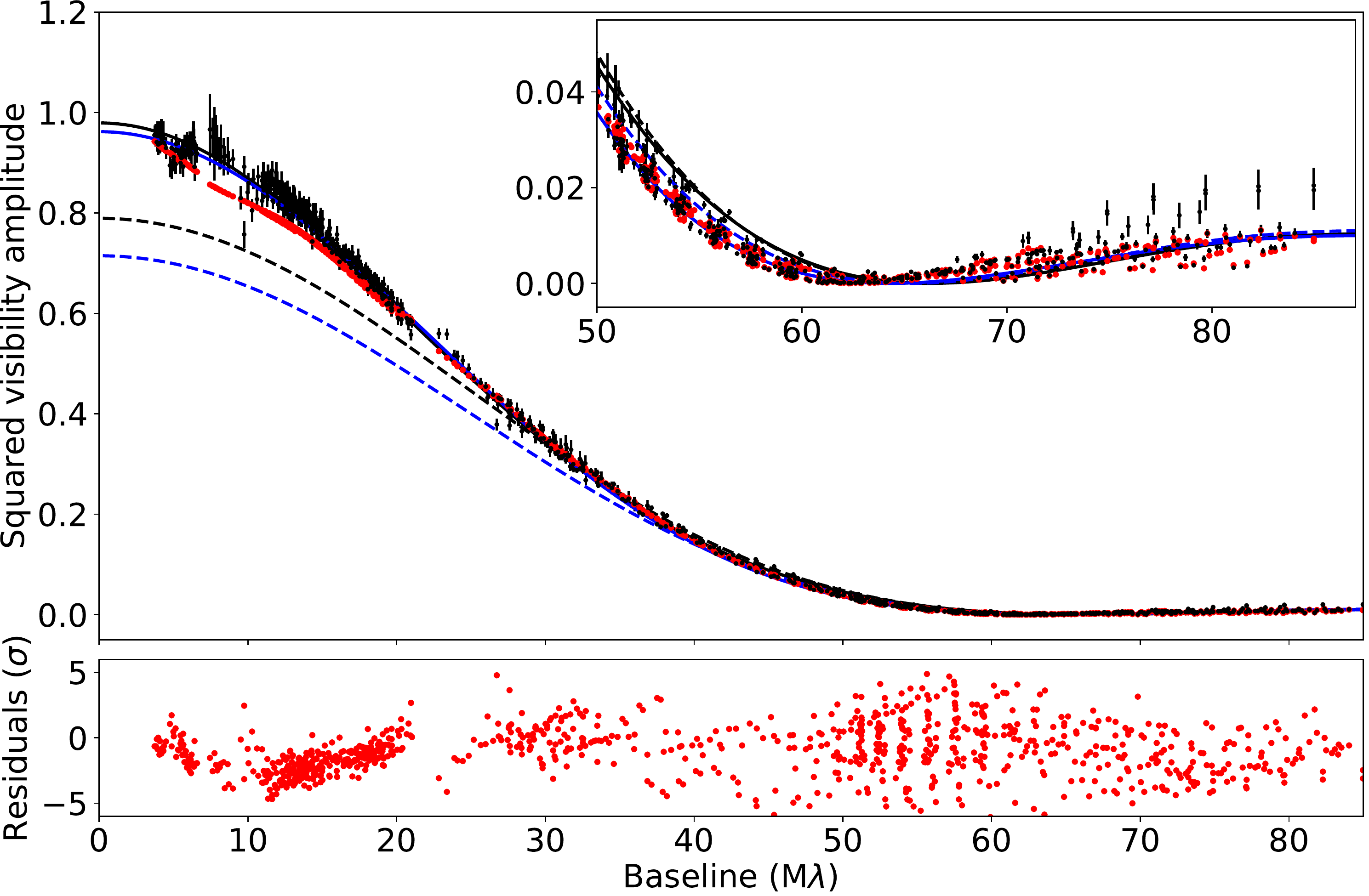}
    \includegraphics[width=0.49\linewidth]{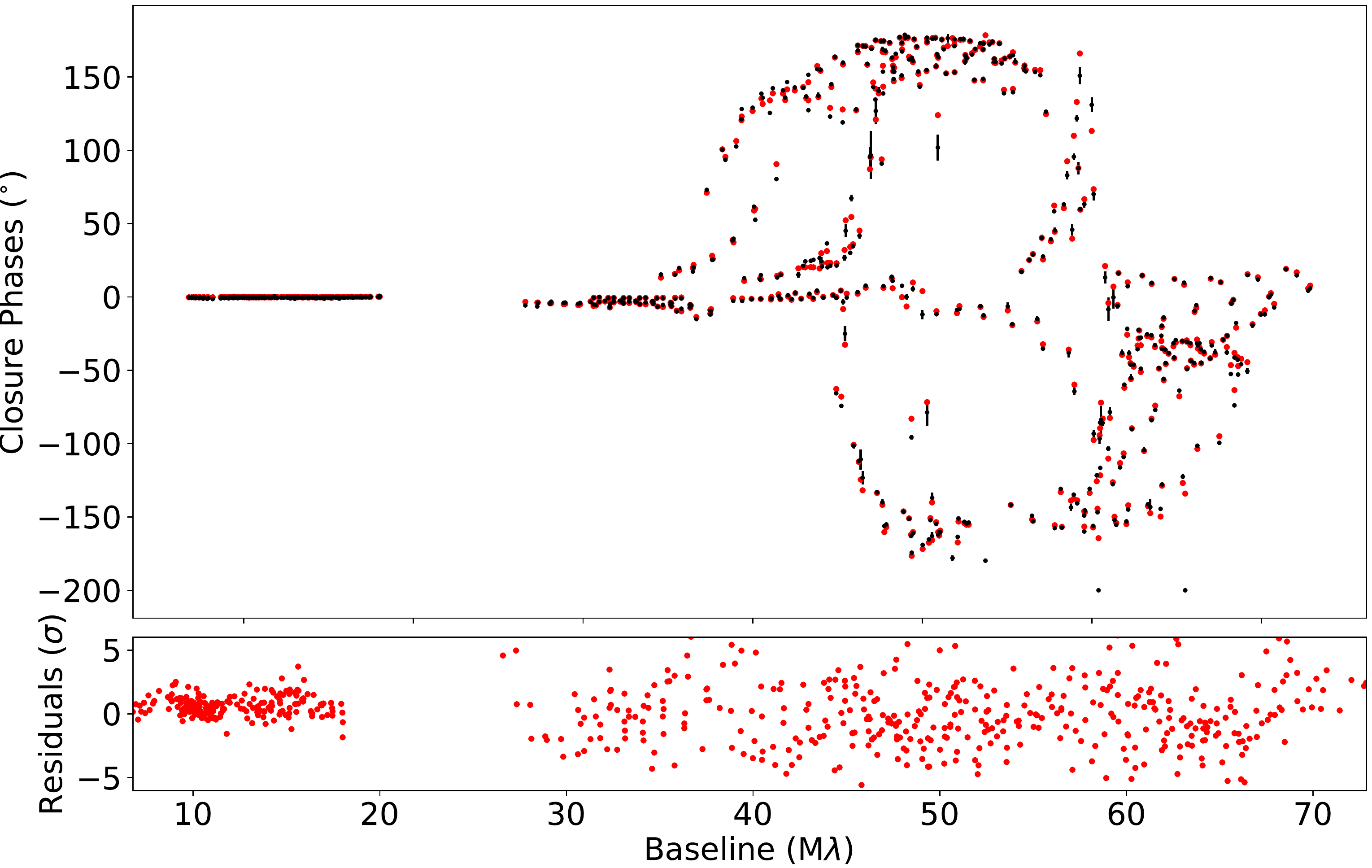}
    \caption{
            PIONIER visibility results of V602 Car of 2016 (top) and 2019 (bottom) as a function of baseline length. The left panel shows the squared visibility amplitudes, where the inlay enlarges the part of the low values. The right panel shows the closure phases. The vertical bars indicate the symmetric error bars. The black solid line denotes our visibility model including the stellar photosphere, represented by a PHOENIX model atmosphere, and a larger uniform disk indicating the extended atmosphere or MOLsphere. The black dashed line indicates the part of the PHOENIX model atmosphere alone without the added uniform disk (MOLsphere). The blue solid and dashed  lines represent the same parts, but for the selected 3D RHD snapshots for each epoch instead of the PHOENIX model (see Sect.~\ref{comparison_generalsection}). The synthetic values based on the reconstructed images are shown in red (SQUEEZE algorithm). The  small panels below the main panels provide the residuals between observations and reconstructed images. 
    }
    \label{fig:visandclos}
\end{figure*}

We reduced and calibrated the data with the $\textit{pndrs}$ package \citep{2011A&A...535A..67L}. The resulting visibility data of our observations can be found in Fig. \ref{fig:visandclos},
together with model fits and synthetic visibilites of our image reconstructions, as discussed below.

\section{Data analysis}\label{sect:data_analysis}

The visibility data in Fig. \ref{fig:visandclos} indicate an overall spherical stellar disk. However, deviations from a continuously decreasing visibility in the first lobe and closure phases different from 0/180$^{\circ}$ at higher spatial frequencies indicate the presence of inhomogeneities.

As detected in previous K-band observations \citep{2015A&A...575A..50A}, V602~Car possesses an extended molecular layer, in the near-IR most importantly of H$_2$O and CO, also called MOLsphere \citep{2000ApJ...540L..99T}. These same molecules are also present in the H-band, and such extended layers have been detected in the H-band, for example for the AGB star R~Aquarii \citep{2008ApJ...679..746R}. We should thus   expect the MOLsphere of V602~Car to be seen in our H-band data as well.
In order to describe the stellar photosphere and this MOLsphere, we used a two-component model: a PHOENIX model atmosphere \citep{1999JCoAM.109...41H} for the stellar photosphere and a uniform disk (UD) describing the MOLsphere, as  was done in \citet{2015A&A...575A..50A}. We chose a PHOENIX  model from  the grid of \citet{2013A&A...554A..76A} with parameters close to the established values for V602~Car by \citet{2015A&A...575A..50A}: 20 M$_{\rm \odot}$, T$_{\rm eff}$ = 3400 K, and log(g) = -0.5. The fit was performed in the same way as in \citet{2017A&A...597A...9W} and separately for each spectral channel. We treated the flux fractions $f_\mathrm{Ross}$ and $f_\mathrm{UD}$  as free parameters to allow for an additional over-resolved background component. 

Table~\ref{table:fits_parameters} lists the resulting best-fit parameters, together with the values averaged over the spectral channels. 
As expected for long-period variables, the flux contribution of the 
molecular layer is stronger in the water vapor bands toward the edges of 
the H-band. The angular diameter of the MOLsphere may not correlate well with 
its flux contribution, and may be less well constrained, in particular 
for low flux contributions.
For the 2016 epoch, the best fit was found to be a photosphere with an angular diameter $\Theta_\mathrm{Ross}$ of 4.4$\pm$0.2\,mas and a MOLsphere contributing on average $\sim$10\% of the total flux with an angular diameter of $\sim$ 8\,mas.
For the 2019 epoch, we obtained consistent values with a $\Theta_\mathrm{Ross}$ of 4.5$\pm$0.2\,mas and a MOLsphere with the same parameters as for 2016. For both epochs the flux fraction of a larger unresolved component was negligible, and our values of the Rosseland angular diameter are consistent with the estimate of $\Theta_\mathrm{Ross}$ = 5.08 $\pm$ 0.75 mas by \citet{2015A&A...575A..50A}.

The synthetic squared visibility values are plotted in Fig. \ref{fig:visandclos}. The PHOENIX plus MOLsphere model successfully describes our visibility data. The effect of the UD representing the MOLsphere is clearly visible, since the PHOENIX model alone is unable to reproduce the measured shape of the visibility function.

A close inspection of the visibilities at baselines 50-90 M$\mathrm{\lambda}$ (Fig.~\ref{fig:visandclos}), in particular for the 2016 epoch, reveals the presence of more than one visibility minimum along different baseline angles, where visibility minima are separated by about 5\% in baseline length. As discussed for Betelgeuse \citep{2010A&A...515A..12C,2016A&A...588A.130M}, this feature may indicate that the star is seen by the interferometer as an overall slightly elongated disk, with differences of about 5\% in radius across different angles, instead of a perfectly spherical disk. However, as shown by \citet{2009A&A...506.1351C,2010A&A...515A..12C}, big intense convection cells within an overall spherical stellar disk can also be the origin of such dispersion of the spatial frequency at the visibility null. In order to probe this possibility, we reconstructed the observational images from these visibilities.

\begin{figure*}
    \centering
    \includegraphics[width=0.495\linewidth]{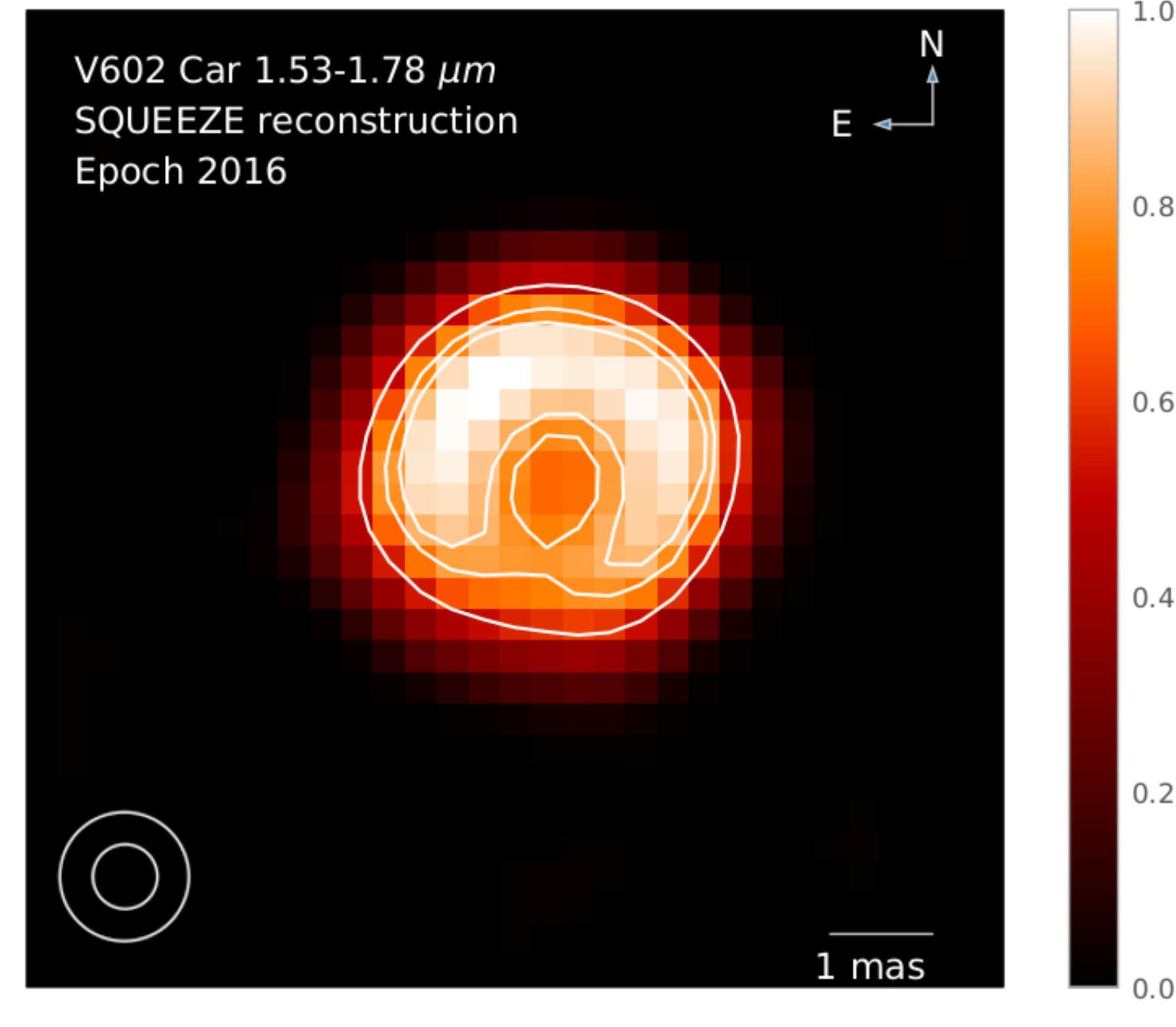}
    \includegraphics[width=0.495\linewidth]{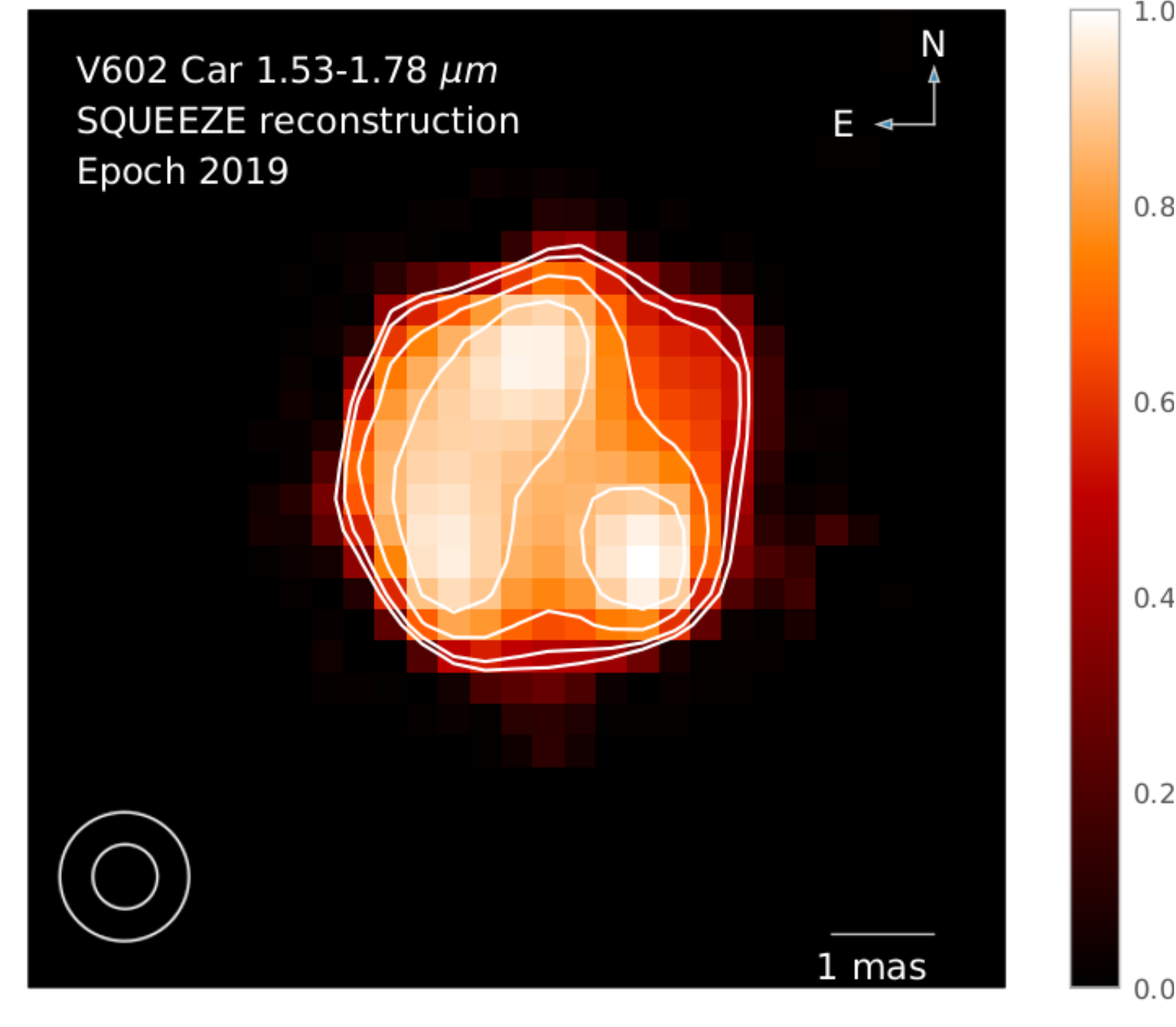}
    \caption{SQUEEZE reconstruction of V602 Car with spectral channels combined.
    (\textit{Left}): 2016 data set. Contours are drawn at levels 55\%, 77\% and 85\% of the peak intensity.
    (\textit{Right}). 2019 data set. In this case, contours are drawn at levels 40\%, 50\%, 70\%, and 87\% of the peak intensity.
    Here and hereafter, the pixel scale is 0.3 mas/pixel. The size of the circles in the lower left corners indicate our nominal angular resolution of 1.2 mas and the smallest circle represents our best estimate of the real resolution obtained, 0.6 mas (see Appendix \ref{sect:errors_uv}).
    }
    \label{fig:allchannels}
\end{figure*}

\section{Aperture synthesis imaging}\label{sect:aperture}

We used the reconstruction package SQUEEZE \citep{2010SPIE.7734E..2IB} to obtain aperture synthesis images.

We expect an overall (star plus stellar environment) source size beyond 8 mas in angular diameter based on the analysis in Sect.~\ref{sect:data_analysis}. As defined in \citet{2003RPPh...66..789M}, the nominal spatial resolution of our imaging observations $\lambda/(2B_\mathrm{max})$ is 1.2\,mas.
We tested three different scenarios to select the best pixel size and field of view (FOV) for our image reconstructions: 
1) 0.6 mas/pixel with 64x64 pixel FOV; 2) 0.3 mas/pixel with 64x64 pixel FOV; 3) 0.3 mas/pixel with 128x128 pixel FOV. When restricting the field of view (comparing scenarios 2 and 3), the $\chi^2$ of the reconstructed images clearly favored scenario 3, that with a larger FOV. This is a good indication of extended flux that scenario 2 is not able to recover. The visual comparison of scenarios 1 and 3 (i.e., the same FOV with different resolutions) showed the same features in both images.  No new information was created with the pixel size of 0.3 mas/pixel, but $\chi^2$ improved, so we kept this as our final pixel size and FOV.

We performed SQUEEZE reconstructions for two main regularizations: Laplacian ($la$) and Total Variation ($tv$). For both of these regularizations we also included an L0-norm regularization ($l0$) to decrease the number of spurious point-like sources in the FOV. We also used a transpectral regularization ($ts$) to center all the images in the bandpass at the same position of the FOV (when working with combined channels). 
We tested two combinations of regularizations: $la+l0+ts$ and $tv+l0+ts$. 
The optimum value of a given regularization's hyperparameter ($\mu$) was selected in the following way.
We created an L-curve characterizing the response of the prior term versus the $\chi^2$ value of the image solution for several values of $\mu$. 
The optimum value of $\mu$ is associated with the elbow of the L-curve. This procedure\footnote{See \textit{Reconstruction test report and data processing cookbooks} by Sanchez-Bermudez et al., available at \url{http://www.jmmc.fr/oimaging.htm}} was followed first for the $l0$ regularization, then for the $ts$ regularization, and finally for the $la$ and $tv$ regularizations. 

We also tested the possible influence of an initial
model on the image reconstruction process. The image reconstructions without an initial model were obtained by employing the procedure explained in \citet{2018Natur.553..310P}; the steps are as follows: 
i) create a reconstructed image with a resolution of one-quarter the number of  pixels and four times the mas/pixel of the final image, with a simple Dirac delta function as a start image;
ii) use this image as the initial guess for creating another one with one-half the number of pixels and two times the mas/pixel of the final image; iii) using this intermediate image as initial model, reconstruct the final image at full resolution. 
The reconstructions with initial model used the best-fit models from the PHOENIX + UD model discussed in Sect. \ref{sect:data_analysis}. The difference between these two methods of reconstruction (initial model versus no initial model) was negligible.

We first reconstructed images at the six spectral channels individually. When comparing the images, we did not find significant differences across spectral channels. The structural similarity index (SSIM; see Sect.~\ref{sect:comparison_morphology}) showed a very high similarity 
(0.99) across the spectral channels. Therefore, we combined the data of all spectral channels covering wavelengths of 1.53\,$\mu$m to 1.78\,$\mu$m.

We selected the SQUEEZE images with lowest $\chi^2$;  in the case of combined channels for 2016 this corresponds to $\chi^2$ = 1.57 with $\mu_{tv}$ = 500, $\mu_{l0}$ = 3, $\mu_{ts}$ = 1, using the initial model described in Sect. \ref{sect:data_analysis}. The 2019 combined channels image has a $\chi^2$ = 7.09 with $\mu_{tv}$ = 2000, $\mu_{l0}$ = 30, $\mu_{ts}$ = 1, also using  an initial model. 
The reason of the larger $\chi^2$ of the 2019 image compared to the 2016 
image is not clear. It may be related to the smaller estimated errors of 
the measured visibility and closure phases in the 2019 data, so that 
systematic absolute calibration uncertainties have a larger relative 
contribution. Our image reconstruction tests (see Sect.~\ref{sect:recerror}) confirm that the 2019 image reconstruction is as at least as reliable as the 
2016 image reconstruction.
The reconstructed images were not further convolved beyond the chosen pixel scale of 0.3 mas/pixel, as discussed above.

\subsection{Final reconstructed images}\label{sect:final_image}

Figure~\ref{fig:allchannels} shows the final reconstructed images for the 2016 and 2019 epochs. The images of the individual spectral channels are shown in Appendix~\ref{sect:2016_channels} to illustrate that they are very similar across all the spectral channels.
We obtained the synthetic visibilities of the final reconstructed images at our \textit{uv} observational points using the OITOOLS package\footnote{Available at \url{https://github.com/fabienbaron/OITOOLS.jl}}.
The comparison of the interferometric observables from the experimental data with those extracted from the reconstructed images (Fig. \ref{fig:allchannels}) shows a very good agreement (Fig. \ref{fig:visandclos}). This confirms that extended flux caused by the MOLsphere (Sect. \ref{sect:data_analysis}) is present in the reconstructed images, as already indicated by the improved $\chi^2$ values with increased FOV (Sect.~\ref{sect:aperture}). For the sake of clarity, we show in Fig. \ref{fig:visandclos} only the visibility values based on the wavelength- and time-averaged reconstructions, while the observed visibilities are shown for individual observing dates and spectral channels. Some of the residual differences in Fig. \ref{fig:visandclos} may be caused by this effect.

The reconstructed image of the 2016 epoch shows the stellar disk with an intriguing, bright arc-like feature toward the northern rim of the stellar surface. In 2019, the orientation of the arc-like feature is different and a new peak of emission is detected on the opposite side of the stellar surface. 
The extended molecular layer or MOLsphere, although present in the reconstructed images, lies close to our achieved dynamic range (of about 1:10 to 1:20), 
so that it is not clearly visible. While the parameters of the MOLsphere show a dependence on wavelength, the photospheric structure is not expected to be wavelength dependent, which explains why the reconstructed images appear to be very similar across spectral channels.

The double visibility null, as seen in the observed visibilities in Fig. \ref{fig:visandclos} and described in Sect. \ref{sect:data_analysis}, is reproduced by the image reconstructions, and is thus most likely caused by the surface features and not by an overall elongated stellar disk. 
 
\subsection{Error estimates of the final reconstructed image}
\label{sect:recerror}

We characterized possible errors that may be introduced by the reconstruction process to assess the soundness of the detected surface features.
We also used  the IRBis reconstruction package \citep[Image Reconstruction software using the Bispectrum]{2014A&A...565A..48H} to test the dependency of our results on the reconstruction package employed.
A detailed explanation of these tests can be found in Appendix \ref{sect:allapendix}. Our analysis revealed the following: 
i) No new features are introduced within SQUEEZE when altering the final reconstructed images by one standard deviation;
ii) Synthetic observational data based on 3D snapshots
at our $uv$ points and with our level of noise recovers 
the substructure present in the original image, with maximum intensity losses of 26\% for the 2016 and 30\% for the 2019 epochs; 
iii) The difference between SQUEEZE and IRBis reconstructed images shows that the same structures are present in both image reconstructions.
Therefore, we conclude that the detected structure is most likely real and not due to any artificial effect. 

\begin{figure*}[h]
    \centering
    \includegraphics[width=0.49\linewidth]{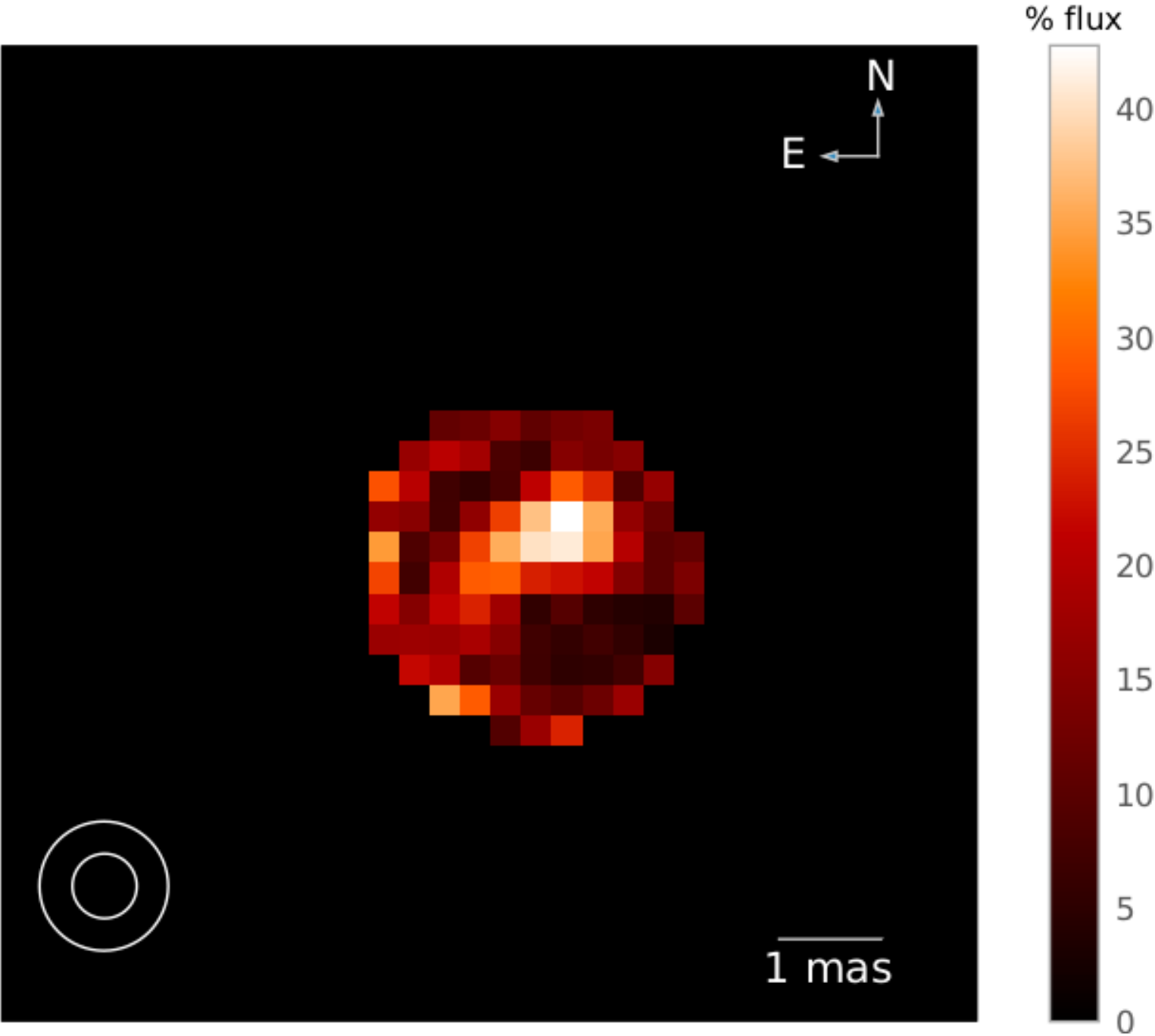}
    \includegraphics[width=0.49\linewidth]{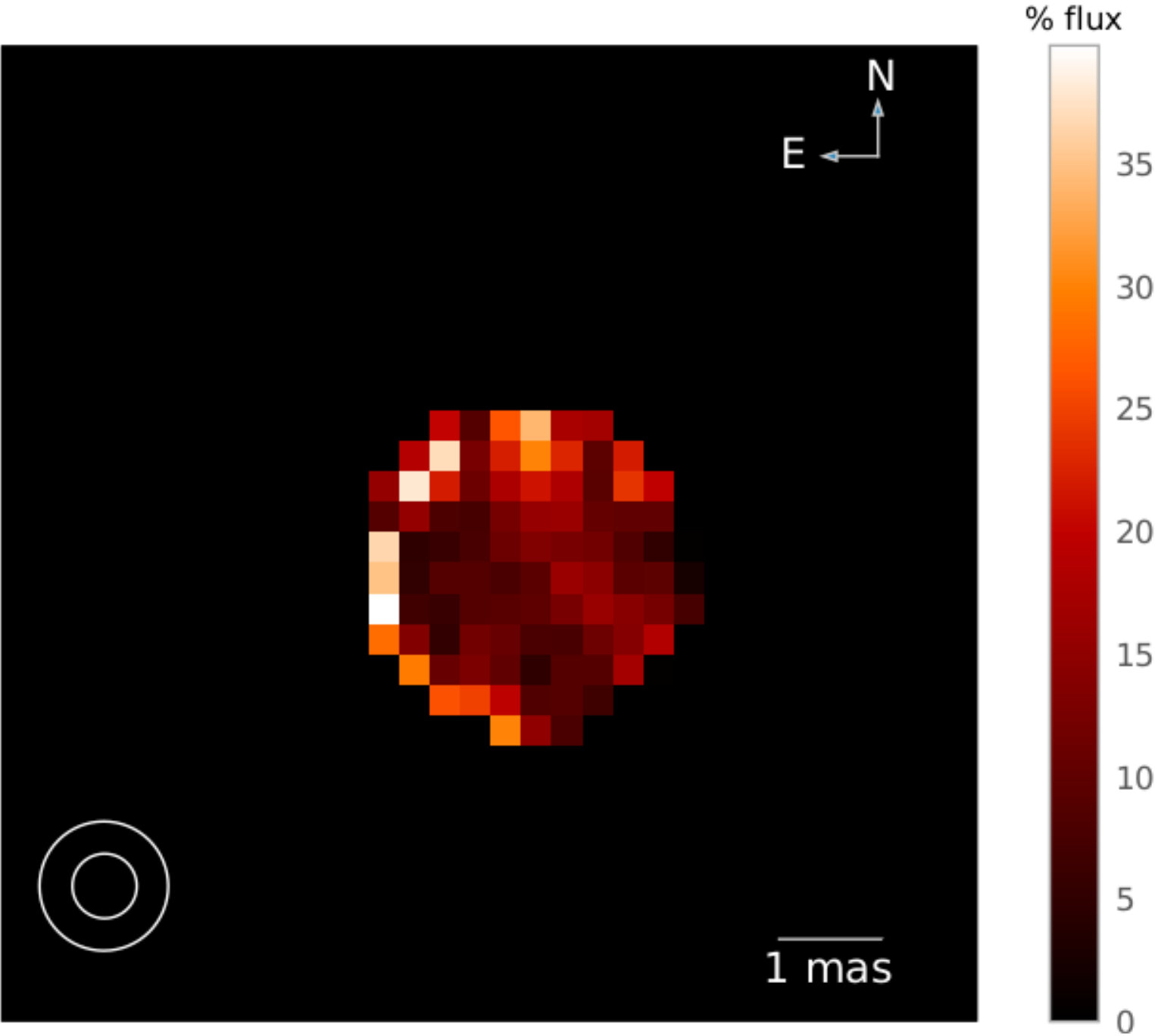}
    \caption{
    Error images for the  2016 epoch (left) and  2019 epoch (right) resulting conservatively from the addition of three possible sources of errors 
    discussed in Appendix \protect\ref{sect:allapendix}.
    The errors are expressed in terms of the original image flux. Here and hereafter, the 3D-RHD-model-related images have being convolved with a 0.6 mas beam, that is, our best estimate of the real resolution obtained (see Appendix \ref{sect:errors_uv}).
    }
    \label{fig:total_error}
\end{figure*}

Figure \ref{fig:total_error} shows the total error map, conservatively taking into account all these possible error sources, as described in detail in Appendix \ref{sect:allapendix}.
The average errors, in terms of original image flux, are 17\% and 14\% for 2016 and 2019, respectively. Most of these error sources are systematic,  extending across the images, so that the pixel-to-pixel error is 
significantly smaller.

Our tests using reconstructions of synthetic data (based on 3D RHD models and with our $uv$ coverage and observational errors) with different convolution kernels (see Appendix \ref{sect:errors_uv}) revealed that original images and reconstructions match best with a convolution kernel of 0.6\,mas. This suggests that we reach with our data and $uv$ coverage a super-resolution of $\sim$ 0.6\,mas compared to the nominal resolution $\lambda/(2B_\mathrm{max})$ of 1.2\,mas.  

\section{Comparison with 3D RHD simulations}\label{comparison_generalsection}

In order to compare our PIONIER data of V602 Car to theoretical models, we used numerical 3D RHD simulations obtained with the CO$^5$BOLD code \citep[COnservative COde for the COmputation of COmpressible COnvection in a BOx of L Dimensions;][]{2012JCoPh.231..919F}. The simulation used was st35gm04n38 (401$^3$ grid points, T$_{\rm eff}$ = 3414 $\pm$ 17 $K$, log $g$ = -0.39 $\pm$ 0.01, 5 M$_{\rm \odot}$, 582 $\pm$ 5 R$_{\rm \odot}$). The grid resolution is 4.055 R$_{\rm \odot}$ with a total field of view of 1626 R$_{\rm \odot}$. This model shows an effective temperature and surface gravity as established for V602 Car (see Sect.~\ref{sec:intro}), while it has a smaller radius and a lower mass compared to the observational parameters. Due to the limited number of currently available 3D simulations of RSG stars, and in particular to the computationally demanding calculation of higher mass stellar models, 3D models of a current mass of
10--13\,M$_\odot$, as expected for V602 Car, and  larger radii are not yet available. Nevertheless, this 3D model represents typical properties of an RSG star, and distinctively different dynamical properties than lower-mass pulsating AGB star models \citep[cf. the discussion by][]{2019A&A...632A..28K}.
This simulation reproduces the effects of convection and, additionally, non-radial waves \citep{2011A&A...535A..22C}. This model was first used by \citet{2018A&A...610A..29K} and \citet{2019A&A...632A..28K}, and a detailed discussion on the model can be found therein. We computed 81 temporal snapshots about 23 days apart and covering a stellar time of about 1863 days in total.
Intensity images were then computed using the pure-LTE radiative transfer Optim3D \citep{2009A&A...506.1351C}
at the bandpass of our PIONIER observation of 1.65$\pm$0.15\,$\mu$m (averaged over 56 maps across the range 1.5-1.8 $\mu$m).

\begin{figure*}
    \centering
    \includegraphics[width=0.49\linewidth]{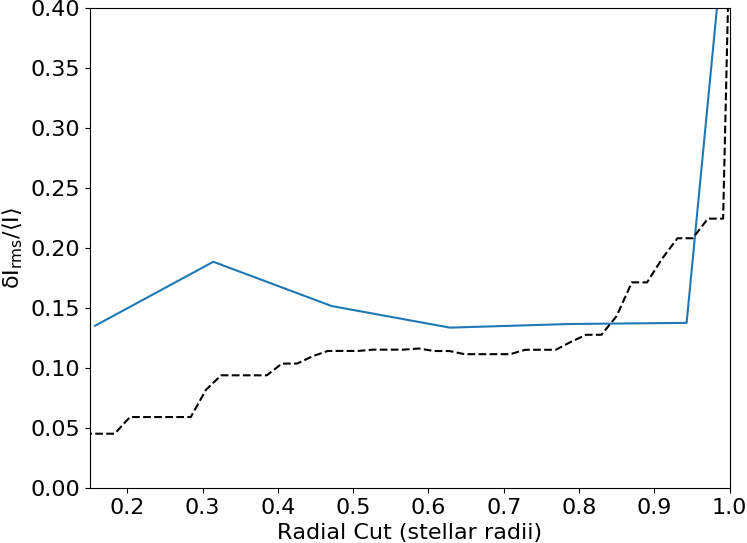}
    \includegraphics[width=0.49\linewidth]{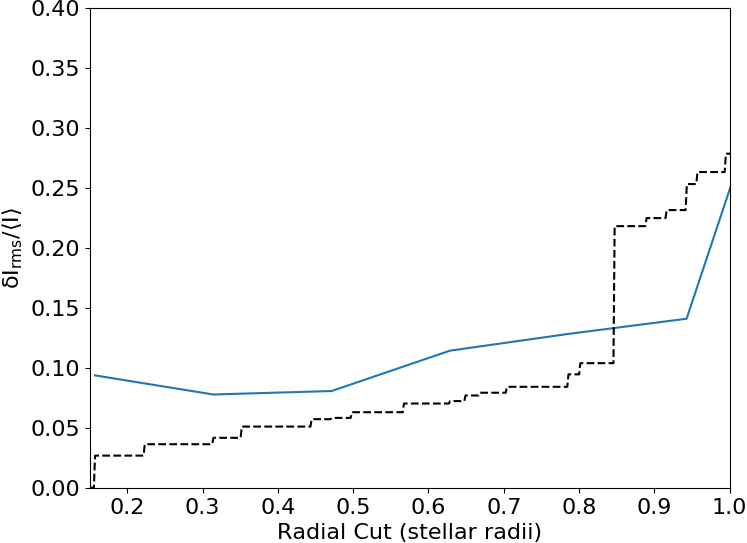}
    \caption{Contrast of the data set as a function of the radial cut considered. The black dashed line indicates the contrast of the SQUEEZE reconstructed image (2016 in the left panel and 2019 in the right panel) and the blue solid line is one of the best simulated snapshot for each data set (snapshot 65 for 2016 data and snapshot 67 for 2019 data).}
    \label{fig:contrast}
\end{figure*}

\begin{figure}
    \centering
    \includegraphics[width=\linewidth]{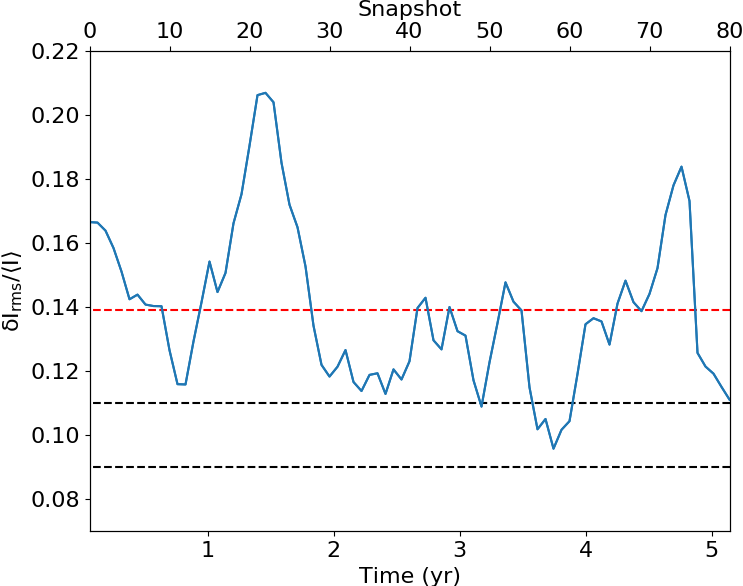}
    \caption{The blue line represents the contrast of the simulated snapshots after being corrected for limb-darkening (LD) with a radial cut of 0.75 stellar radii. The black dashed lines represent the contrasts measured in the final reconstructed images (11 $\pm$ 2\% in 2016, 9 $\pm$ 2\% in 2019), while the orange dashed line represents the average value over all the snapshots (14 $\pm$ 2\%). The LD was corrected using the best-fit model image from Sect. \ref{sect:data_analysis}. 
    }
    \label{fig:contrastvssnap}
\end{figure}

\subsection{Comparison in terms of contrast}
\label{sect:contrast}
We estimated the contrast of our reconstructed images, $\delta I_{rms}$/$\langle I \rangle$, as defined in \citet{2013A&A...557A...7T} to compare them to  3D RHD simulations. 
The contrast on the stellar surface is affected both by surface features and by the
limb-darkening (LD) effect. We are interested in the contrast of the surface features themselves.
In order to correct for the LD effect, we used two independent methods: i) dividing the reconstructed image by the best-fit model image described in Sect. \ref{sect:data_analysis};   ii) applying Equation 2 found in \citet{2009A&A...506.1351C} with parameters in Table 2 of the same text. In this second method we created a high resolution image of the LD model (401$\times$401 pixels) with the same field of view (FOV) as the reconstructed image. 
Then we re-binned it to the same resolution and pixel size of our reconstucted image,
so that   both images (LD model and reconstructed) possessed the same FOV and resolution. Finally, we divided pixel by pixel in a similar fashion to the first method. Both of these methods resulted in a very similar correction. From now on the results exposed are valid for both of them.

We need to ensure that we do not include the stellar limb in our estimate of the contrast of surface features.
We define a cut-off radius, i.e., the maximum radius adopted from the center of the star and for all angles, from which the outer pixels are not considered
to compute the contrast.
Figure~\ref{fig:contrast} shows the contrast as a 
function of the chosen cut-off radius.
For the 2016 data, the contrast increases with increasing radial cut up to $\sim$0.5 stellar radii. This could be explained by an increasing number of 
 image patterns included as the cut-off radius gets larger. For larger cut-off radii, between about 0.75 and 0.95 stellar radii, the contrast again shows a fast increase. This may be an effect of the limb-darkening that may not be perfectly corrected. When the radial cut surpasses the value of about 0.95 stellar radii we see a rapid and steep increase, which may be representative of the contrast between the stellar disk and the outside of the disk, which may also not be perfectly circular. A similar behavior is seen for the 2019 data, where more patterns and structures are included as the radial cut gets larger, until a rapid increase occurs at $\sim$0.85 stellar radii.

We  only consider the contrast below a radial cut of 0.75 stellar radii to avoid the bias of the uncertainty of the limb-darkening correction.The lower cut must include the arc-like feature present in Fig. \ref{fig:allchannels} for 2016 and the more complex features of the 2019 data. Therefore, we establish a lower cut of 0.5 stellar radii in the 2016 data, implying a contrast value of 11\% $\pm$ 2\%, which corresponds to the plateau found between these radii. Due to shape of the substructures present in 2019 data, no lower cut value can be easily determined for this epoch. Therefore, we assume as a radial cut the maximum value considered here (i.e., 0.75 stellar radii, with a contrast value of 9\% $\pm$ 2\%). Although the average pixel errors of the images were 17\% and 14\% for 2016 and 2019, as described in Sect.~\ref{sect:final_image}, and thus larger than the surface feature contrasts, we note that the pixel errors are conservative values that take into account multiple sources of errors, as outlined in Appendix~\ref{sect:allapendix}, several of which are systematic. This does not mean that the pixel-to-pixel uncertainties and the contrast uncertainties are as large as this error map.

Following the same 
procedure, we calculated the contrast for the 81 snapshots of the simulated 3D RHD snapshots.
Figure~\ref{fig:contrastvssnap} shows how the contrast varies across different snapshots, assuming a radial cut of 0.75 stellar radii.
The average contrast value over the 81 snapshots is 14\% $\pm$ 2\%. This value is slightly larger than that of our image reconstructions, but with individual snapshots that have consistent contrast values. 

Considering that our observational epochs lasted about 70 days, we made a test in which we first averaged consecutive model snapshots over this time span (2--3 snapshots) and then computed the contrasts of the averaged 
snapshots. The snapshot images were similar over these time spans, resulting in only marginal differences in the contrast curve shown in Fig. \ref{fig:contrastvssnap}.

Previous works have found similar contrast values in RSGs. \citet{Wittkowski_V766Cen} consistently reported  a contrast of 10\%$\pm$4\% for the RSG V766 Cen, while \citet{Montarges_CETAU} found a lower contrast of 5--6\% $\pm$1\% for the RSG CE Tau.
Both estimates were based on similar imaging of data obtained with the PIONIER instrument in the near-IR H-band.

\subsection{Comparison in terms of morphology}\label{sect:comparison_morphology}
We then investigated whether the 3D RHD simulations could reproduce the observed morphology of our reconstructed images, such as the arc-like feature discussed in Sect. \ref{sect:final_image}.

The calculated snapshots of the 3D RHD simulations represent the stellar convection dynamics every $\sim$23 days, covering a total of 1863 days. 
There is a fundamental problem in direct comparisons of such simulated snapshots to our data: the surface pattern changes in a stochastic way and never repeats itself. With a finite number of simulated snapshots, we cannot expect any snapshot to coincide exactly with the pattern at one of our observed epochs. If we cannot expect a model to describe the observational data, a formal $\chi^2$ comparison between model and observation is not appropriate and may lead to spurious results. As a solution to this fundamental problem, we introduce the use of the SSIM \citep[][]{wang2004image}
to find the most similar of the 81 model snapshots compared to our reconstructed image (Fig. \ref{fig:allchannels}). The SSIM of a pair of images represents a superior method for image comparison. It is typically used in order to quantify the differences between a distorted image and a reference image. It is based on the perceived change in the structural information from one image to the other, and ranges from -1 to 1, where 1 indicates perfect similarity (Eq. \ref{SSIM_formula}). 

\begin{figure*}
    \centering
    \includegraphics[width=0.38\textwidth, keepaspectratio]{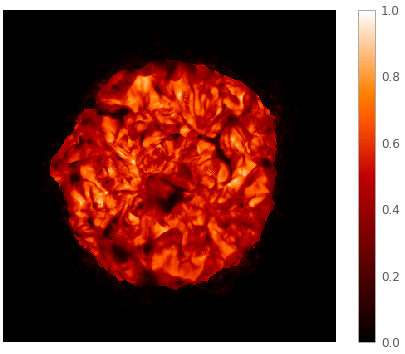}
    \includegraphics[width=0.38\textwidth, keepaspectratio]{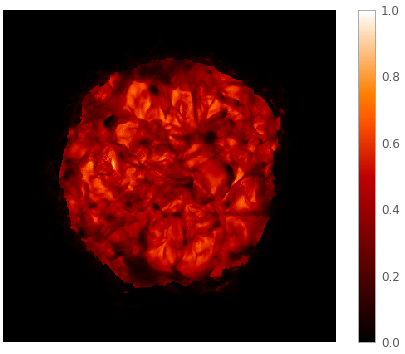}
    \includegraphics[width=0.38\textwidth, keepaspectratio]{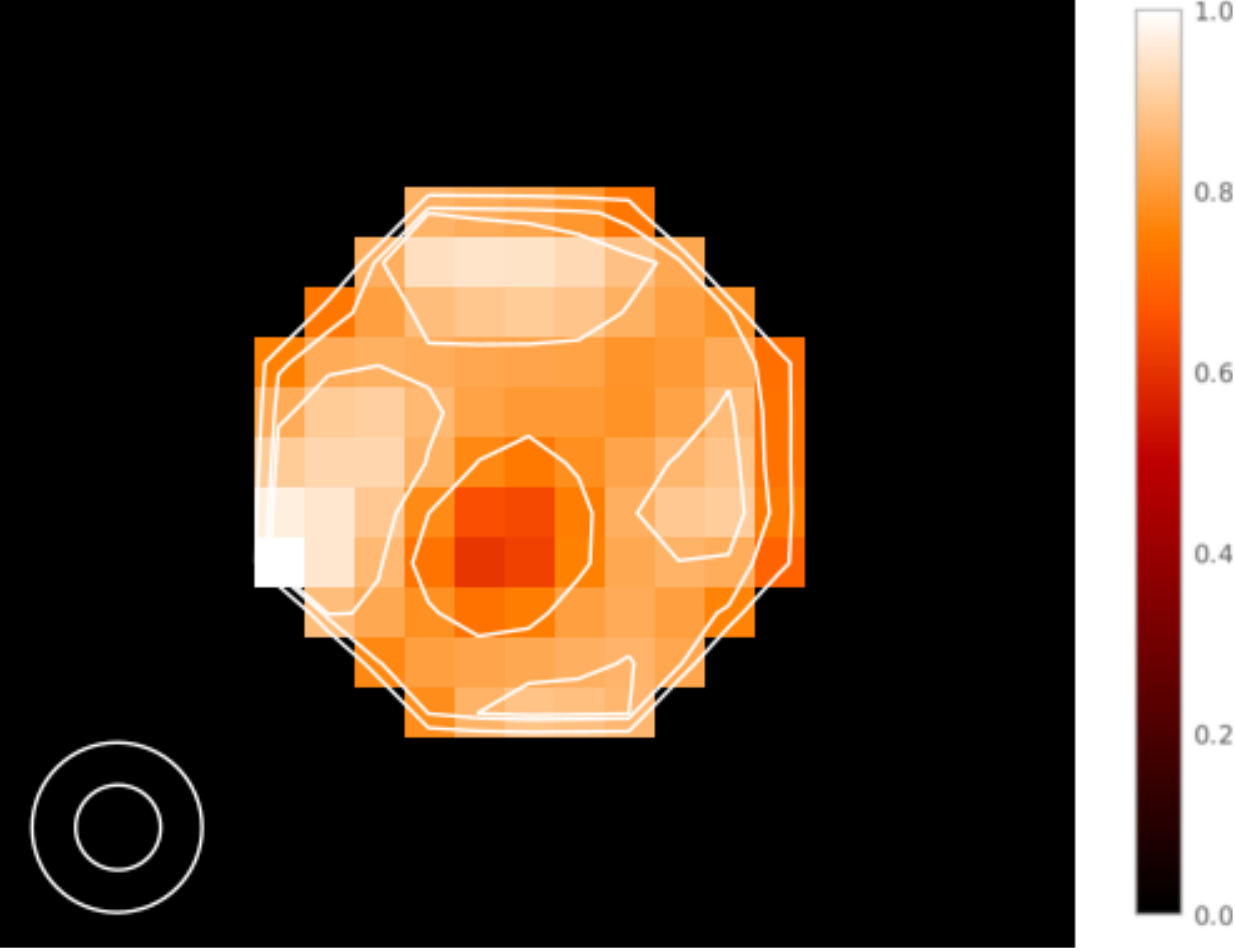}
    \includegraphics[width=0.38\textwidth, keepaspectratio]{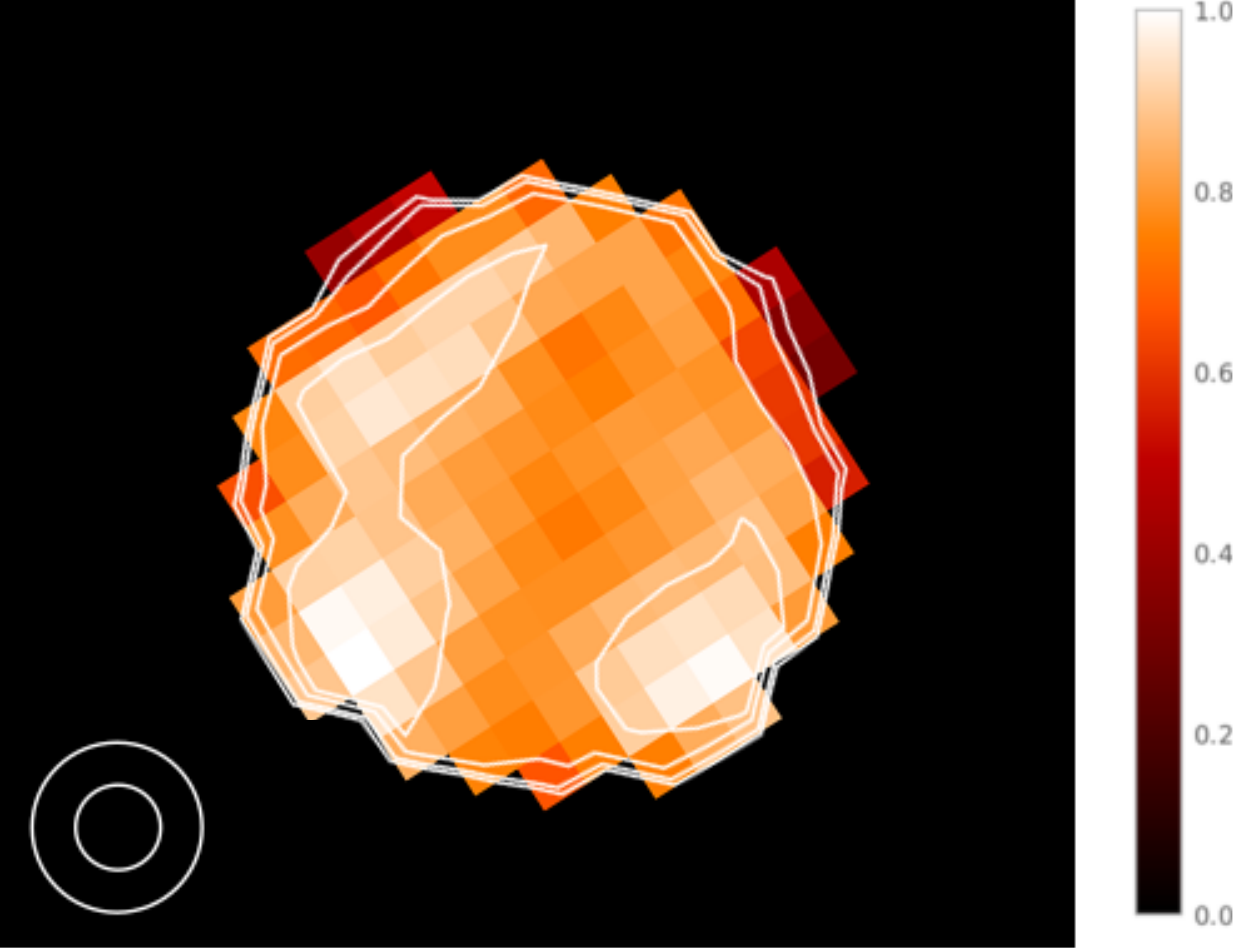}
    \includegraphics[width=0.38\textwidth, keepaspectratio]{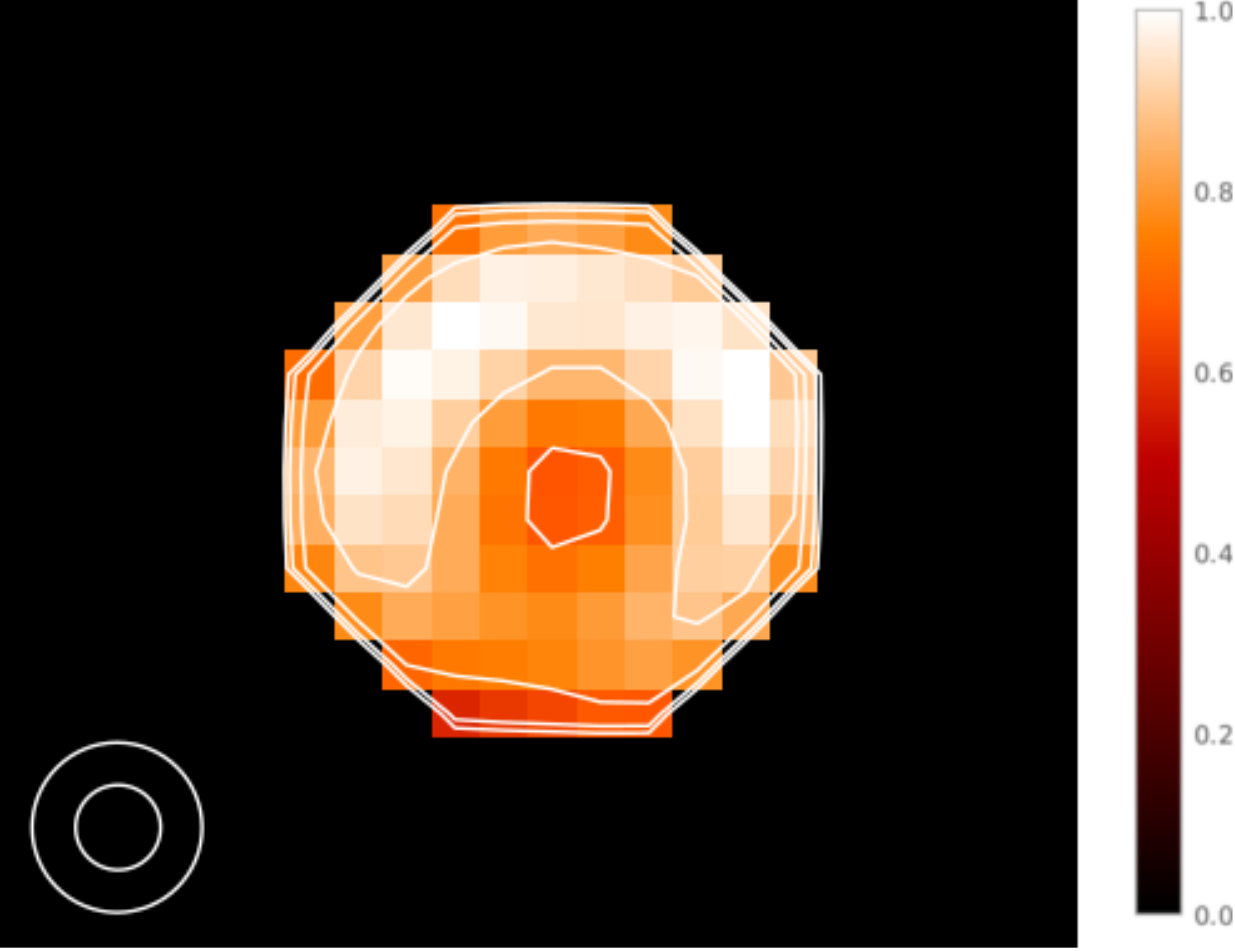}
    \includegraphics[width=0.38\textwidth, keepaspectratio]{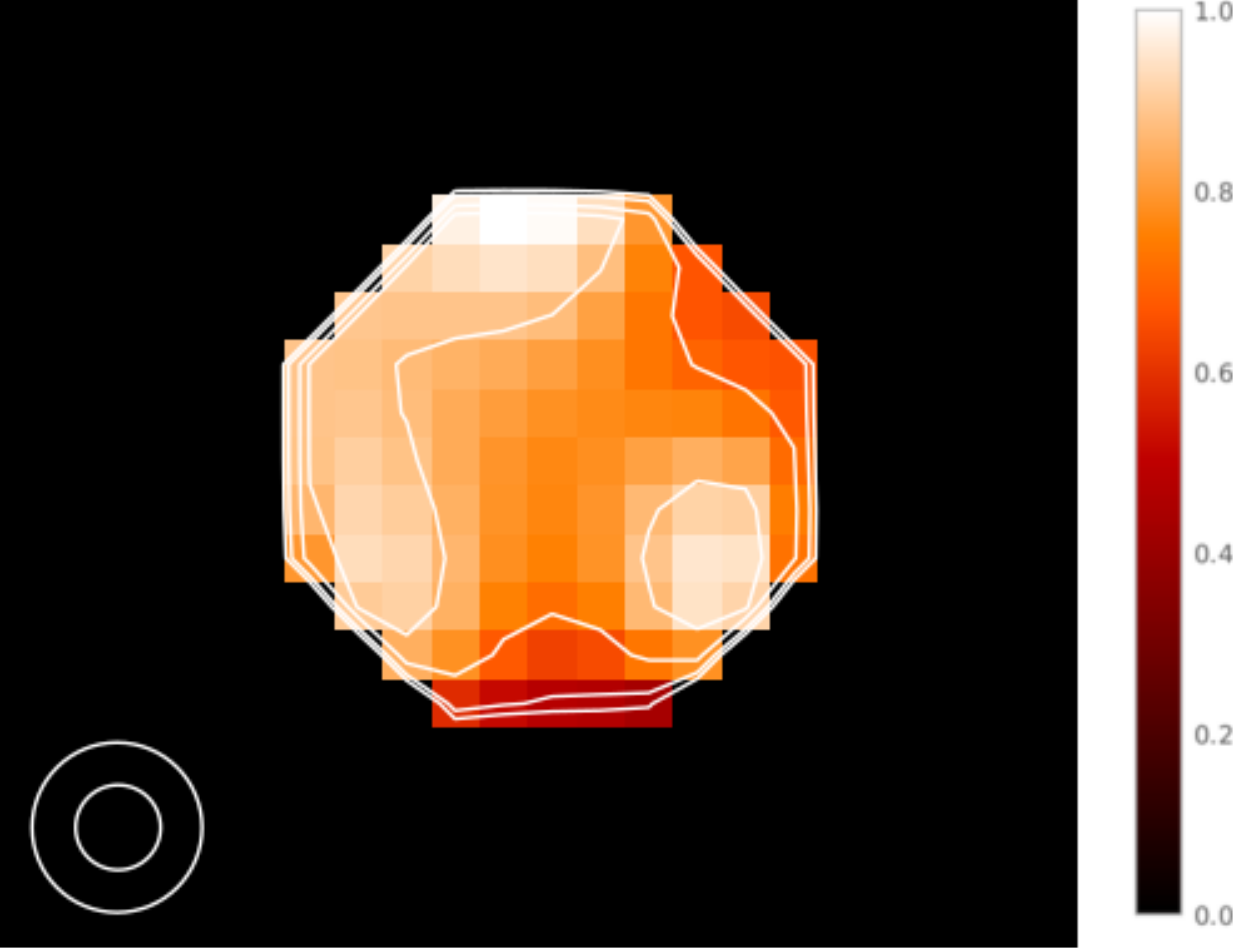}
    \caption{Comparison in terms of morphology between model snapshots and reconstructed images.
    (\textit{Top row}): Intensity image of one of the selected best snapshots (number 065) for 2016 in relative intensity (\textit{left}), and for snapshot 067, which is the best choice for the 2019 data (\textit{right}).
    (\textit{Middle row}): The same snapshot images as in the \textit{upper row} after being convolved with a 0.6 mas beam, rotated to match the observed morphology, and corrected for the limb-darkening effect with a cut-off radius of 0.75 stellar radii.
    (\textit{Bottom row}): Reconstructed observational images after LD correction and with a cut-off radius of 0.75 stellar radii. In the \textit{middle} and \textit{lower rows}, the contours are drawn at levels of 55\%, 77\%, and 85\% of the peak intensity for 2016 (\textit{left column}) and at 40\%, 50\%, 70\%, and 87\% of the peak intensity for 2019 (\textit{right column}).
    }
    \label{fig:model_65}
    \label{fig:LD_comparison_2016}
    \label{fig:LD_comparison_2019}
\end{figure*}

We first convolved the model images to our best estimate of the real resolution obtained, 0.6 mas (see Appendix \ref{sect:errors_uv}), re-sized the images
to the pixel scale and field of view of our reconstructed images, and applied a cut-off radius
of 0.75 stellar radii as in Sect.~\ref{sect:contrast}.
To account for the unknown orientation on the plane of the sky, we rotated the model images every 5$^{\circ}$ around its center and estimated the SSIM for each rotation angle.
As expected, none of the snapshots coincided with our observed epochs perfectly, but several were equally similar.
For the 2016 data, the most similar snapshots were (rotation angle in parenthesis): 003 (325$^{\circ}$ $\pm$ 5$^{\circ}$), 004 (325$^{\circ}$ $\pm$ 5$^{\circ}$), 005 (320$^{\circ}$ $\pm$ 5$^{\circ}$), 006 (325$^{\circ}$ $\pm$ 5$^{\circ}$), 064 (10$^{\circ}$ $\pm$ 5$^{\circ}$), 065 (10$^{\circ}$ $\pm$ 5$^{\circ}$), 066 (10$^{\circ}$ $\pm$ 5$^{\circ}$), and 067 (10$^{\circ}$ $\pm$ 5$^{\circ}$), all with SSIM = 0.85. All of these snapshots coincided with those that we had previously selected to be most similar to our observed image (via visual inspection). 
We selected snapshot 065 as a representative of this subset of snapshots, in which the arc-like structure is visually clearest.
Figure~\ref{fig:model_65} (upper left panel) shows this model image at the original model resolution.
Figure~\ref{fig:LD_comparison_2016} (left middle and lower panels) shows a comparison of this model image to our image reconstruction after adjusting it to the pixel size of the reconstruction, and rotating it to best match the reconstruction.

The same procedure was followed for the 2019 data selecting, in this case, snapshots 067 (-30$^{\circ}$ $\pm$ 5$^{\circ}$), 068 (-30$^{\circ}$ $\pm$ 5$^{\circ}$), 069 (-30$^{\circ}$ $\pm$ 5$^{\circ}$), and 070 (-35$^{\circ}$ $\pm$ 5$^{\circ}$), all with SSIM = 0.87. As a representative of this subset of snapshots, we selected snapshot 067 (see Fig. \ref{fig:model_65} right panels). 
Adjacent snapshots for the 2016 and 2019 data are very similar to these selected representatives. Although visually not identical to each other, our limited spatial resolution and dynamic range render them equally similar to the observational image. Finally, we  tested the uncertainties when computing the SSIM by adding and subtracting the intensity error image (Fig.~\ref{fig:total_error}) to the observational images (Fig.~\ref{fig:allchannels}) and computing the SSIM between these resulting images and the selected 3D RHD model images corresponding to each epoch. The SSIM value differs only 0.02 with respect to the case when no error image is considered.

The surface features seen in the model snapshots (Fig. \ref{fig:model_65}, top row) are 
unlikely individual deep convection cells that reach out to the surface
layers, as the timescale on which the structure changes in the models 
is too fast. The features in the model snapshots are therefore likely
related to instationary convection, i.e., to pressure fluctuations 
(non-radial waves that do not exist long enough to produce a clear mode
visible in a power spectrum) that are caused by sonic convective 
motions and are able to affect  a single surface granule and also a 
group of neighboring granules. With the limited spatial resolution of 
our observation (Fig. \ref{fig:model_65}, middle and bottom row), the structure is further convolved to larger observed patches on the stellar surface. 
This means that we cannot determine the sizes of individual granules or convection cells \citep[see also][]{2003SPIE.4838..348F}.

Each subset of the most similar snapshots to our 2016 data, 003 to 006 and 064 to 067, represents a time span of 69.5 days, indicating that the structure remains similar on timescales of $\sim$2 months. 
A similar result is found in the 2019 analysis where snapshots 067 to 070 also represent  69.5 days. This confirms, with our accuracy and spatial resolution, that it was a valid approach to combine data obtained 
over about 70 days (see Sect. \ref{sect:observations}).

The agreement between our image reconstructions and the most similar snapshots of the 3D RHD simulations may indicate, within our angular resolution and achieved dynamic range, that the observed stellar surface features of V602 Car at two individual epochs can be reproduced by the physics accounted for in the simulations we considered (i.e., non-local radiation transport, shock waves, gray and non-gray opacities; see \citealt{2011A&A...535A..22C}).

We then computed azimuthally averaged intensity profiles and synthetic visibility values for the selected snapshots and compared them to our observed visibility spectra. We computed the intensity profiles using rings regularly spaced in $\mu$, related to the impact parameter by $r/R_\mathrm{star} = \sqrt{1-\mu^{2}}$. We used 56 spectral maps between 1.5 $\mu m$ and 1.8 $\mu m$ at a spectral resolution of 300.
Synthetic visibility values were then derived following the same procedure as in \citet{2017A&A...597A...9W}, and in the same way as for the fit of a PHOENIX model in Sect.~\ref{sect:data_analysis}.
Table \ref{table:fits_parameters} shows the resulting best-fit parameters. The synthetic visibilities based on this model are included in Fig.~\ref{fig:visandclos}.
In both the 2016 and 2019 data, the best-fit parameters are close to those obtained from the fit of the PHOENIX model together with a
UD representing the MOLsphere.
We adopted a resulting photospheric angular diameter of
V602~Car of 4.4~$\pm$~0.2\,mas. This is the 
average of the photospheric angular diameters for 
2016 and 2019, and are based on the fits including the PHOENIX model.

The best fits were achieved with an additional UD component and a free zero visibility scale, as for the 1D PHOENIX model atmosphere. The visibility plot of the best model for each epoch is included in Fig.~\ref{fig:visandclos}.
This shows that the 3D RHD simulations alone cannot reproduce the observed visibility values, but that an additional, more extended component is still required to reproduce the observed data. The presence of a MOLsphere on top of the photosphere may alter the contrast and morphology of the photospheric features to some extent (below about 4\% in pixel value). This may explain a part of the residual differences between image reconstructions and 3D model images in terms of contrast and morphology.

\begin{table*}
    \centering
    \caption{Fit parameters to the PHOENIX model and to the best snapshots from 3D RHD simulations, for each epoch.}
    \label{table:fits_parameters}
    \begin{tabular}{cccccccc} 
    \hline\hline
    Model & Epoch & Channel ($\mu m$) & $\Theta_{1}$ (mas) & $f_{1}$ (\%) & $\Theta_\mathrm{UD}$ (mas) &  $f_\mathrm{UD}$ (\%) & $f_\mathrm{free}$ (\%)\\
    \hline
    PHOENIX & 2016 & 1.53 & 4.4 $\pm$ 0.2 & 84.9 & 6.4 $\pm$ 0.2 &  12.3 &  2.8\\
    & & 1.58 & 4.4 $\pm$ 0.2 & 86.6 & 6.3 $\pm$ 0.2 &  10.8 &  2.6 \\
    & & 1.63 & 4.4 $\pm$ 0.2 & 88.3 & 6.1 $\pm$ 0.2 &  8.4 &  3.3 \\
    & & 1.68 & 4.5 $\pm$ 0.2 & 94.2 & 12.3 $\pm$ 0.2 &  5.4 &  0.4 \\
    & & 1.72 & 4.4 $\pm$ 0.2 & 89.5 & 8.2 $\pm$ 0.2 &  9.8 &  0.7 \\
    & & 1.77 & 4.5 $\pm$ 0.2 & 88.7 & 8.3 $\pm$ 0.2 &  10.6 &  0.7 \\
    & & Average & 4.4 $\pm$ 0.2 & 88.7 $\pm$ 3.2 & 7.9 $\pm$ 2.3 &  9.5 $\pm$ 2.4 &  1.0 $\pm$ 1.2 \\
    \hline
    PHOENIX & 2019 & 1.53 & 4.5 $\pm$ 0.2 & 93.0 & 11.1 $\pm$ 0.2 &  7.2 &  0.0\\
    & & 1.58 & 4.6 $\pm$ 0.2 & 94.2 & 11.7 $\pm$ 0.2 &  6.0 &  0.0 \\
    & & 1.63 & 4.5 $\pm$ 0.2 & 91.2 & 6.7 $\pm$ 0.2 &  6.5 &  2.2 \\
    & & 1.68 & 4.5 $\pm$ 0.2 & 90.5 & 6.3 $\pm$ 0.2 &  7.5 &  2.0 \\
    & & 1.72 & 4.4 $\pm$ 0.2 & 82.7 & 5.7 $\pm$ 0.2 &  16.0 &  1.3 \\
    & & 1.77 & 4.4 $\pm$ 0.2 & 80.8 & 6.2 $\pm$ 0.2 &  17.4 &  1.8 \\
    & & Average & 4.5 $\pm$ 0.2 & 88.8 $\pm$ 5.7 & 8.0 $\pm$ 2.7 &  10.1 $\pm$ 0.5 & 1.1 $\pm$ 1.0\\
     \hline
    3D RHD 065 & 2016 & 1.53 & 4.1 $\pm$ 0.2 & 79.3 & 6.0 $\pm$ 0.2 & 17.9 & 2.8\\
    & & 1.58 & 4.1 $\pm$ 0.2 & 82.2 & 6.1 $\pm$ 0.2 &  15.3 &  2.5\\
    & & 1.63 & 4.2 $\pm$ 0.2 & 84.2 & 6.1 $\pm$ 0.2 &  12.8 &  3.0\\
    & & 1.68 & 4.2 $\pm$ 0.2 & 88.2 & 7.2 $\pm$ 0.2 &  10.1 &  1.7\\
    & & 1.72 & 4.2 $\pm$ 0.2 & 87.8 & 7.9 $\pm$ 0.2 &  11.5 &  0.7\\
    & & 1.77 & 4.2 $\pm$ 0.2 & 86.7 & 7.9 $\pm$ 0.2 &  12.5 &  0.8\\
    & & Average & 4.2 $\pm$ 0.2 & 84.7 $\pm$ 3.5 & 6.9 $\pm$ 0.9 &  13.4 $\pm$ 2.8&  1.9 $\pm$ 1.0\\
     \hline
    3D RHD 067& 2019 & 1.53 & 4.2 $\pm$ 0.2 & 86.3 & 5.9 $\pm$ 0.2 & 10.8 & 2.9 \\
    & & 1.58 & 4.3 $\pm$ 0.2 & 88.7 & 6.0 $\pm$ 0.2 &  8.9 &  2.4 \\
    & & 1.63 & 4.3 $\pm$ 0.2 & 89.7 & 6.7 $\pm$ 0.2 &  8.7 &  1.6 \\
    & & 1.68 & 4.3 $\pm$ 0.2 & 88.8 & 6.3 $\pm$ 0.2 &  9.3 &  1.9 \\
    & & 1.72 & 4.2 $\pm$ 0.2 & 76.4 & 5.5 $\pm$ 0.2 &  22.3 &  1.3 \\
    & & 1.77 & 4.2 $\pm$ 0.2 & 77.4 & 6.0$\pm$ 0.2 &  21.1 &  1.5 \\
    & & Average & 4.2 $\pm$ 0.2 & 84.5 $\pm$ 6.0 & 6.1$\pm$ 0.4 &  13.5 $\pm$ 6.4 & 1.9 $\pm$ 0.6 \\
    \hline
    \end{tabular}
    \begin{flushleft}
    \footnotesize{\textbf{Notes.} $\Theta_{1}$ represents the Rosseland angular diameter in the case of the PHOENIX model, and the layer where r/R$_{star}$ = 1 in the 3D RHD model. $\Theta_\mathrm{UD}$ is the angular diameter of the uniform disk describing the MOLsphere. Finally, $f_{1}$, $f_\mathrm{UD}$, and $f_\mathrm{free}$ describe the relative flux of the PHOENIX/3D RHD, MOLsphere, and  free zero visibility scale components.
    }
    \end{flushleft}
\end{table*}

\section{Conclusions}

Our new VLTI/PIONIER visibility data sets of V602 Car indicate an overall spherical stellar disk and an extended molecular layer, similarly to what has been detected in previous observations. The same data also indicate the presence of substructures within the stellar disk at the epochs 2016 and 2019. 
In order to further probe the stellar surface of V602 Car we obtained aperture synthesis images
using two different reconstruction packages: SQUEEZE and IRBis. Both packages resulted in very similar results.  
The reconstructed images revealed a bright arc-like feature toward the northern rim of the stellar surface of the RSG V602 Car in 2016. Three years later, in 2019, an arc-like feature appeared at a different orientation and a new peak of emission emerged on the opposite side of the stellar surface.
The flux contribution caused by the extended molecular layer is present in the reconstructed images, but it is not clearly visible because it lies close to our achieved dynamic range. We can therefore not constrain its morphology.

We compared the reconstructed images to the latest 3D RHD simulations of RSGs. There is a fundamental problem in direct comparisons of such simulated snapshots to our data: the surface pattern changes in a stochastic way and never repeats itself. With a finite number of available simulated snapshots, we cannot expect any snapshot to coincide exactly with the pattern at one of our observed epochs. A classic $\chi^2$ comparison between model and observation is thus not appropriate and may lead to spurious results. As a solution to this problem, we introduced the use of the SSIM to find the most similar of the model snapshots compared to our reconstructed image. 
This comparison resulted in the identification of 8 and 4 adjacent snapshots (out of 81 total) that are equally similar to the observational data obtained in 2016 and 2019, respectively. The SSIM was 0.85 and 0.87, respectively, indicating that none of the snapshots coincides perfectly with our observed epochs, but some show a high degree of similarity. We concluded that, within our limitations in angular resolution and dynamic range, the observed stellar surface features of V602 Car can be 
reproduced by the physics accounted for in the simulations we considered at two individual epochs. Further observations at higher spatial and temporal resolution are needed to confirm the agreement.  We interpreted the observed surface features to be related to 
instationary convection. The structure is further convolved to larger 
observed patches on the stellar surface with our observational spatial 
resolution. As a result, we lose information on the sizes of individual granules
or convection cells.
The time during which the structure of the most similar snapshots remains stable is $\sim$70 days.
As a more quantitative method of comparing observational data and simulations, we computed the contrast of the best snapshots and found agreement with the contrast of the reconstructed images (within the associated errors).

Although the observed stellar surface structure can be nicely explained by the 3D RHD models, the simulations alone are not able to reproduce the observed visibility data. An additional extended molecular component is still needed, pointing to the current limitations of RHD simulations of RSG stars, as found in \citet{2015A&A...575A..50A}. 
While the effects of convection on the stellar surface may be nicely described by current 3D simulations and the physics they contain, convection alone may not be the only relevant process to levitate the atmosphere, which is the first step of the mass-loss process.   

\begin{acknowledgements}
      J.B.C., J.M.M, and J.C.G. were partially supported by the Spanish MINECO projects AYA2012-38491-C02-01 and AYA2015-63939-C2-2-P and by the Generalitat Valenciana project PROMETEO/2009/104 and PROMETEOII/2014/057. We thank Karl-Heinz Hofmann for his help using the IRBis image reconstruction package.
\end{acknowledgements}

\bibliographystyle{aa_url} 
\bibliography{references}

\newpage
\onecolumn
\begin{appendix}

\section{Comparison of the pre-NAOMI 2016 and post-NAOMI 2019 PIONIER data}
\label{sec:compNAOMI}

We obtained pre-NAOMI PIONIER data in 2016 and post-NAOMI PIONIER data in 2019 during NAOMI Science Verification.
The two data sets are very similar in terms of $uv$ coverage, time span, and atmospheric conditions. 
The total number of $uv$ points is 156 for both data sets, and their distribution is very similar 
(see Fig.~\ref{fig:uvplane}). The 2016 data span  81 days, while those of 2019 span  70 days.
The atmospheric conditions were similar (see Table~\ref{table:logs}) with an average seeing of 0.56\,\arcsec and
average coherence time of 4.3\,msec in 2016, and of 0.64\arcsec and 5.7 msec in 2019. Both data sets
used the same interferometric calibrator and the same observational and data reduction strategies.

The average error of the squared visibility amplitudes $\sigma(V^2)/V^2$ is 10.2\% in 2016 and 5.6\% in 2019.
The average error of the closure phases is 1.8$^{\circ}$ in 2016 and 0.9$^{\circ}$ in 2019. 
The scatter of the visibility points, in particular at short baselines is much reduced in the 2019 data set
compared to the 2016 data set (cf. Fig.~\ref{fig:visandclos}).
As otherwise the two data sets are very comparable, we attribute this improvement in precision and accuracy
of the visibility data to the addition of the adaptive optics system NAOMI.
Consequently, our average estimated pixel errors of the reconstructed images, based on different tests as outlined
in detail in Appendix ~\ref{sect:allapendix}, has improved from 17\% to 14\% for 2016 and 2019, respectively.

\section{Estimation of image errors}\label{sect:allapendix}

In order to ensure the validity of the substructures found in Fig. \ref{fig:allchannels}, we probed different systematic errors due to three effects:

i) Errors inside the SQUEEZE reconstruction package (see details in  Appendix \ref{sect:errors_squeeze});

ii) Errors due to the reconstruction package used (see details in Appendix \ref{sect:irbis});

iii) Errors due to effects of our limited \textit{uv} coverage (see details in Appendix \ref{sect:errors_uv}).

\subsection{Errors within SQUEEZE}\label{sect:errors_squeeze}

\begin{figure}[h]
    \centering
    \includegraphics[width=0.33\linewidth]{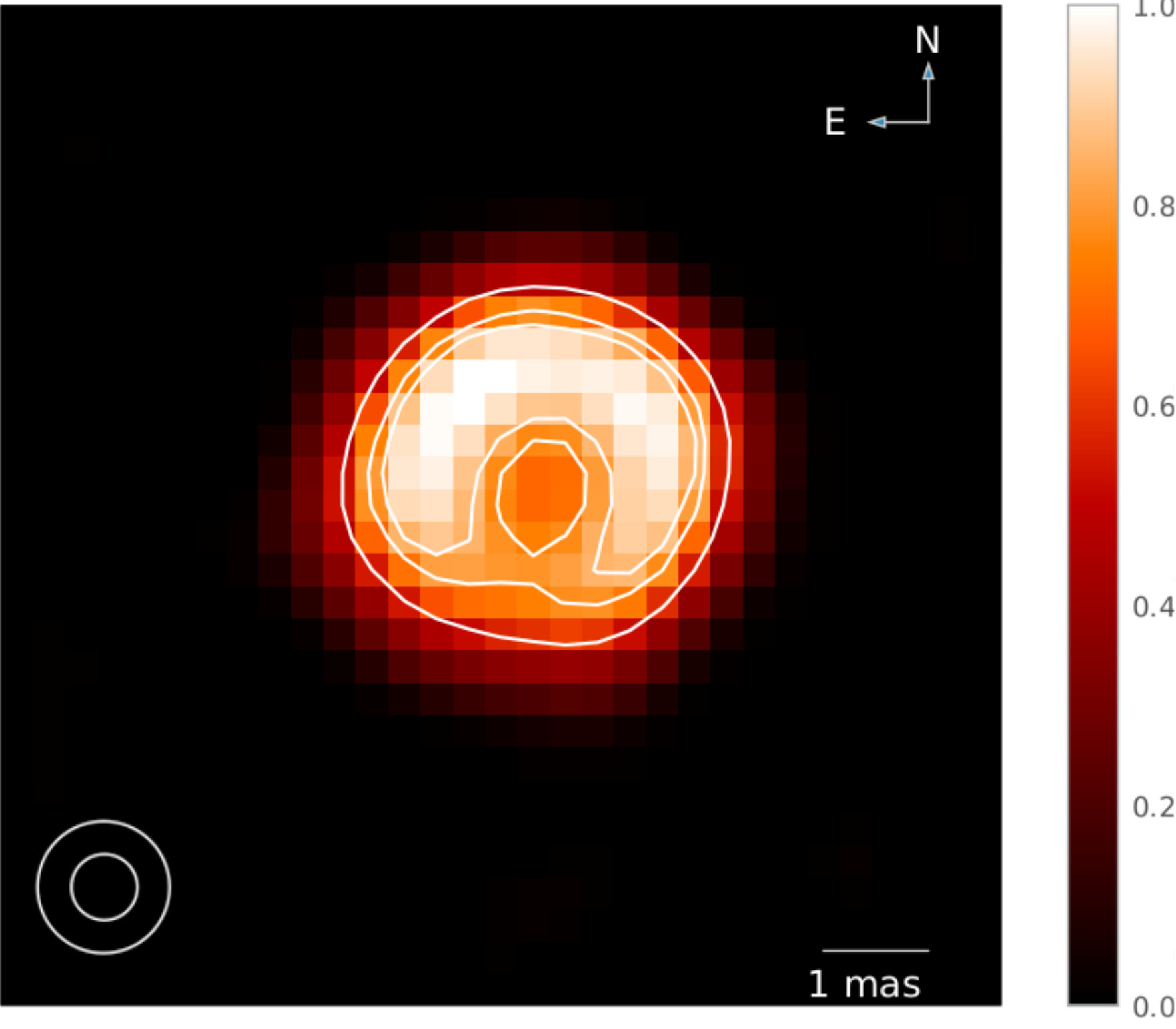}    \includegraphics[width=0.33\linewidth]{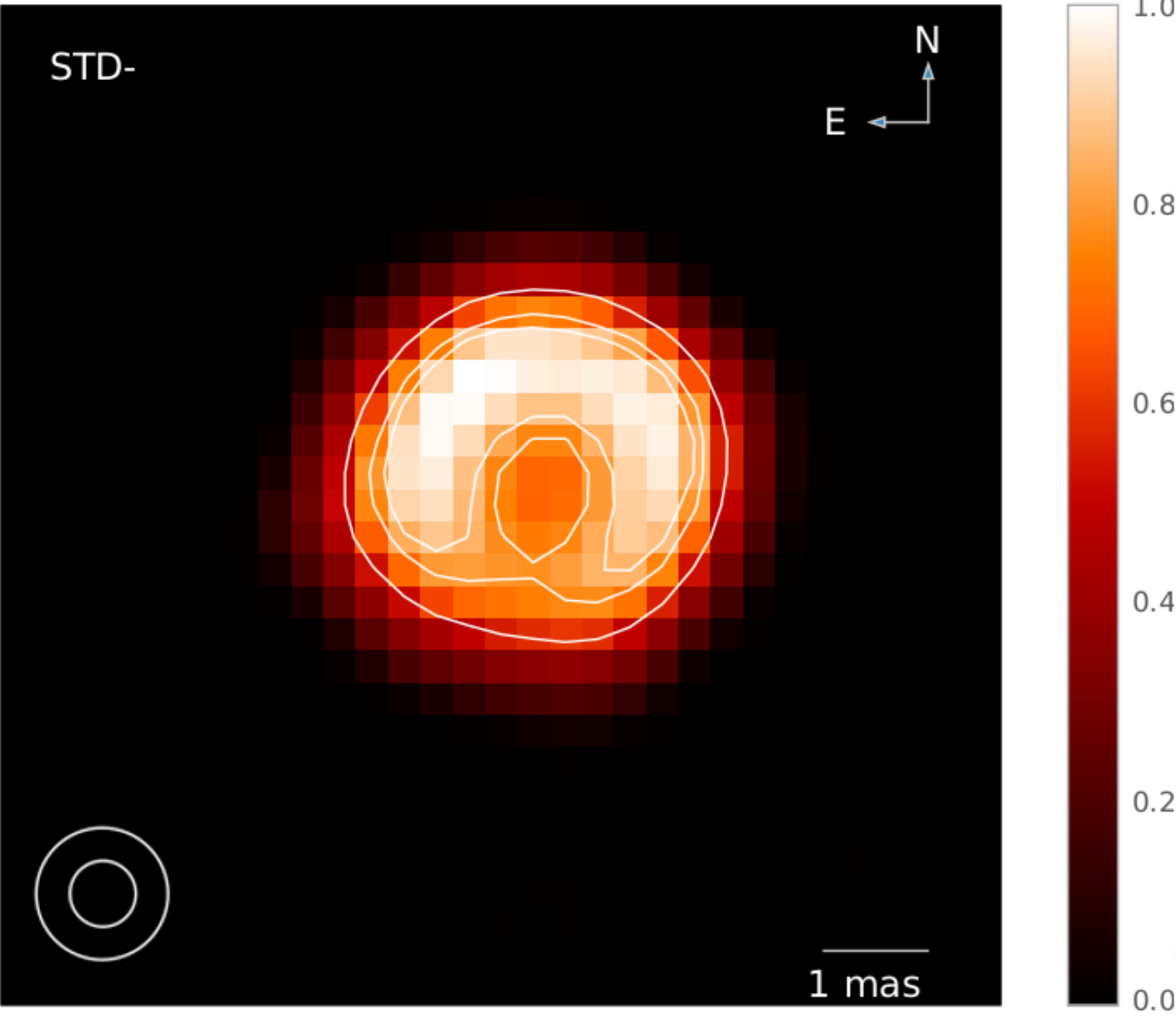}
    \includegraphics[width=0.33\linewidth]{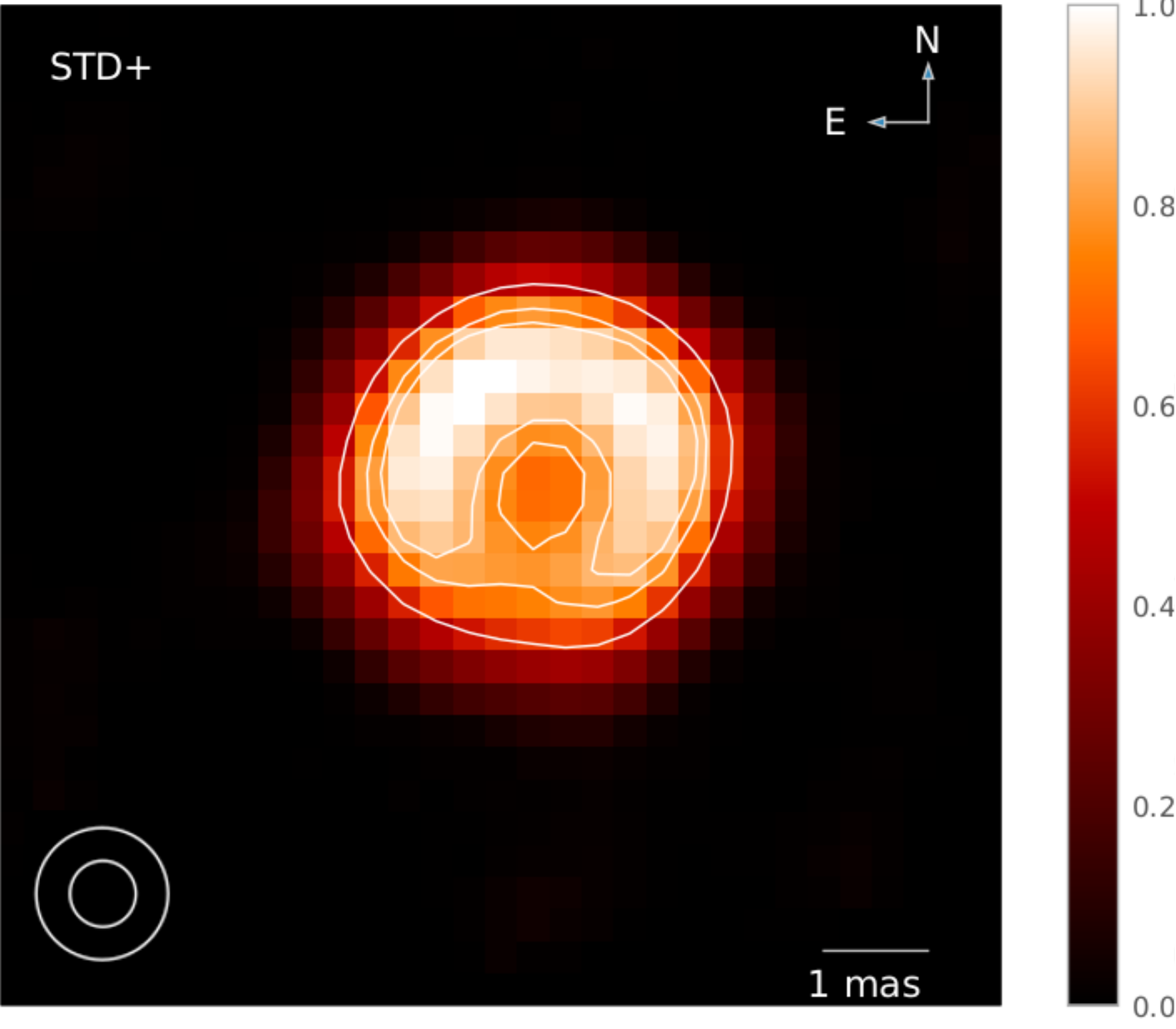}
    \includegraphics[width=0.33\linewidth]{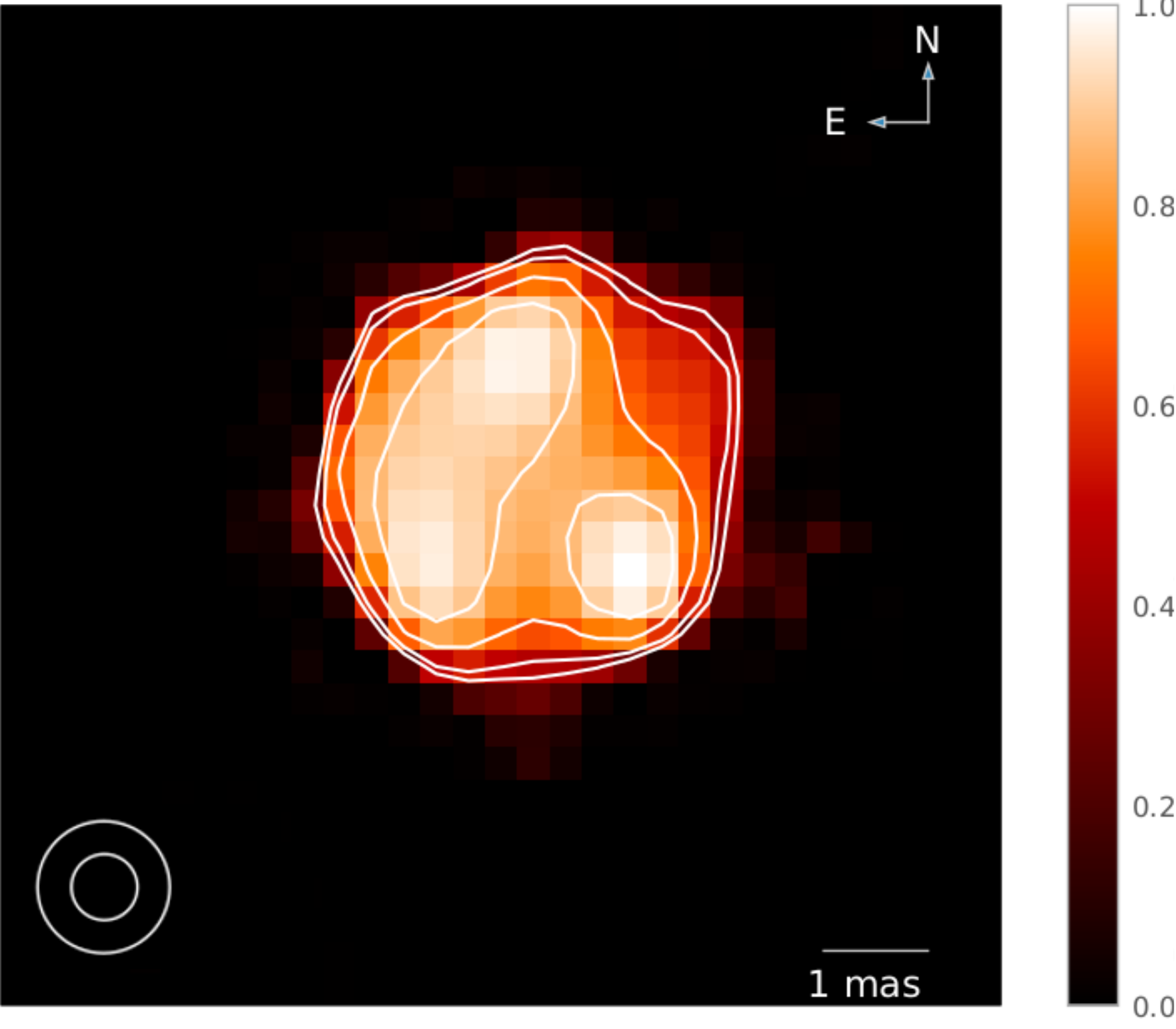}    \includegraphics[width=0.33\linewidth]{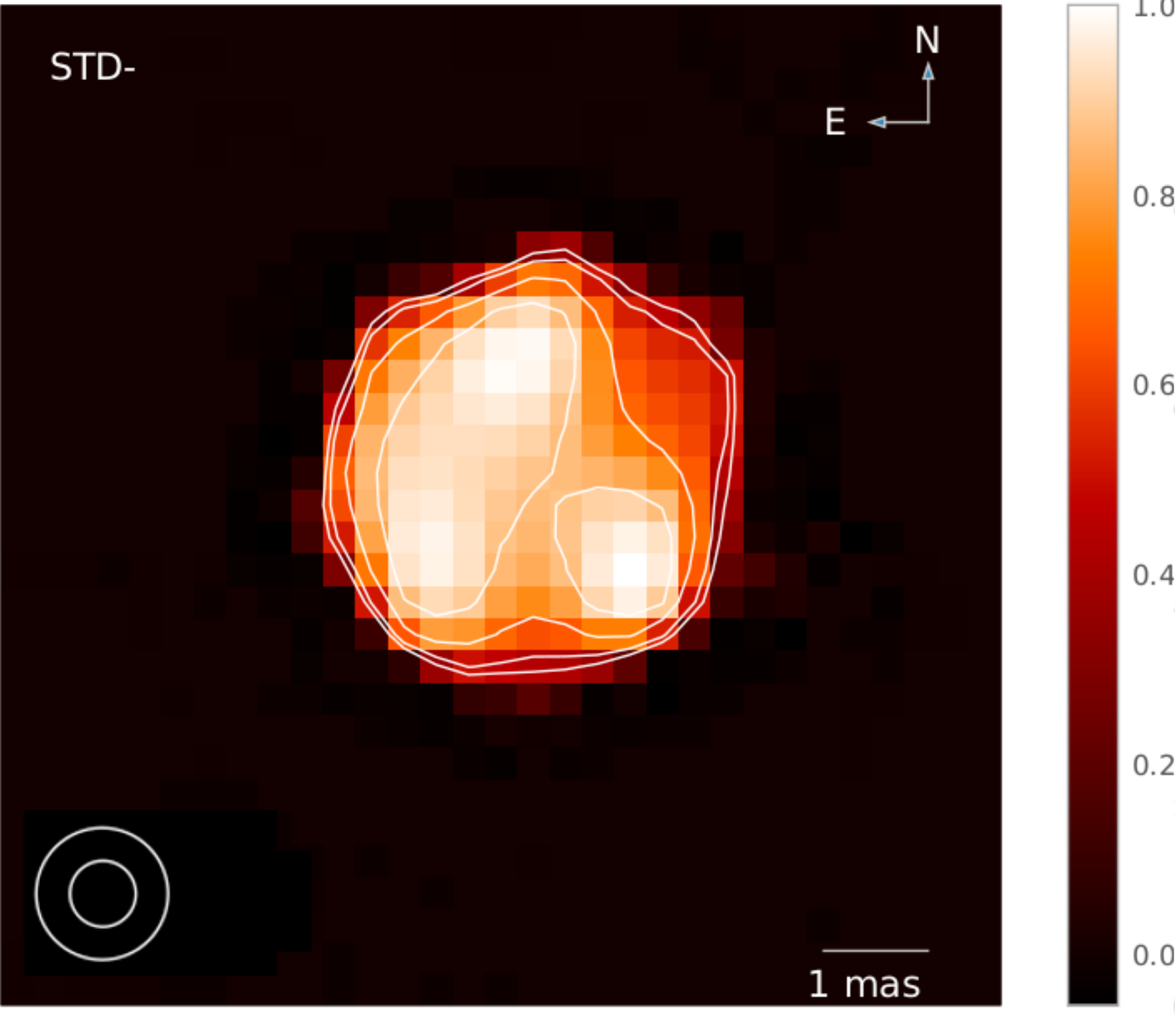}
    \includegraphics[width=0.33\linewidth]{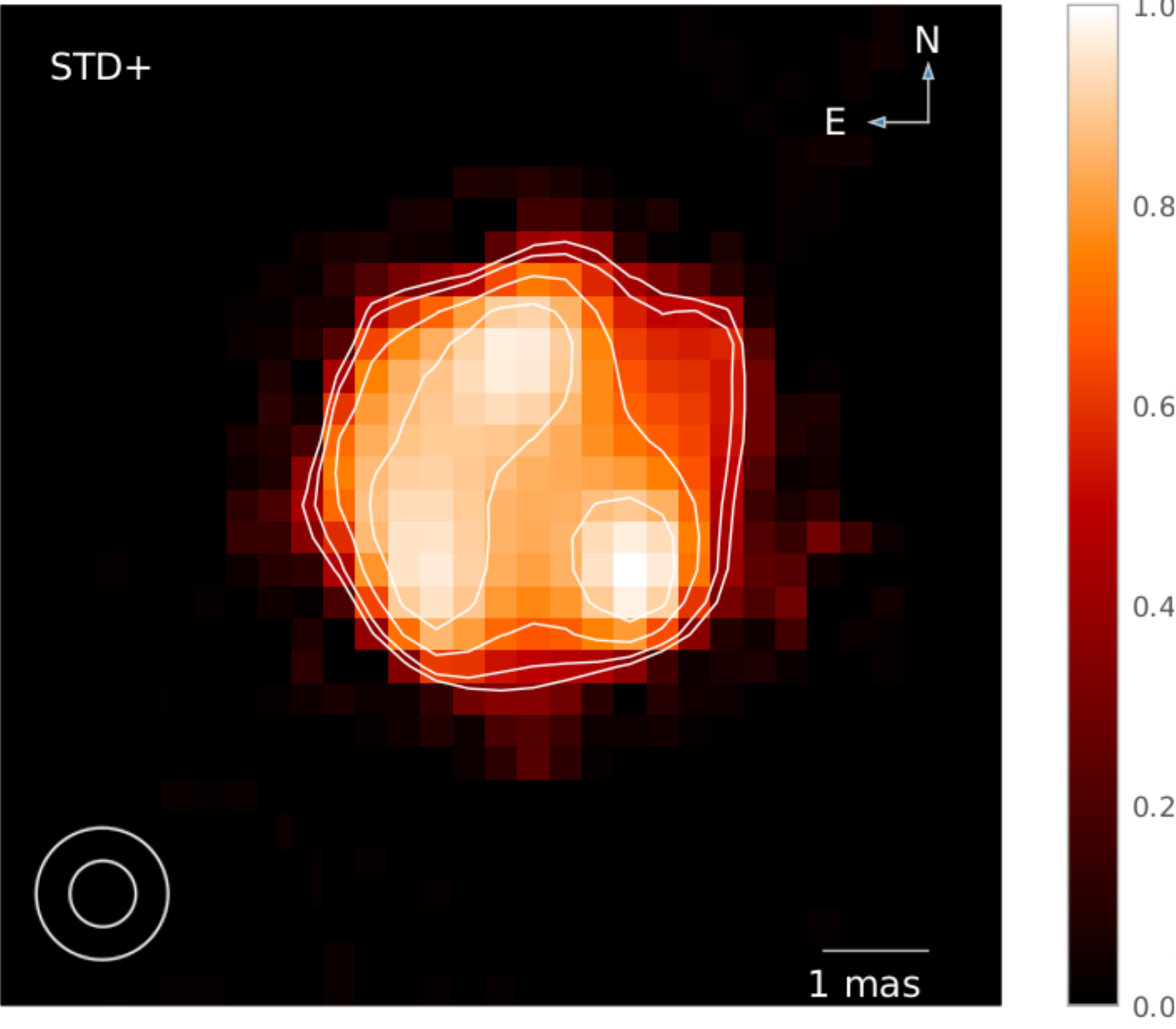}
    \caption{
    Squeeze average images (left column), images one standard deviation below the average (middle column), and images one standard deviation above the average (right column). The top row represents 2016 data (contours at 55\%, 77\%, and 85\% of peak intensity), while the bottom row represents the 2019 data (contours at 40\%, 50\%, 70\%, and 87\% of peak intensity). 
    }
    \label{fig:standard_deviation}
    \end{figure}

To obtain the image that is shown in Fig. \ref{fig:allchannels}, we computed 50 different SQUEEZE images and averaged them. 
In a similar fashion to that used by \citet{2018Natur.553..310P}, Fig. \ref{fig:standard_deviation} shows the images one standard deviation above (and below) the average image for 2016 and 2019 data. The persistence of the same features in these error images indicates that the substructures do not originate from the averaging procedure we followed.

\subsection{IRBis reconstruction and comparison with SQUEEZE}\label{sect:irbis}

Following a similar method to that used  for the carbon AGB star R Scl \citep{2017A&A...601A...3W}, we used the IRBis reconstruction package as follows: 

i) We selected as start images the best-fit model from the PHOENIX + UD model discussed in Sect. \ref{sect:data_analysis}. 

ii) We used a flat prior and the six available regularization functions of IRBis. For each regularization we tested reconstructions with decreasing values of the hyperparameter $\mu$ and increasing radii of the object mask.  

iii) We chose as the final image that with the best quality derived from the $\chi^2$ values and residual ratio values of the visibility and closure phases ($q_\mathrm{rec}$ value). This image is based on regularization function 4 (edge preservation)   in the 2016 and 2019 data. 

\begin{figure}[h]
    \centering
    \includegraphics[width=0.49\linewidth]{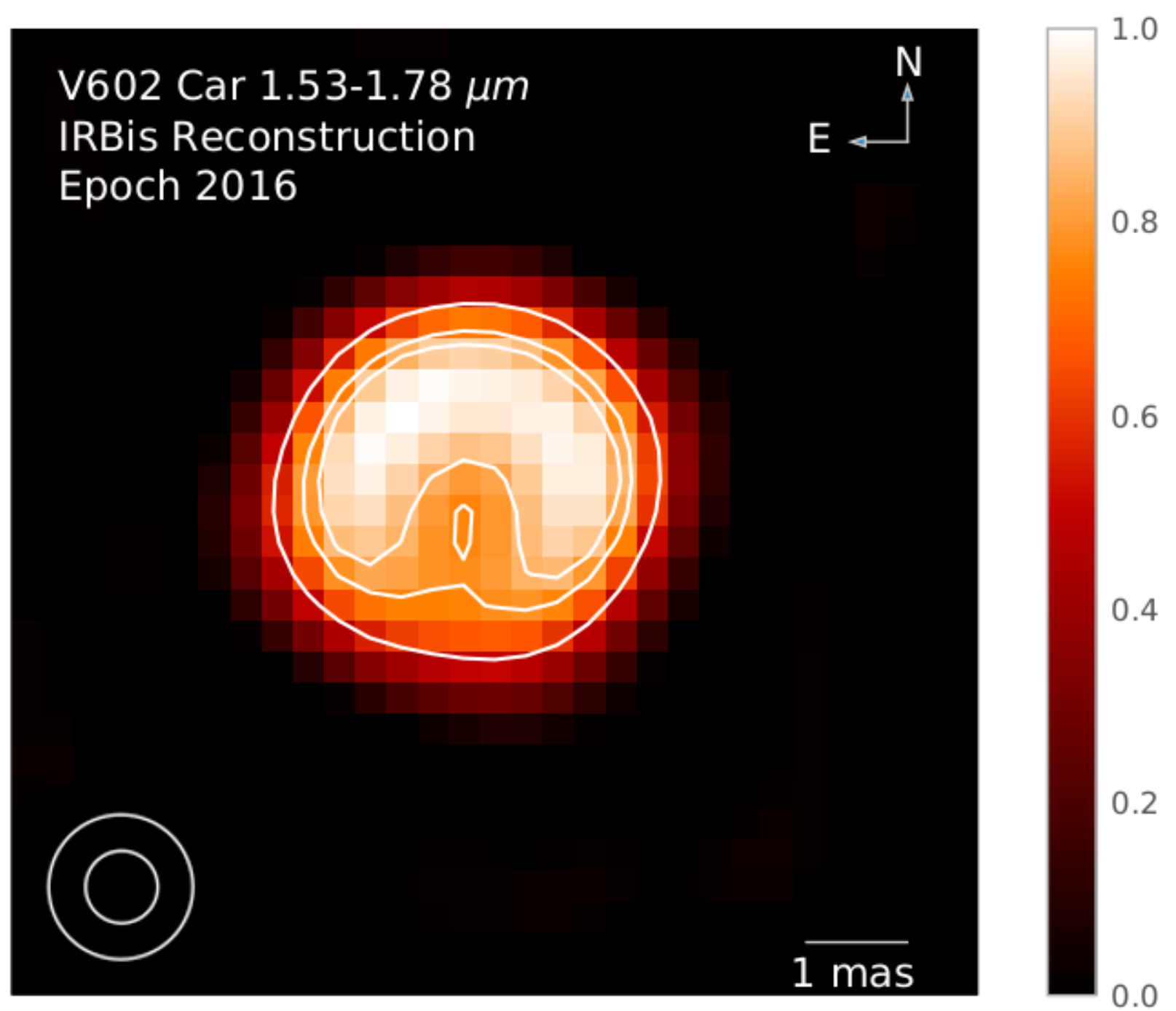}
    \includegraphics[width=0.49\linewidth]{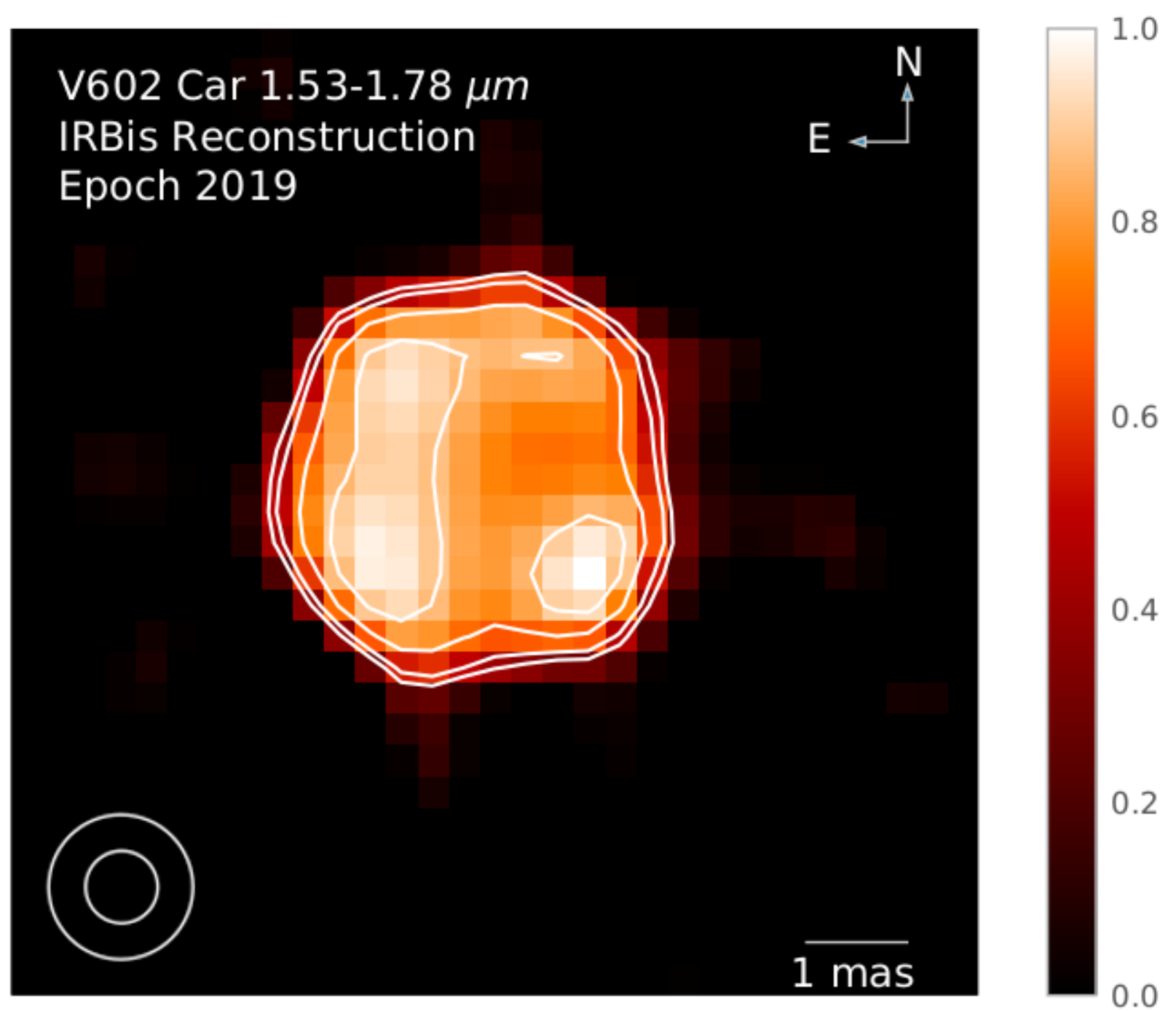}
    \caption{
    Same as Fig. \ref{fig:allchannels}, but for the IRBis reconstruction package. Contours are drawn at levels 55\%, 77\%, and 85\% of the peak intensity in 2016 and at 40\%, 50\%, 70\%, and 87\% of peak intensity in 2019.    }
    \label{fig:irbis_image}
    \end{figure}

Images obtained with regularization functions 1 (compactness), 3 (smoothness), 5 (smoothness), and 6 (quadratic Tikhonov), resulted in very similar images of similar quality parameters. Function 2 (maximum entropy) resulted in poorer reconstructions. The final images can be found in Fig.~\ref{fig:irbis_image}.

\begin{figure}[h]
    \centering
    \includegraphics[width=0.44\linewidth]{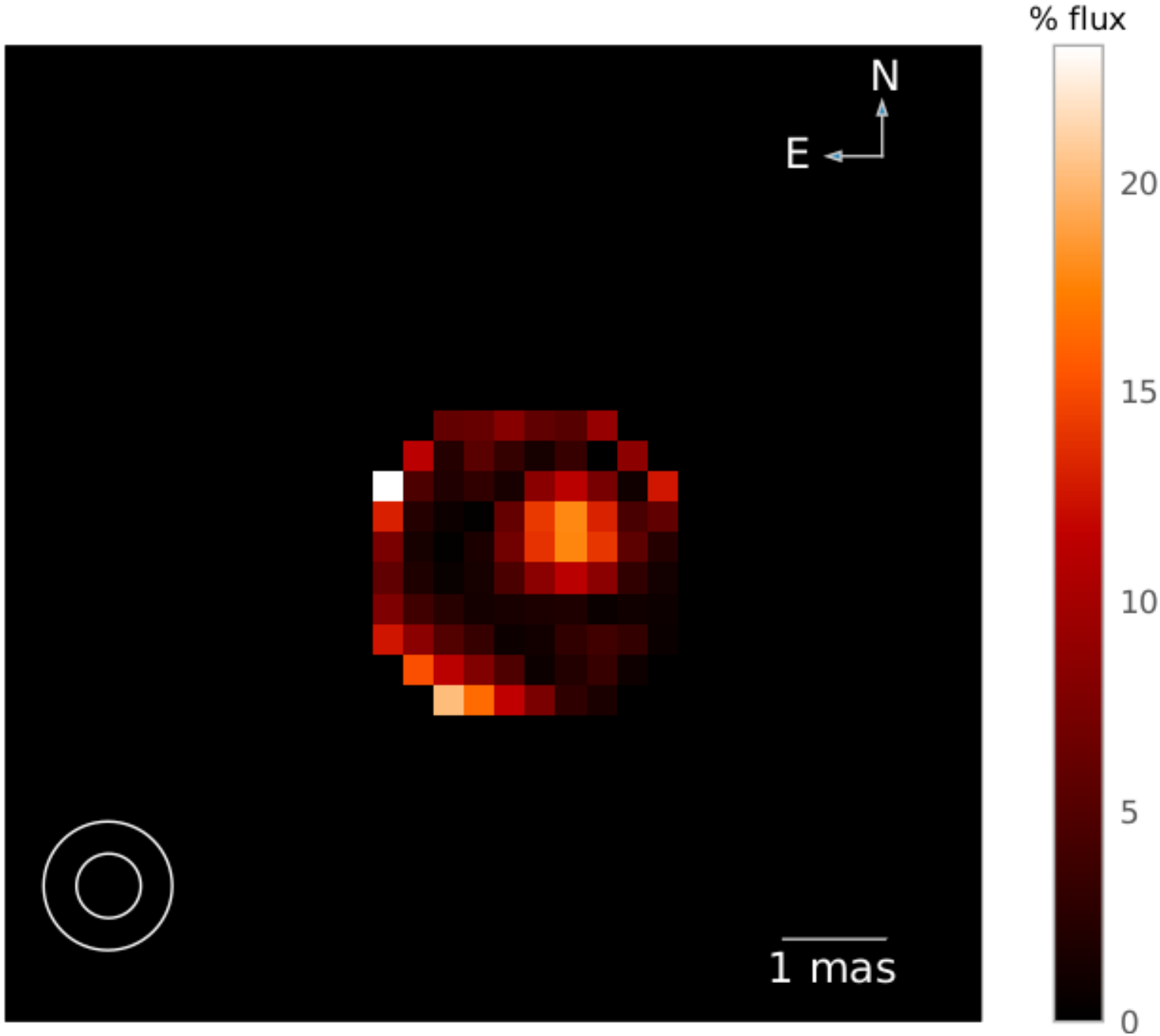}
    \includegraphics[width=0.44\linewidth]{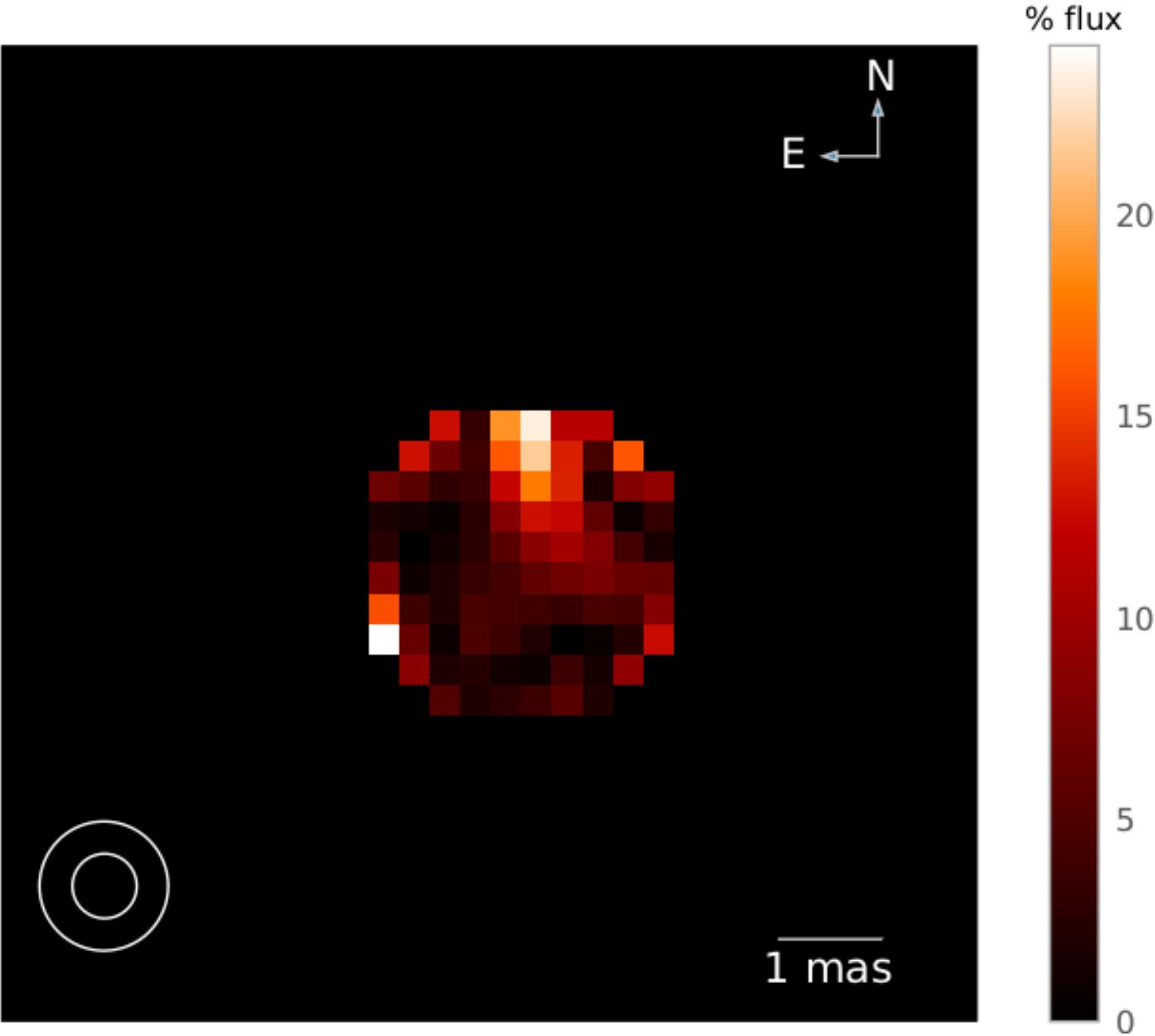}
    \caption{
    Image resulting from the pixel-by-pixel difference between SQUEEZE and IRBis reconstructions shown in Fig. \ref{fig:allchannels} and Fig. \ref{fig:irbis_image}. (\textit{Left}): 2016 epoch. (\textit{Right}): 2019 epoch. The scale represents the percentage of flux of the original SQUEEZE image. The same cut-off radius is applied as presented in the main text:  0.75 stellar radii.
    }
    \label{fig:resta}
    \end{figure}

For the same epoch, IRBis and SQUEEZE result in very similar images. Figure \ref{fig:resta} shows the difference between the two images (SQUEEZE - IRBis) evaluated pixel by pixel in terms of the flux of the SQUEEZE image. The results indicate that the same structures are present  in the SQUEEZE and the IRBis reconstruction images, although some pixels differ in intensity value by up to 24\%. 
We applied a cut-off radius of 0.75 stellar radii (as stated in the main text) in order to avoid larger, non-physical errors near the limb of the star. 

When comparing two astronomical images of the same object that only vary in the reconstruction method employed, the image differences need to originate in these methods. This scenario is equivalent to that which the SSIM was constructed for: a reference image (e.g., SQUEEZE reconstruction) and a distorted image with respect to its reference (e.g., IRBis reconstruction).  
The value obtained for these two reconstructed images is SSIM = 0.99  in the 2016 and the 2019 data. 

\subsection{Errors due to limited uv coverage}\label{sect:errors_uv}

A limited \textit{uv} coverage might produce artificial effects that could be concealed in the final image. In order to test for these effects, we simulated visibilities from the 3D RHD model snapshots using OITOOLS at the same \textit{uv} points of our observations. We  added a typical amount of noise to the image and then reconstructed it using SQUEEZE in the usual way. Since we are interested in the validity of the surface features, no MOLsphere was added. We verified that an addition of a uniform MOLsphere, as modeled in Table~2, does not change the result.

\begin{figure}[h]
    \centering
    \includegraphics[width=0.33\linewidth]{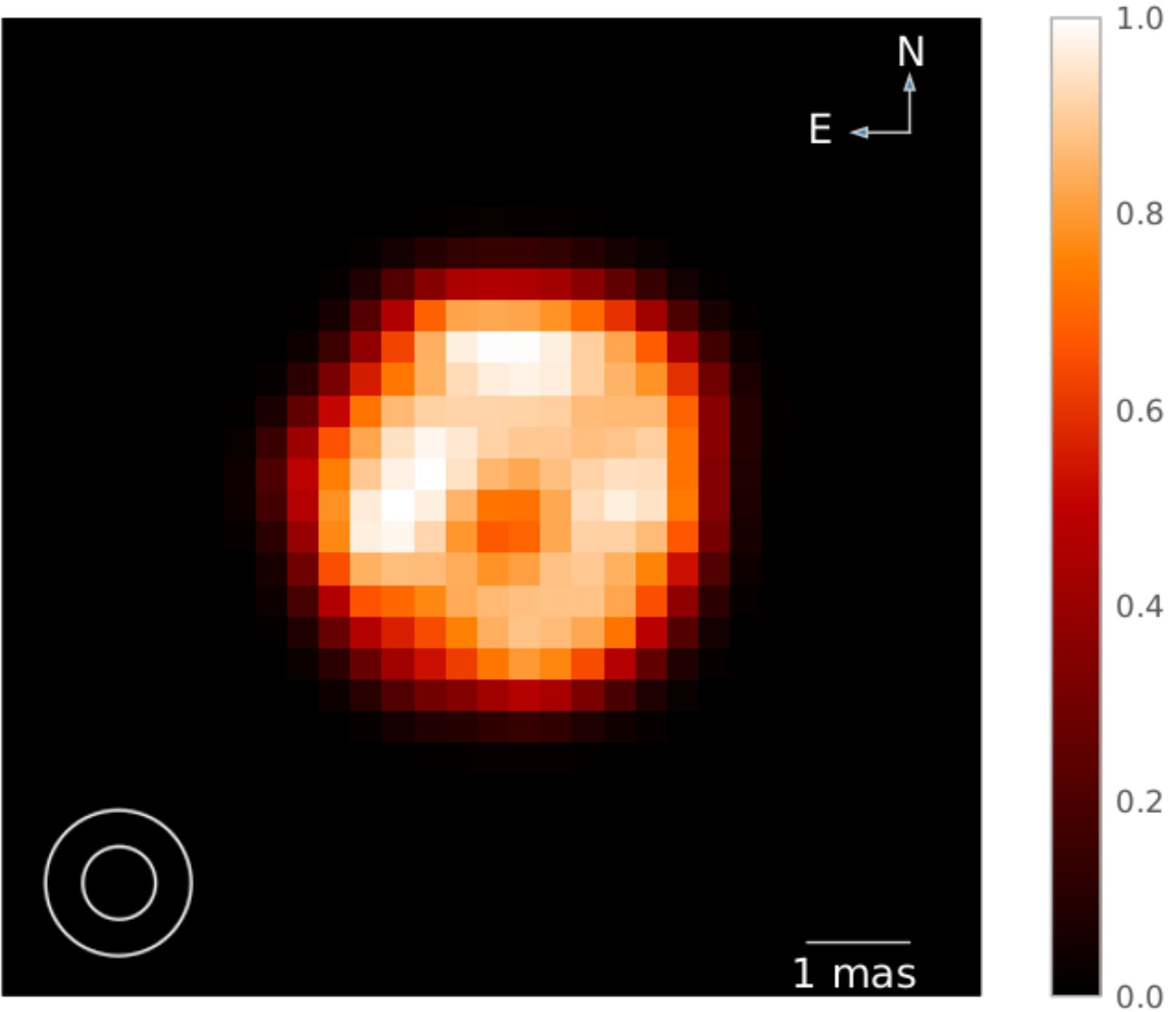}
    \includegraphics[width=0.33\linewidth]{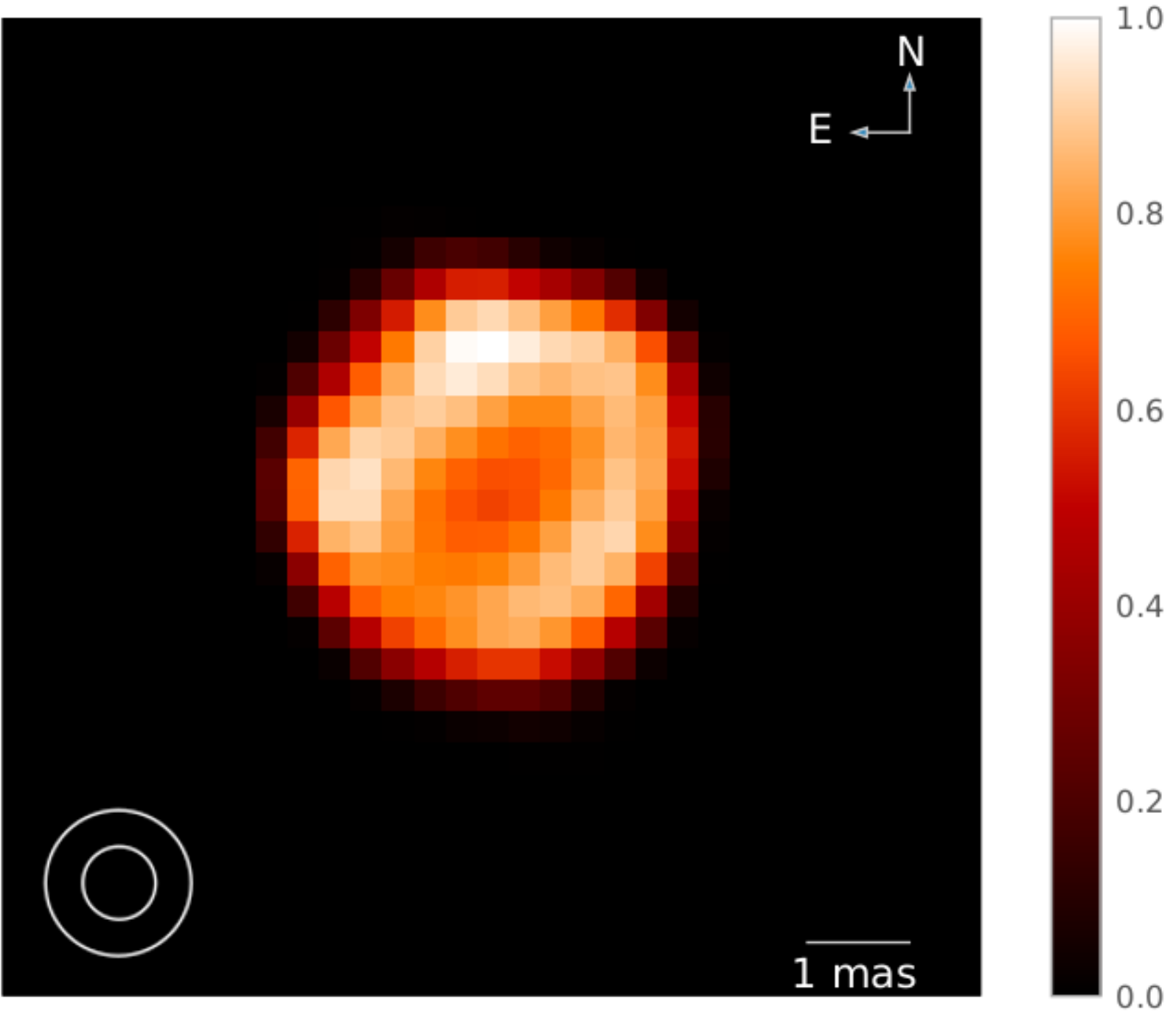}
    \includegraphics[width=0.33\linewidth]{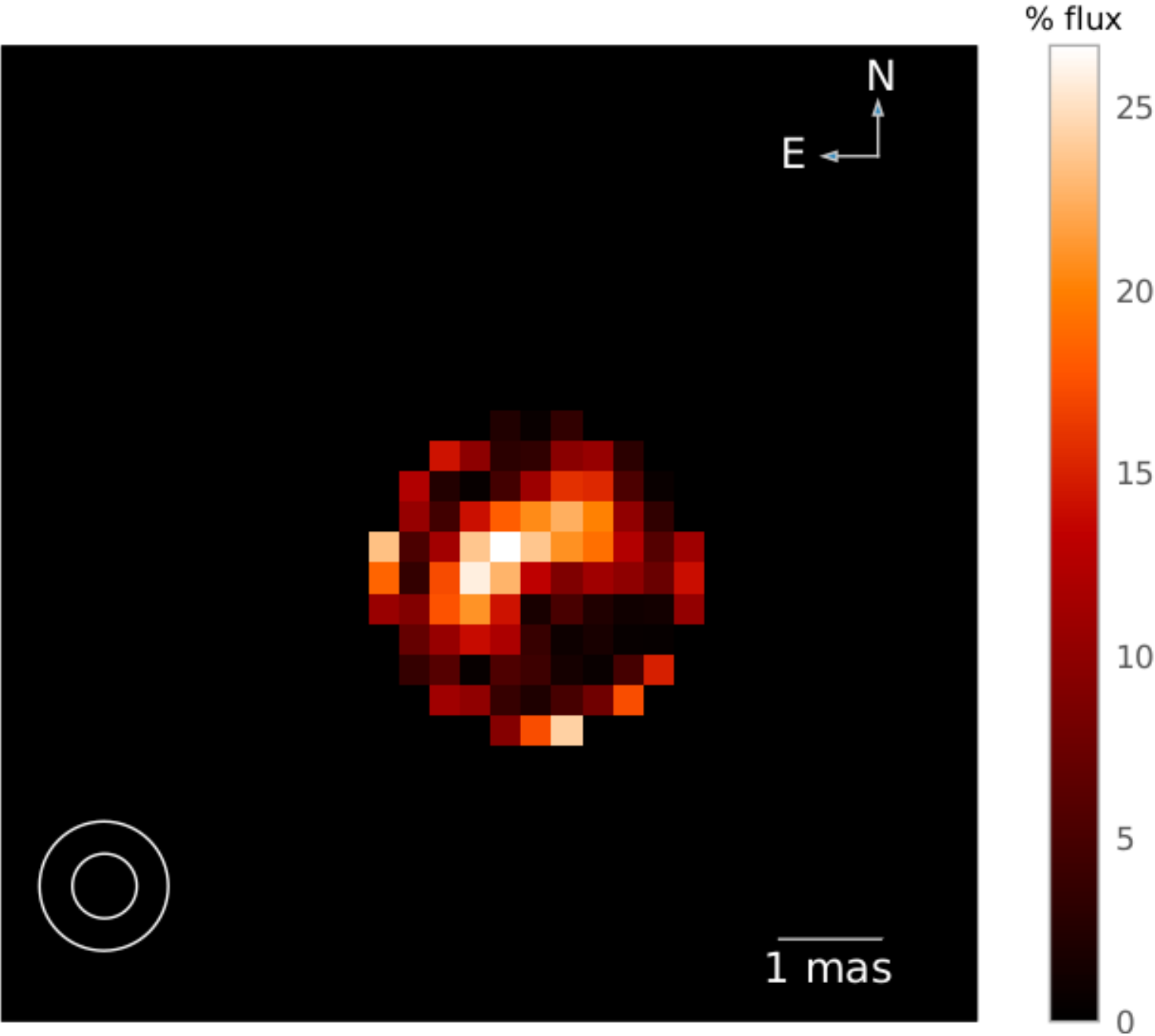}
    \includegraphics[width=0.33\linewidth]{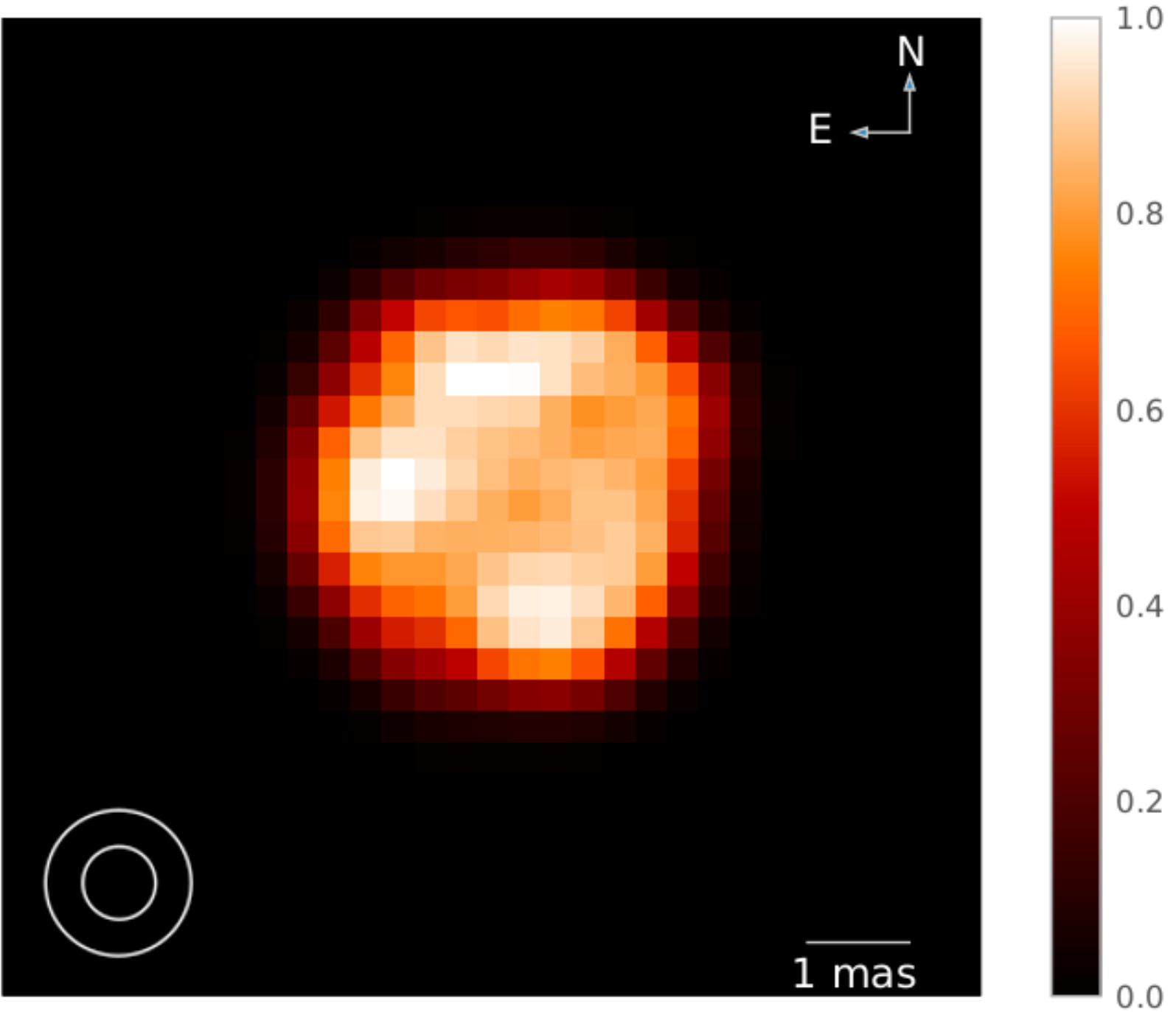}
    \includegraphics[width=0.33\linewidth]{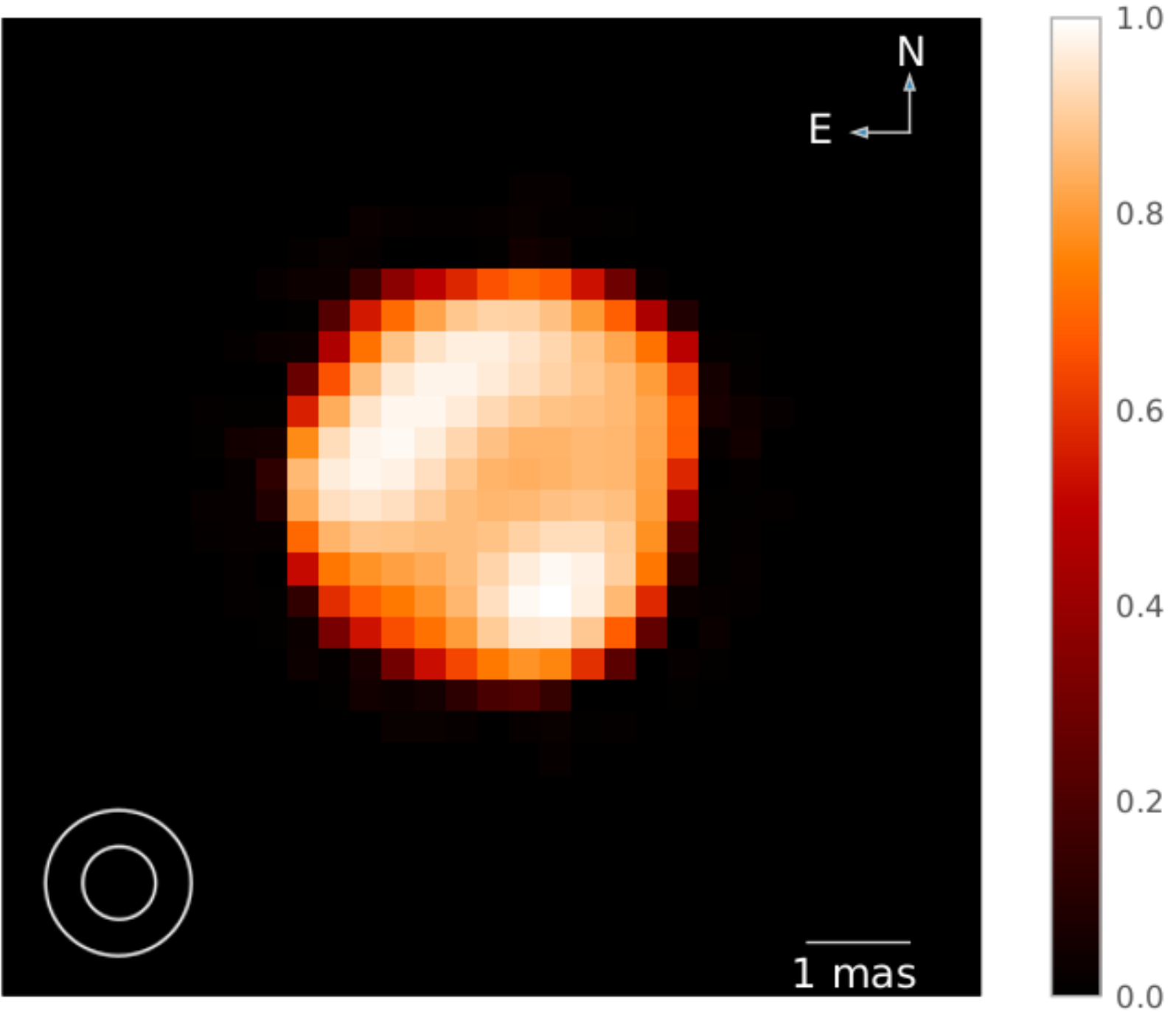}
    \includegraphics[width=0.33\linewidth]{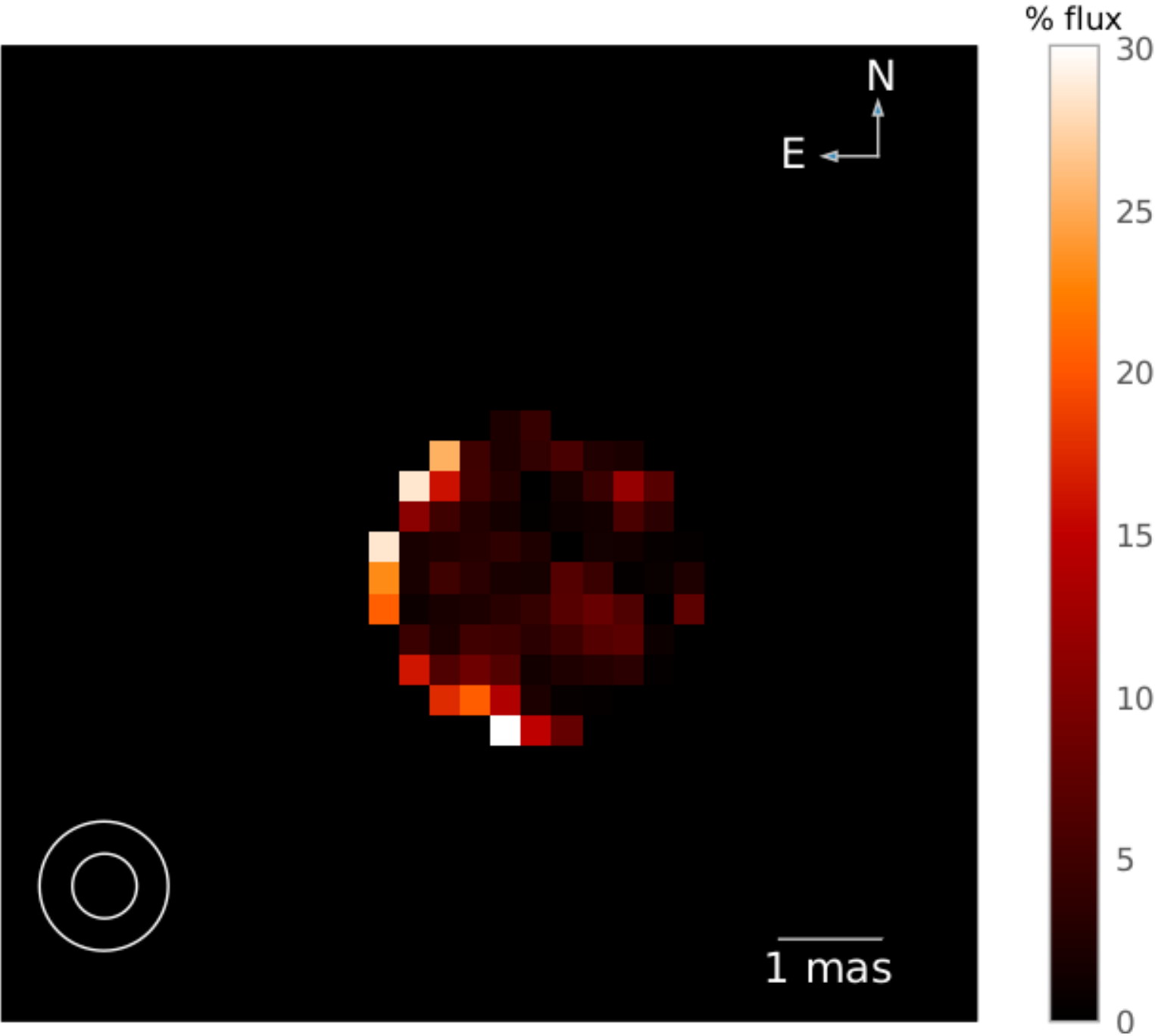}
    \caption{Original snapshot image convolved to 0.6 mas resolution (left column), SQUEEZE reconstructed image from the synthetic visibilities of the left image (middle column; not further convolved  beyond the pixel scale of 0.3 mas/pixel), and pixel-by-pixel difference image between \textit{left column} and \textit{middle column} images in terms of the original flux (right column). Top panel for snapshot 65 and bottom panel for snapshot 67. No rotation of the images was applied here. The same cut-off radius
is applied as presented in the main text:  0.75 stellar radii.
    }
    \label{fig:error_uv}
    \end{figure}

Our results (see Fig. \ref{fig:error_uv}) show that the reconstructions of the simulated visibilities (middle column of Fig. \ref{fig:error_uv}) result in a very similar image to the original  (left column of Fig. \ref{fig:error_uv}), with an SSIM = 0.90 in 2016 and SSIM = 0.89 in 2019. 
We computed these difference images with a convolved beam of 0.3, 0.6, and 1.2 mas and found that the resolution that best kept all the information about the substructures while reducing the errors is 0.6 mas, which is shown in Fig. \ref{fig:error_uv}.
Based on this analysis, we can estimate the super-resolution that we achieve with our $uv$ coverage and our noise to 0.6\,mas,  compared to the nominal resolution $\lambda/(2B_\mathrm{max})$ of 1.2\,mas.

From this discussion, we conclude that the substructures found in Fig. \ref{fig:allchannels} are probably not caused or altered by the limited \textit{uv} coverage of our observations.

\newpage
\section{Spectral channel images of V602 Car}\label{sect:2016_channels}\label{sect:2019_channels}

\begin{figure*}[h]
    \centering
    \includegraphics[width=0.49\linewidth]{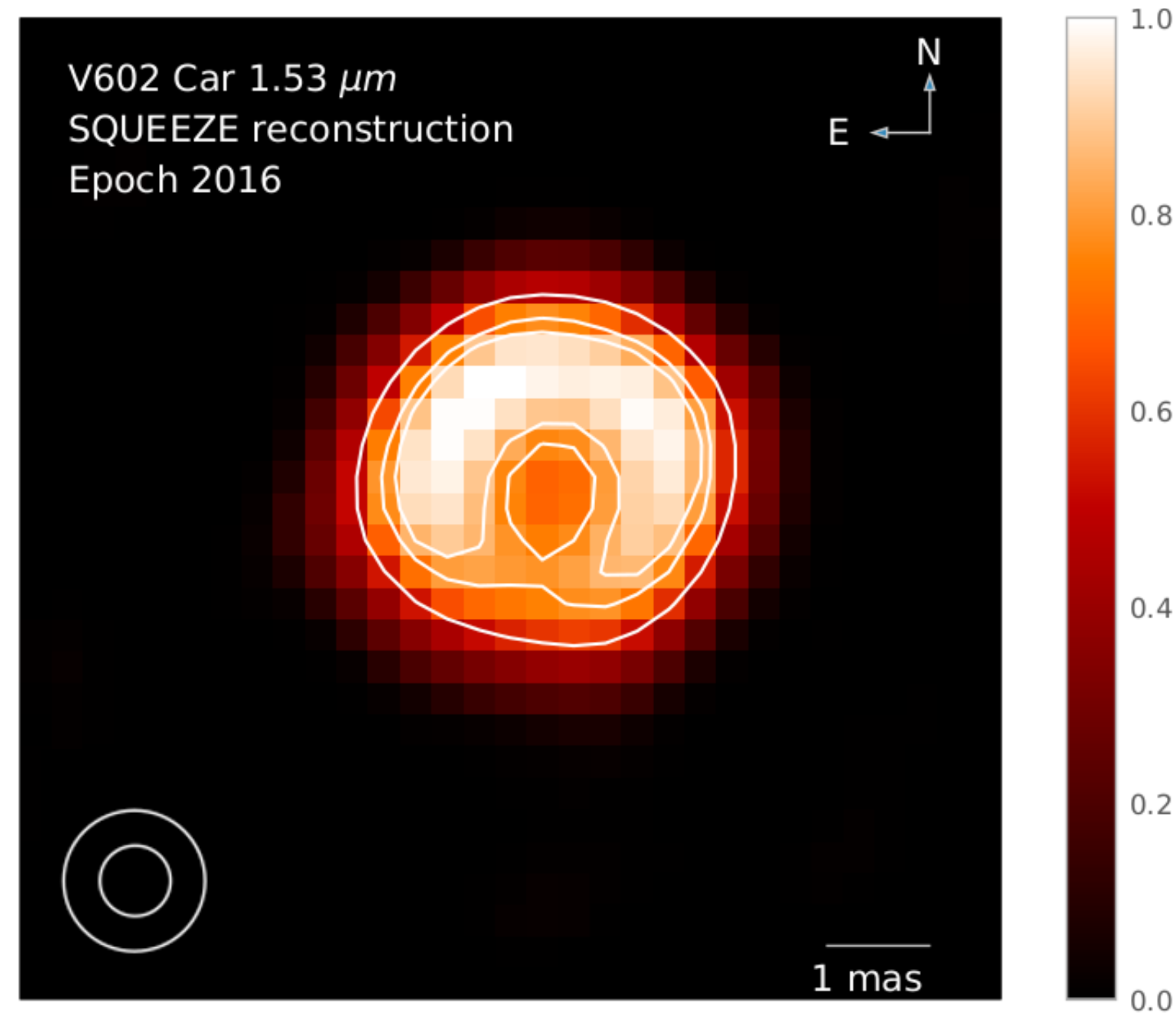}
    \includegraphics[width=0.49\linewidth]{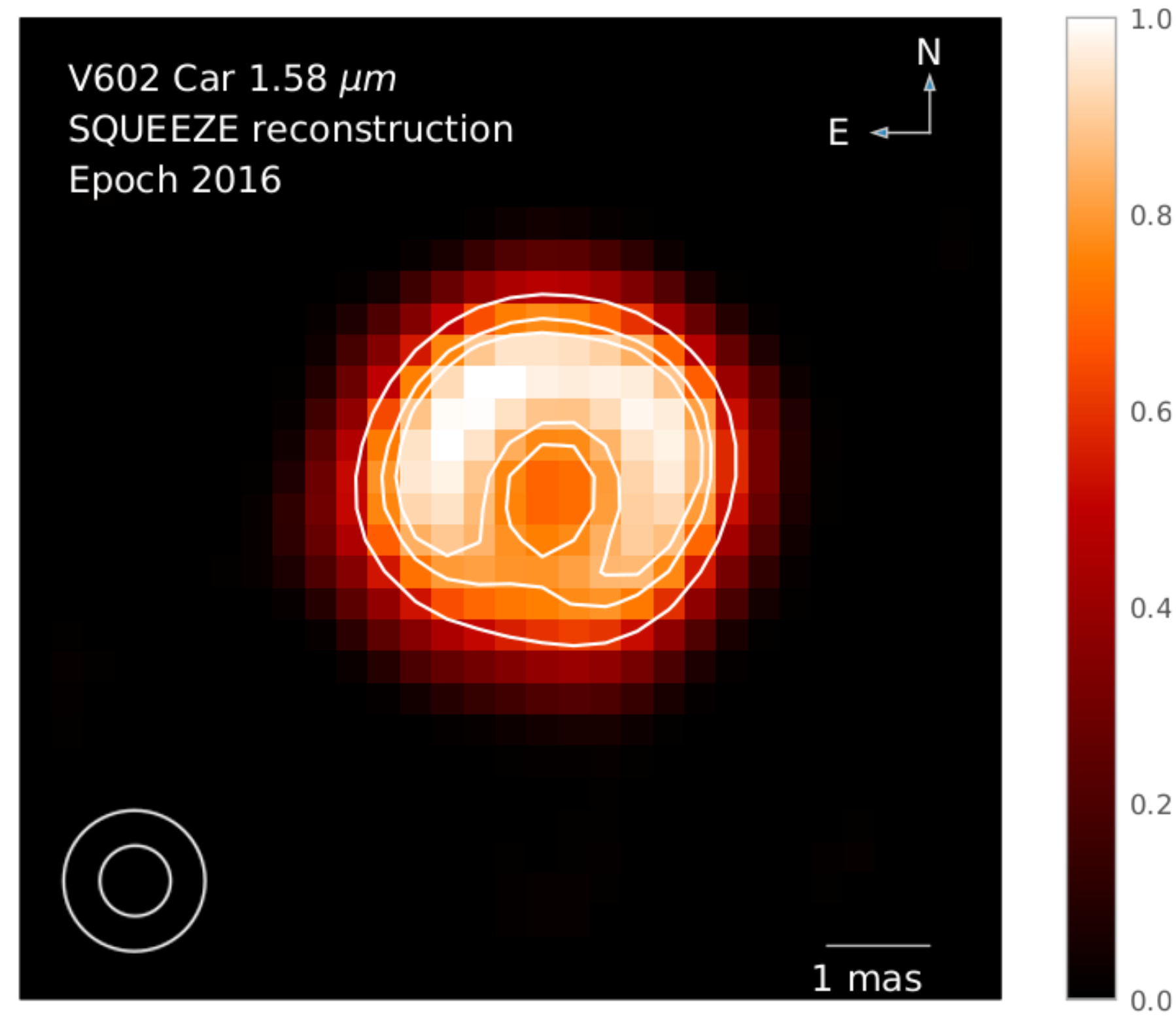}
    \includegraphics[width=0.49\linewidth]{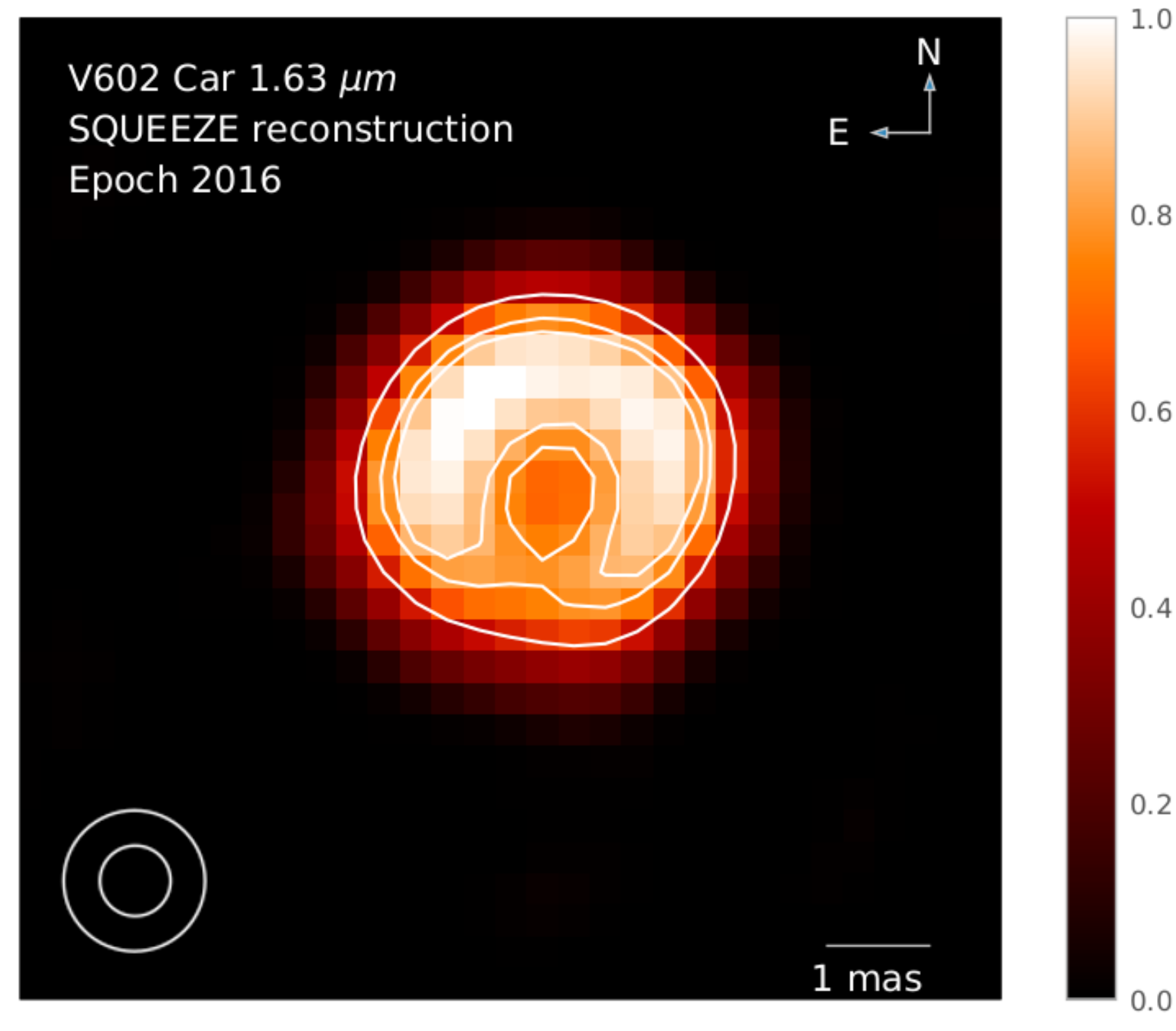}
    \includegraphics[width=0.49\linewidth]{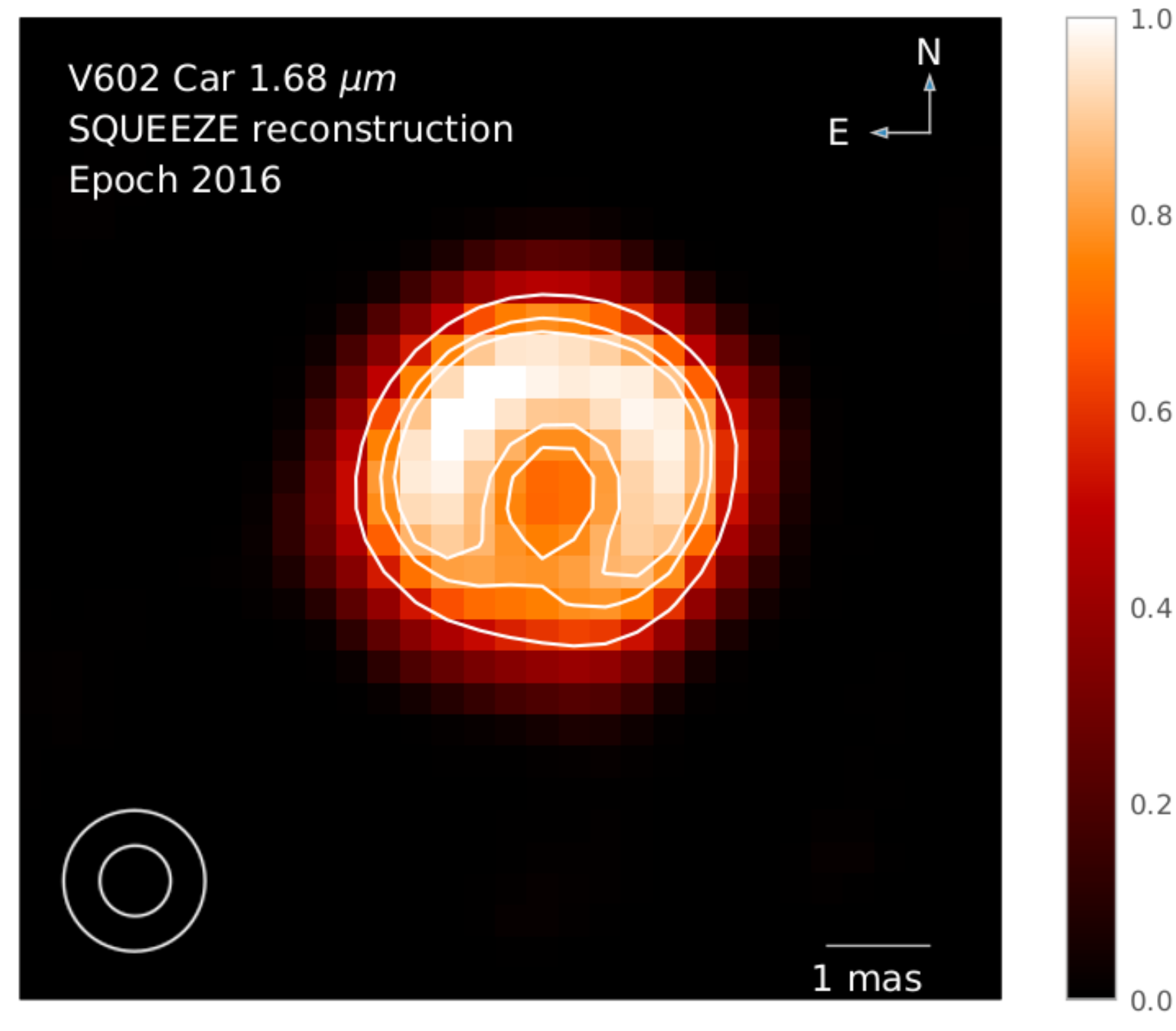}
    \includegraphics[width=0.49\linewidth]{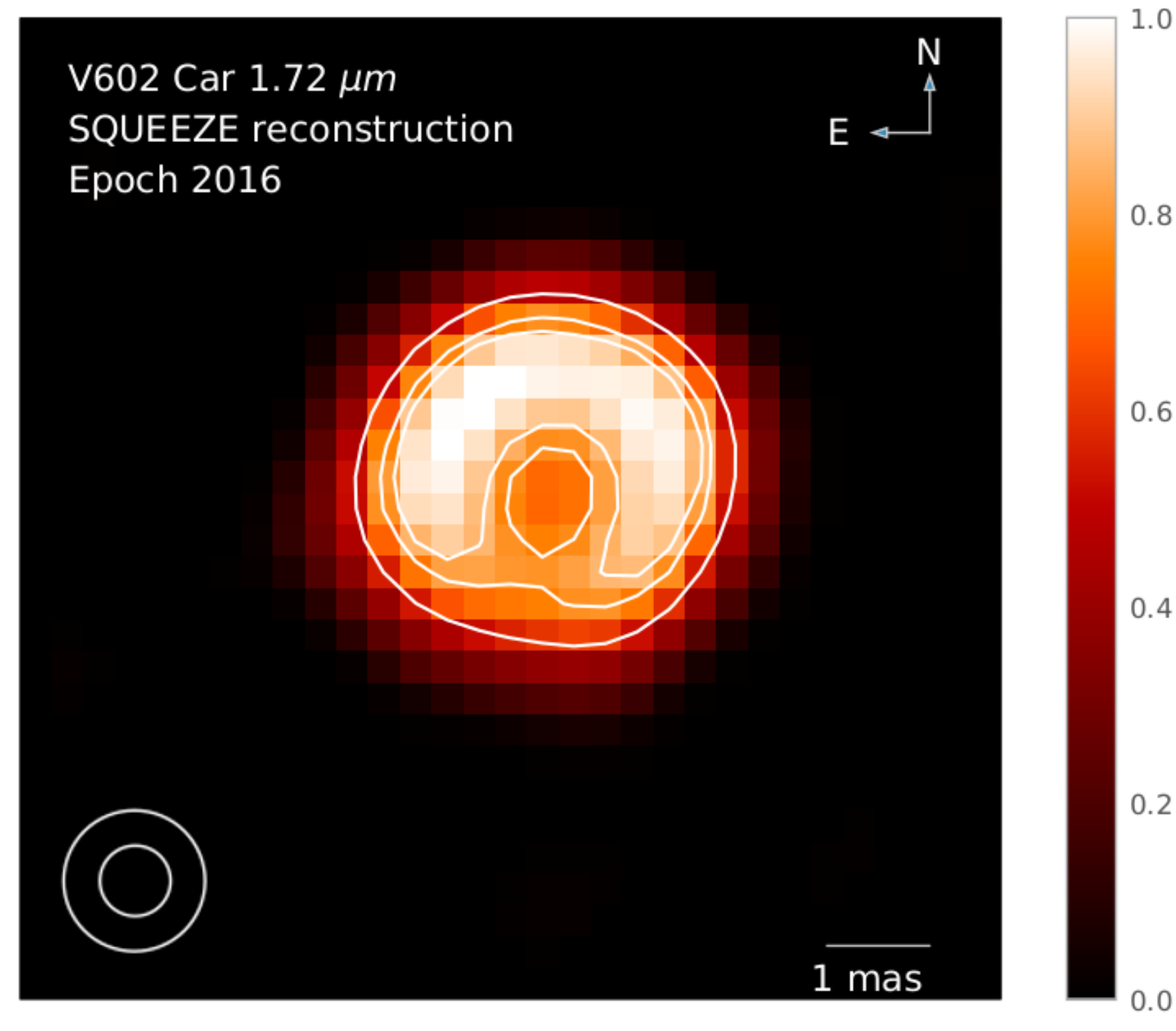}
    \includegraphics[width=0.49\linewidth]{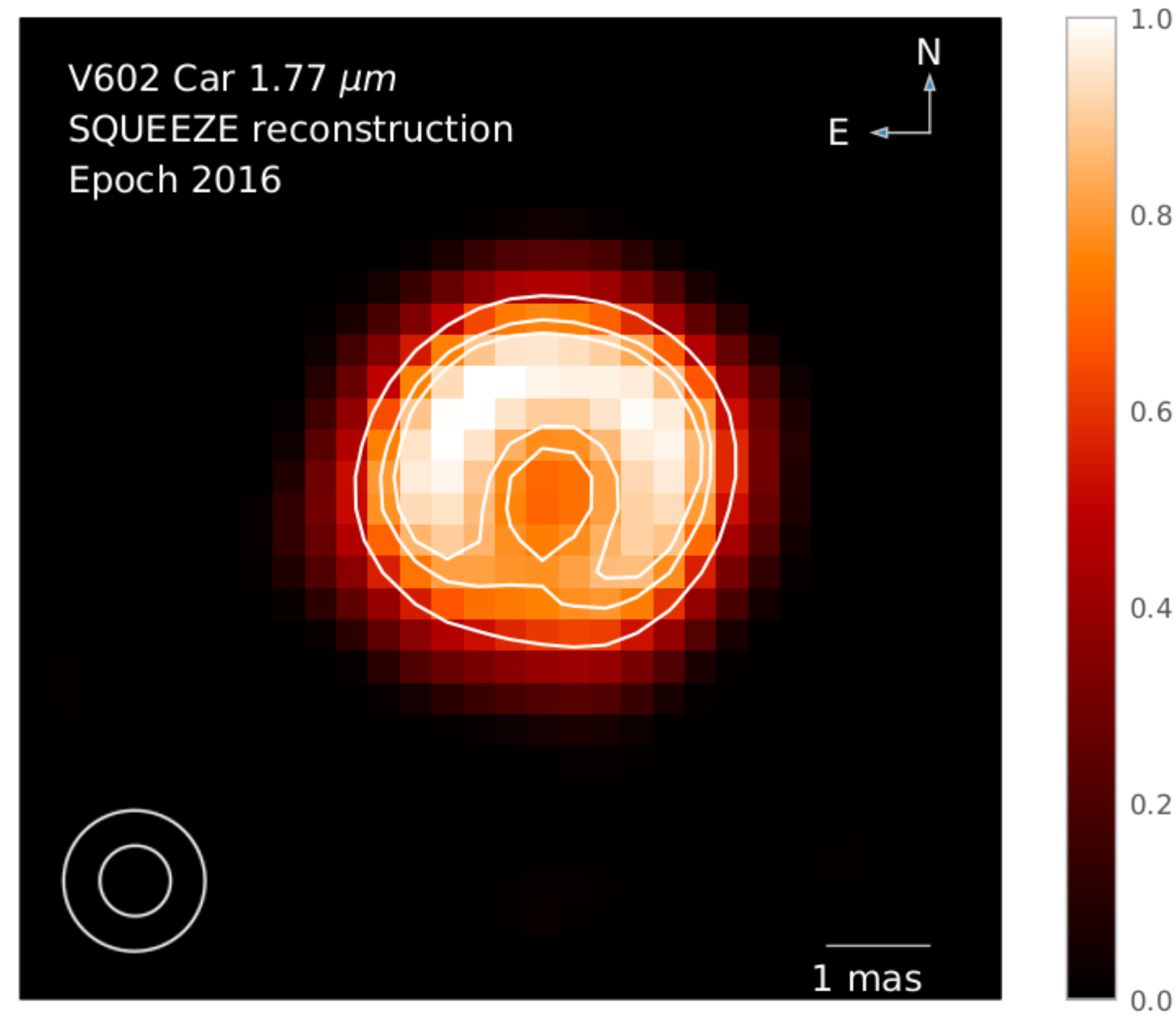}
    \caption{SQUEEZE reconstruction of 2016 V602 Car data at each spectral channel.
    Contours are drawn at levels 55\%, 77,\% and 85\% of the peak intensity.
    }
    \label{fig:2016_channels}
\end{figure*}

\begin{figure*}[h]
    \centering
    \includegraphics[width=0.49\linewidth]{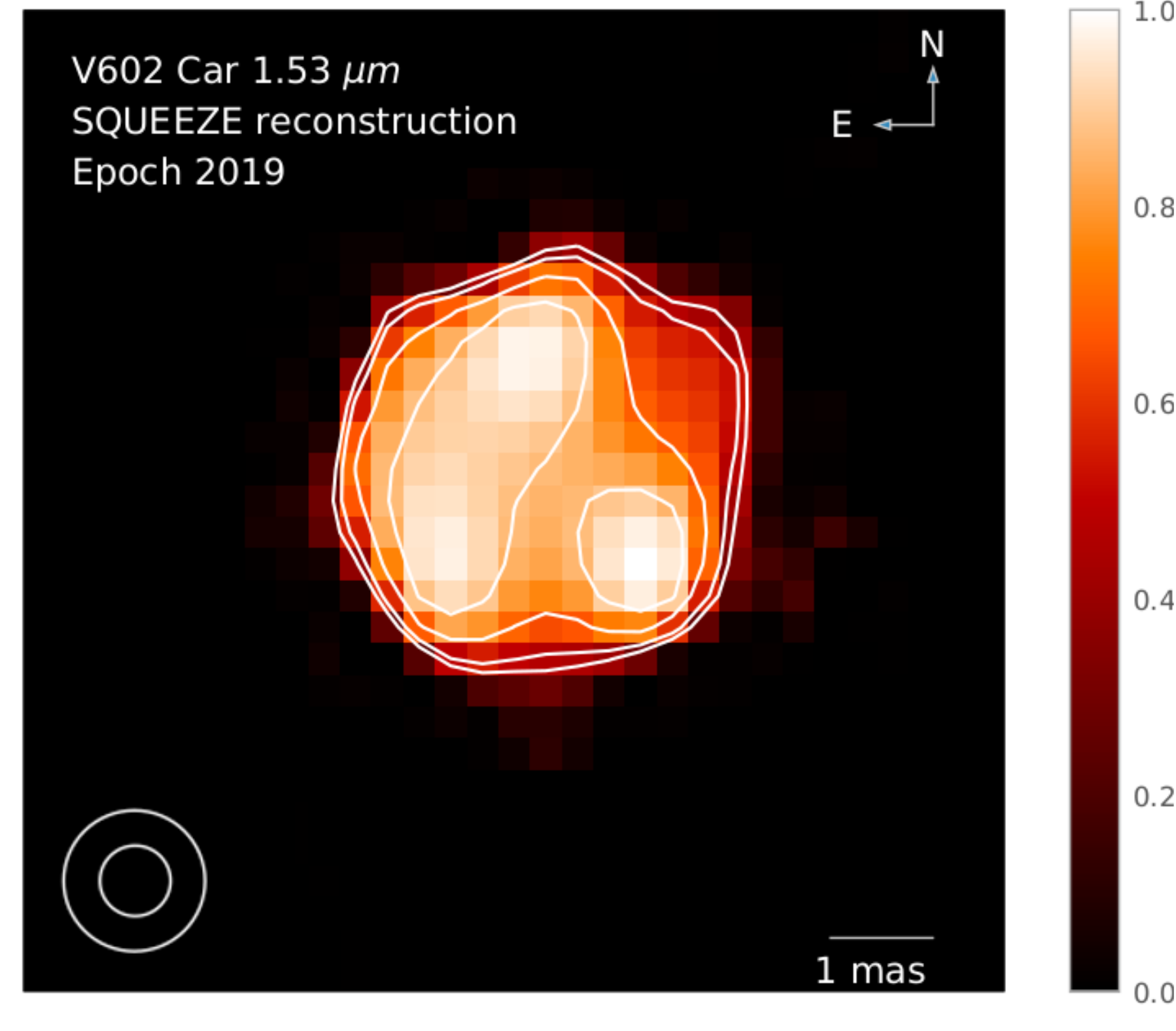}
    \includegraphics[width=0.49\linewidth]{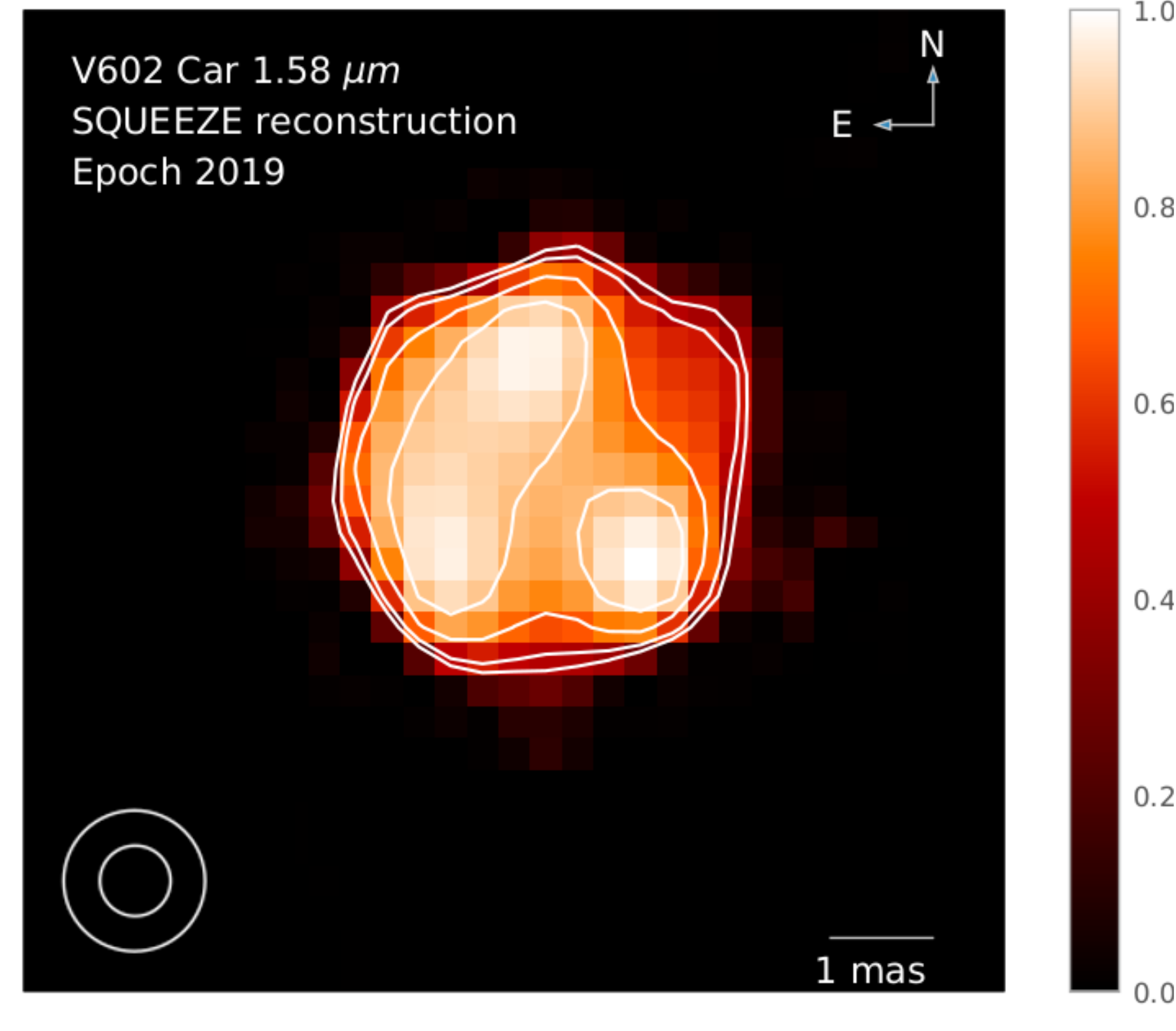}
    \includegraphics[width=0.49\linewidth]{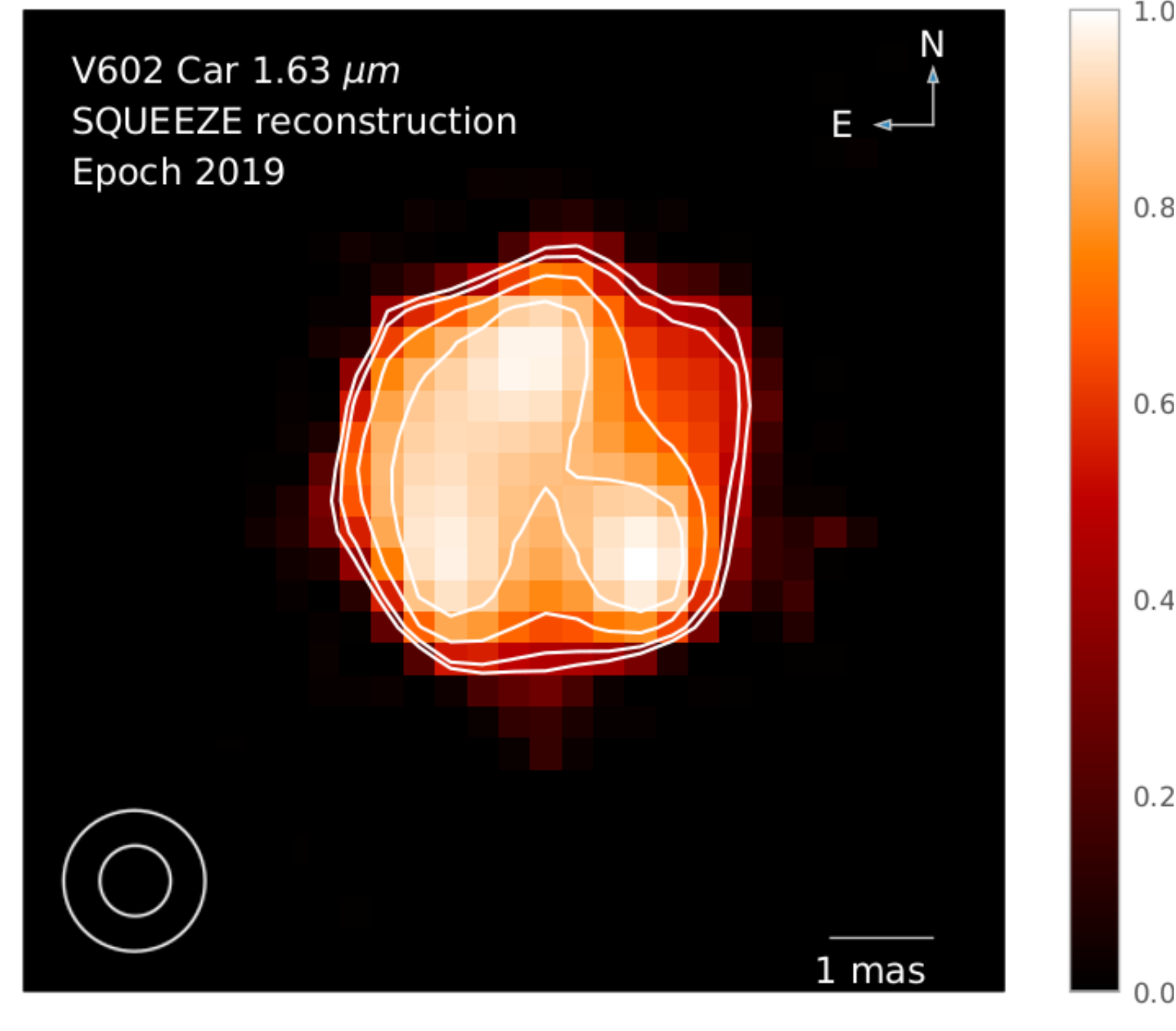}
    \includegraphics[width=0.49\linewidth]{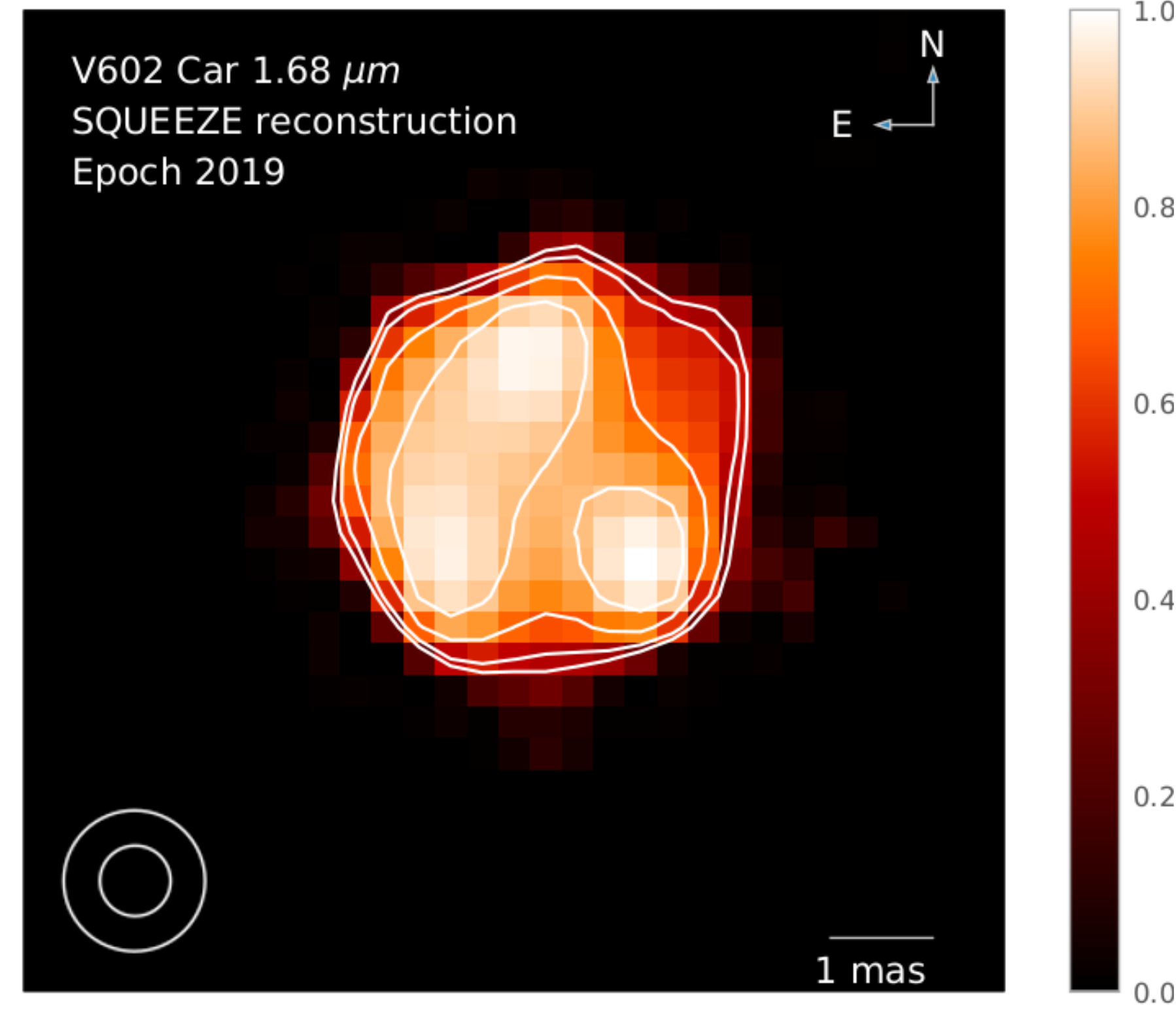}
    \includegraphics[width=0.49\linewidth]{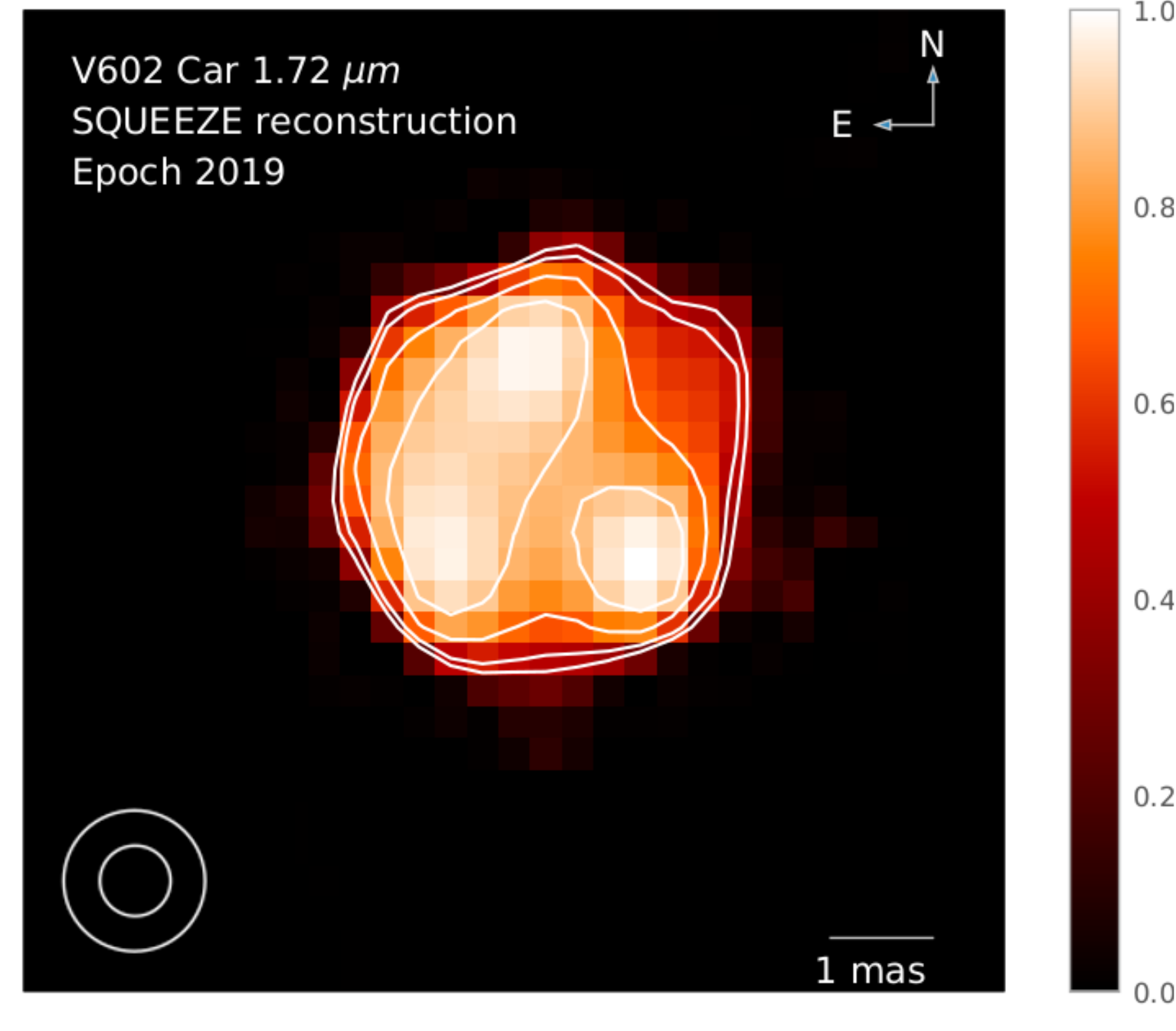}
    \includegraphics[width=0.49\linewidth]{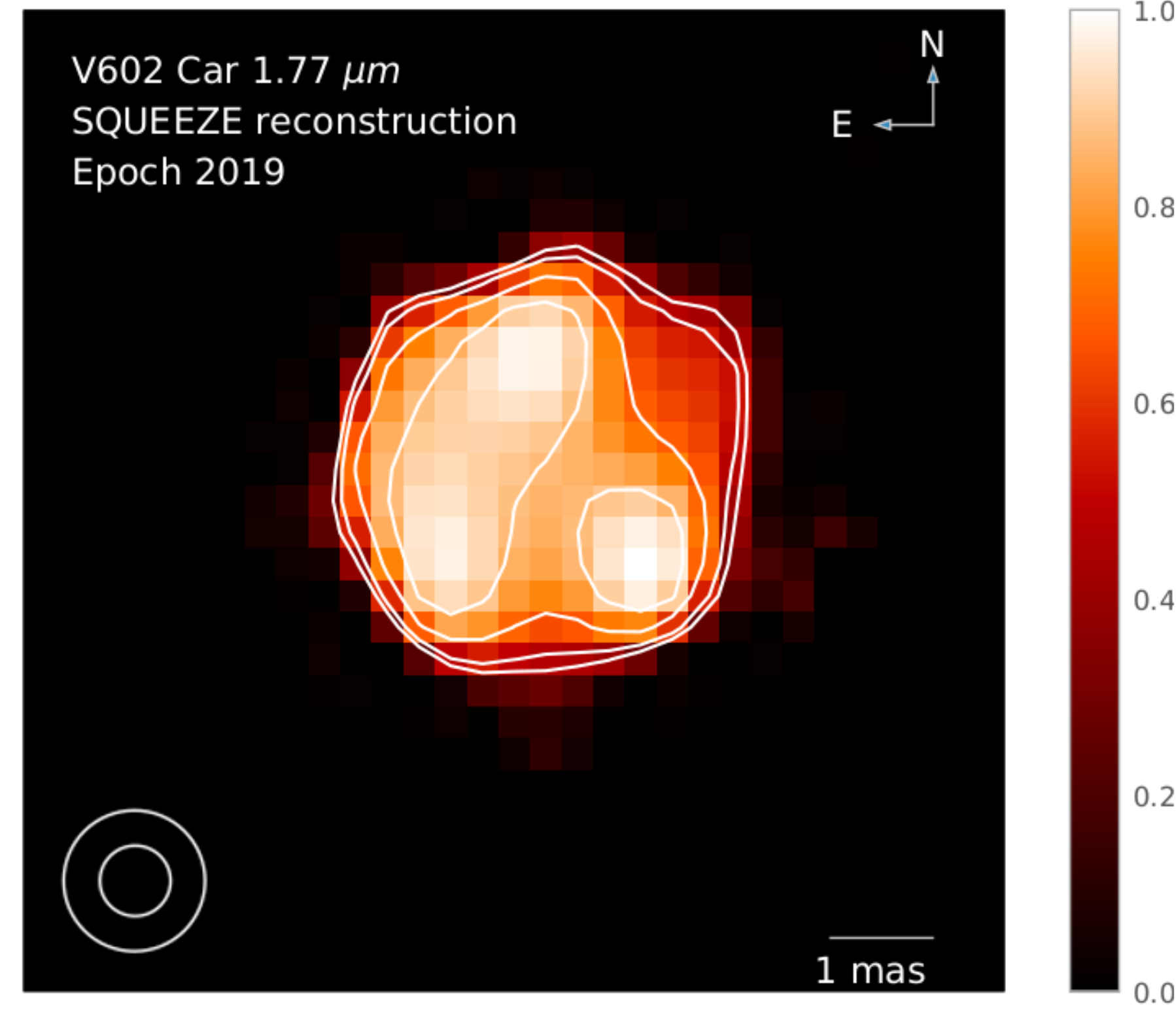}
    \caption{SQUEEZE reconstruction of 2019 V602 Car data at each spectral channel.
    Contours are drawn at levels 40\%, 50\%, 70\%, and 87\% of the peak intensity.
    }
    \label{fig:2019_channels}
\end{figure*}

\section{Mathematical definition of the SSIM}

When calculated on various image windows ($x$ and $y$) of the same size (N$\times$N), the SSIM index is computed:

    \begin{equation}
    \hbox{SSIM}(x,y) = \frac{(2\mu_x\mu_y + c_1)(2\sigma_{xy} + c_2)}{(\mu_x^2 + \mu_y^2 + c_1)(\sigma_x^2 + \sigma_y^2 + c_2)}
    \label{SSIM_formula}
    ,\end{equation}
    
where

$\mu_x$ is the average value of x;

$\mu_y$ is the average value of y;

$\sigma_x^2$ is the variance of x;

$\sigma_y^2$ is the variance of y;

$\sigma_{xy}$ is the covariance of x and y;

$c_1$ = $(k_1L)^2$, $c_2$ = $(k_2L)^2$ are two variables to stabilize the division with weak denominator;

$L$ represents the dynamic range of the pixel values and is determined by the number of levels of luminance per pixel.

By default, $k_1$ = 0.01 and $k_2$ = 0.03.

\end{appendix}
\end{document}